%% file: prx.tex
\documentclass[aps,prx,twocolumn,superscriptaddress]{revtex4-2}
\usepackage{amsmath,amsthm,amssymb,pifont}
\usepackage[utf8]{inputenc}
\usepackage[T1]{fontenc}
\usepackage[american]{babel}
\usepackage{graphicx,bbold}
\usepackage{braket}
\usepackage{MnSymbol}
\usepackage{hyperref}
\usepackage[dvipsnames]{xcolor}
\definecolor{BlueRome}{HTML}{4287f5}
\definecolor{C1}{RGB}{52, 89, 149}
\definecolor{C2}{RGB}{251, 77, 61}
\definecolor{C3}{RGB}{3, 206, 164}
\definecolor{C4}{RGB}{202, 21, 81}
\hypersetup{colorlinks=true, linkcolor=BlueRome, citecolor=BlueRome, urlcolor=BlueRome}

\usepackage{tikz}
\usepackage[export]{adjustbox}
\usepackage{enumerate}
\usepackage{orcidlink}
\usepackage[normalem]{ulem}
\usepackage{tikzit}

\input{zh.tikzstyles}
\definecolor{zx_green}{HTML}{C1E1C1}
\definecolor{zx_red}{HTML}{FFD1DC}
\definecolor{zx_yellow}{HTML}{FDFD96}

\usepackage{multirow}

\definecolor{markus}{HTML}{006600}

\newtheorem*{thm*}{Theorem}

\theoremstyle{remark}

\newcommand*{\OTOC}{\mathcal{C}}
\newcommand*{\ad}{\mathtt{A}}

\newcommand*{\nn}{\nonumber}

\newcommand*{\id}{\mathbb{1}}

\newcommand*{\mc}{\mathcal}

\DeclareMathOperator{\tr}{tr}

\newcommand{\dist}{\ell} 

\newcommand*{\scalefactor}{1.3}

\hypersetup{
pdftitle={Unitary designs beyond the diagonal approximation},
pdfsubject={Many-body physics, quantum information theory},
pdfauthor={Neil Dowling, Silvia Pappalardi},
pdfkeywords={Many-body physics, quantum information theory}
}

\begin{document}
\title[]{Free Independence and Unitary Design from Random Matrix Product Unitaries}

\author{Neil Dowling}
\email[]{ndowling@uni-koeln.de}
\affiliation{Institut f\"ur Theoretische Physik, Universit\"at zu K\"oln, Z\"ulpicher Strasse 77, 50937 K\"oln, Germany}

\author{Jacopo De Nardis}
\affiliation{Laboratoire de Physique Th\'eorique et Mod\'elisation, CNRS UMR 8089,
CY Cergy Paris Universit\'e, 95302 Cergy-Pontoise Cedex, France}

\author{Markus Heinrich}
\affiliation{Institut f\"ur Theoretische Physik, Universit\"at zu K\"oln, Z\"ulpicher Strasse 77, 50937 K\"oln, Germany}

\author{Xhek Turkeshi}
\affiliation{Institut f\"ur Theoretische Physik, Universit\"at zu K\"oln, Z\"ulpicher Strasse 77, 50937 K\"oln, Germany}

\author{Silvia Pappalardi}
\affiliation{Institut f\"ur Theoretische Physik, Universit\"at zu K\"oln, Z\"ulpicher Strasse 77, 50937 K\"oln, Germany}

\begin{abstract}  
Unitary randomness underpins both fundamental tasks in quantum information and the modern theory of quantum chaos. On one side, a central concept is that of approximate unitary designs: circuits that look random according to small moments and for forward-in-time protocols. In a distinct setting, out-of-time-ordered correlators (OTOCs), intensely studied as a measure of information scrambling, have recently been shown to probe freeness between Heisenberg operators, the noncommutative generalization of statistical independence. Bridging these two concepts, we study the emergence of freeness in a random matrix product unitary ensemble. We prove that, with only polynomial bond dimension, these unitaries reproduce Haar values of higher-order OTOCs for local, finite-trace observables, while traceless observables instead require exponential resources. Indeed, local observables are precisely those predicted to thermalize in chaotic many-body systems according to the eigenstate thermalization hypothesis. Moreover, adding to previous literature, we show how random matrix product unitaries constitute approximate designs: we exactly compute the frame potential of the ensemble, showing convergence to the Haar value with polynomial deviations and so indicating that global observables are freely independent on-average. Our results highlight the need to refine previous notions of unitary design in the context of operator dynamics, guiding us towards protocols for quantum advantage and shedding light on the emergent complexity of chaotic many-body systems.
\end{abstract}

\maketitle

\section{Introduction} 


Understanding when the unitary evolution of an isolated quantum system mimics that of a random unitary
is of foundational and practical importance across several fields of physics. In many-body quantum dynamics, uniformly distributed or Haar-random unitaries serve as a universal benchmark for the complexity of unitary evolution. This perspective, ultimately rooted in Berry’s conjecture that eigenstates of chaotic Hamiltonians resemble random vectors~\cite{BerryTabor1977,bgs1984,Haake2018-cs}, captures key features of quantum thermalization~\cite{dymarsky2016subsystem,huang2019universal,
murthy2019structure,ho2022exact,
cotler2023emergent,
lucas2022generalized, mark2024maximum} and state delocalization~\cite{mace2019multifractal,backer2019multifractal,claeys2022emergent}. Concepts of quantum randomness also find utility for foundational problems in high-energy physics, where random unitaries provide toy models for the information scrambling of black holes~\cite{Hayden_Preskill_2007,Cotler2017,Maldacena2016-rl,cotler2017chaos}. Finally, in quantum information, random circuits underpin a broad range of applications, from device characterization and learning protocols~\cite{Emerson_2005,Knill2008,Huang_2020,elben_randomized_2022,heinrich2023randomiz}, to quantum cryptography~\cite{Ji_2018,Ananth2022,kretschmer2023quantumcr} and complexity theory~\cite{roberts2017chaos,Brandao2021,Haferkamp2022}, with random sampling offering a promising path toward quantum computational advantage~\cite{Boixo2018,Arute2019,bouland_complexity_2019,Morvan2024}.

In general, assessing whether an ensemble of unitaries is genuinely random requires analyzing its moments. This has motivated the study of \emph{approximate unitary $k$-designs}, ensembles that reproduce the correlations of the unitary Haar measure up to order $k$ within a controlled error~\cite{Emerson2003-fk,Gross_2007}; several inequivalent metrics have been proposed to quantify this error~\cite{low2010pseudorando}. 
Recent work focusing on so-called relative-error designs has shown that random circuits can reproduce Haar statistics of wave-function sampling in only logarithmic depth~\cite{schuster2024randomun,laracuente_approximate_2024,Lami2025,DeLuca2024Frame}. However, this approach applies only to the case of \emph{forward-in-time protocols}, missing the richer structure of operator correlations that comes with access to both forward and backward evolution.

\begin{figure*}
    \centering
    \includegraphics[width=0.85\linewidth]{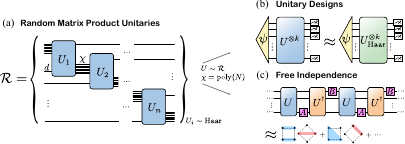}
    \caption{A depiction of the introduced random matrix ensemble and its properties. (a) A matrix product unitary on some $D=d^N$ dimensional system is constructed as a staircase of $n$ unitaries $\{ U_i\}_{i=1}^n$ overlapping on a space of (bond) dimension $\chi$. The random matrix product unitary (RMPU) ensemble $\mc{R}$ is generated by sampling each $U_i$ independently from the Haar measure on the unitary group, for a given $n$ and $\chi$. (b) Previous work shows that for polynomial bond dimension $\chi = \mathrm{poly}(N)$, this ensemble is a relative error unitary design~\cite{schuster2024randomun,laracuente_approximate_2024} and exhibits anticoncentration in the computational basis~\cite{Lami2025}. Roughly speaking, these results mean that the Haar distribution of (possibly correlated) measurements of forward-in-time evolving states are well approximated by the ensemble $\mc{R}$. We extend these results and prove that also $\mc{R}$ is a unitary designs according to the frame potential. (c) We prove that polynomial bond dimension $\chi$ also leads to free independence for local, finite-trace observables, as characterized by the non-crossing partitions. Freeness is witnessed by out-of-time-ordered experiments [Eqs.~\eqref{eq:otoc1}-\eqref{eq:otoc}] and is thus not accounted for by previous results on unitary designs. The red and blue non-crossing partition diagrams here denote the partitioned moments of $A$ [Eq.~\eqref{eq:fact_moms}] and free cumulants of $B$ [Eq.~\eqref{eq:mom-cum}], respectively. See Table~\ref{tab:example} for precise results. 
    }
    \label{fig:rmpu}
\end{figure*}

A complementary perspective is to examine the action
of random unitaries on operators, $ A_U = U^\dagger A U$, and the structure of their higher-order correlations.
In this context, an emblematic object which captures \emph{backward-in time protocols} is the \emph{out-of-time-ordered correlator} (OTOC)~\cite{Shenker2014,roberts2017chaos,cotler2017chaos, tsuji2018bound,cotler_spectral_2020,leone2021quantum}, whose $ 2k $-point form reads
\begin{equation}
\braket{(A_UB)^k} := 
\frac{1}{D} \tr[A_U B \cdots A_U B] \ , \label{eq:otoc1}
\end{equation}
with $A$ and $B$ Hermitian operators in a $D$-dimensional Hilbert space. 
In many-body physics, OTOCs are naturally understood within the full Eigenstate Thermalization Hypothesis (ETH)~\cite{Deutsch1991,Srednicki,Foini2019}, which encode the statistical properties of physical observables in the energy eigenbasis and reveal thermal behavior on various scales~\cite{ foini2019eigenstate2,chan2019eigenstate, murthy2019bounds,richter2020eigenstate, kaneko2020characterizing, wang2021eigenstate,jafferis2022matrix}.
Furthermore, higher-order correlation functions probe quantum information scrambling~\cite{Shenker_Stanford_2014, Maldacena2016-rl, Nahum2018,Keyserlingk2018,dowling2023scrambling}, nonlinear hydrodynamics~\cite{Schweigler2017, Delacretaz2024,wang2025eigenstatethermalizationhypothesiscorrelations}, and generalized quantum Lyapunov exponents \cite{Pappalardi2023quantum, trunin2023quantum, trunin2023refined}, while underpinning protocols to validate quantum advantage~\cite{abanin2025const}.

Recently, OTOCs have been reinterpreted through the lens of {free probability}~\cite{Pappalardi2022-vo}, where it has been noticed that their ensemble-average quantifies \textit{free independence} between operators~\cite{Voiculescu1991, speicher2009free}, the non-commutative analogue of classical statistical independence~\cite{nica2006lectures}. More technically, in the asymptotic limit of large Hilbert-space dimension, two distributions of operators are said to be freely independent (or simply free) if their mixed moments over a given unitary ensemble $\mc{E}$ (i.e., averaged OTOCs), 
\begin{equation}
    {\OTOC}^{(k)}_{\mathcal{E}}(A_U,B) := \int_{U \sim \mathcal{E}} \braket{(A_UB)^k}, \label{eq:otoc}
\end{equation}
factorize in a characteristic combinatorial way governed by \textit{non-crossing partitions}~\cite{speicher1997free,simion2000noncrossing,nica2006lectures}
\begin{equation}
    \lim_{D \to \infty} {\OTOC}^{(k)}_{\mathcal{E}}({A}_U,{B}) = {\OTOC}^{(k)}_{\mathrm{FP}}\left(\braket{A^x},\braket{B^y}\right)\ ,
    \label{eq:freeOTOC}
\end{equation}
where $1 \leq x,y\leq k$ are integers and $\braket{A^x}$, $\braket{B^y}$ are normalized expectation values as in Eq.~\eqref{eq:otoc1}. The function $\OTOC^{(k)}_{\mathrm{FP}}$ is a universal form derivable from free probability, which depends only on the separate normalized moments of $A$ and $B$~\cite{SilviaDesigns}. We define it explicitly in Sec.~\ref{sec:methods}, with its structure depicted in Figs.~\ref{fig:rmpu}(c) and~\ref{fig:nc_lattice}. A defining example is given by Haar-random unitaries, which render Heisenberg operators asymptotically free as the dimension grows~\cite{Voiculescu1991}. Intriguingly, signatures of freeness also emerge dynamically in chaotic many-body systems, where it has been argued that long-time evolution under chaotic Hamiltonians drives local operators toward freeness, reflecting a deep connection between the full ETH and free probability~\cite{Pappalardi2022-vo}. This fact has sparked interest in many-body physics~\cite{wang2023emergence,PappalardiProsen2025, alves2025probes}. To date, the emergence of freeness studied in random matrices~\cite{Voiculescu1991,cipolloni2022thermalisation,camargo2025quant,jahnke2025free}, in few-body systems~\cite{vallini2025longtime}, and in fine-tuned quantum circuit models~\cite{chen2024freeindependen,fritzsch2025freeprob}. Yet important questions remain: \textit{How does freeness relate to established notions of approximate unitary designs? What are the mechanisms for its emergence for finite time and length-scales, particularly in the presence of locality? }\\

\begin{table*}
    \begin{center}
       \begin{tabular}{ c|c|c } 
    \textbf{Quantity} & \textbf{Asymptotic value for RMPU ensemble} & \textbf{Leading order approximation} \\ 
    \hline
    \multirow{3}{*}{Average OTOCs [Eq.~\eqref{eq:multi2}]}
    & 
    \multirow{3}{*}{
    \(
    \Delta \OTOC^{(k)}_{\mc{R}}(A_U,B) =
    \begin{cases}
    \mc{O}(N\, \chi^{-2}), & \text{ for $\delta \leq \{\tr[A],\tr[B]\} \leq D$,} \\
    \mc{O}(d^{2N}\chi^{-4}), & \text{ for $ \tr[A] = \tr[B] =0$,} \\
    \mc{O}(1), & \text{ for $A \equiv A \otimes \dots A$.}
    \end{cases}
    \)
    }
    & \multirow{3}{*}{Free probability [Eq.~\eqref{eq:fp_channel}]} \\ 
    & & \\ 
    & & \\ 
    \hline
    Frame potential [Eq.~\eqref{eq:framepotcalc}] 
    & $\Delta \mathcal{F}^{(k)}_{\mc{R}} ={k(k-1) 2^{-1} }( (N-r)(1-d^{-2})-1  ) \chi^{-2} + \mc{O}(\chi^{-3} )$.  
    & \multirow{3}{*}{Diagonal [Eq.~\eqref{eq:diag}]} \\
    \cline{1-2}
    Anticoncentration (Ref.~\cite{Lami2025}) & $\Delta I_{\mc{R}}^{(k)} = N k (k-1) (d-1)(2d)^{-1} \chi^{-1} + \mc{O}(\chi^{-2} )$. & \\
    \cline{1-2}
    Relative error design (Ref.~\cite{schuster2024randomun}) & $\Delta \tr[\Phi^{(k)}_{\mc{R}}(X)Y ] \leq \mc{O}( N k^2  \chi^{-1}), \quad \text{for } X,Y\geq 0$.  &
\end{tabular}
    \end{center}
    \caption{A summary of our results in comparison to previous work for the random matrix product unitary (RMPU) ensemble $\mc{R}$. The notation``$\Delta_{\mc{R}}$'' in each result refers to the relative multiplicative error in the quantity upon averaging over ${\mc{R}}$ compared to averaging over the Haar ensemble $\mathbb{H}$: $\Delta X_{\mc{R}} := |(X_{\mc{R}}-X_{\mathbb{H}})/X_{\mathbb{H}} |$. The construction of $\mc{R}$ is depicted in Fig.~\ref{fig:rmpu}, and formally defined in Sec.~\ref{sec:rmpu}. The relevant approximation method underlying each result is detailed in the final column, with the diagonal and free probability asymptotics of random unitaries detailed in Sections~\ref{sec:diag} and \ref{sec:fp} respectively. The first row summarizes the results of Sec.~\ref{sec:loRMPU} and Sec.~\ref{sec:subleading}: the value of the ensemble-averaged OTOC [Eq.~\eqref{eq:otoc}] for various classes of observables, while the second row summarizes the results of Sec.~\ref{sec:frame} on the frame potential of ${\mc{R}}$ [Eq.~\eqref{eq:framepotMain}]. In each result for the OTOCs, the shown scaling with $\chi=d^r$ and $N$ is proven analytically, while the scaling with $k$ is dependent on the (normalized) moments of the operators $A$ and $B$. 
    The final two rows on the design properties of the RMPU ensemble are supplied from previous work for comparison. Namely, in the third row we present the $k^{th}$ average inverse participation ratio~\cite{Lami2025}, $I_{\mc{E}}^{(k)}:=D \sum_{x} \int_{U\sim \mathcal{E}} |\braket{x|U|0}|^{2k}$ with $x$ iterating over the computational basis, which measures the phenomena of anticoncentration over the ensemble $\mc{E}$. The final line expresses that $\mc{R}$ forms a relative error design~\cite{schuster2024randomun}, where $A \geq 0$ means $A$ is positive semidefinite. Note that the shown expression is a necessary implication of the somewhat stronger notion of relative error design; see App.~\ref{ap:a}.  }
    \label{tab:example}
\end{table*}

In this work, we introduce the ensemble of \emph{random matrix product unitaries} (RMPUs). This ensemble, denoted by $ \mathcal{R} $ and illustrated in Fig.~\ref{fig:rmpu}(a), is defined on $N$ qudits through a staircase circuit of $n = N-r$ independent random unitaries, overlapping on a Hilbert space of bond dimension $ \chi = d^r$.
When applied to product states, RMPUs generate the so-called random matrix product state ensembles~\cite{Garnerone2009,Garnerone2010,Haag2023}, which have recently been shown to approximate Haar-random states with only $\chi = \mathrm{poly}(N)$ resources~\cite{Lami2025,cheng2025pseudoentanglement}. Here we show that the RMPU ensemble uncovers also physical content in the form of OTOCs. In particular, \emph{RMPUs with polynomial bond dimension generate freeness for relevant classes of operators}. 
More specifically, in Sec.~\ref{sec:rmpu} we compute OTOCs averaged over both $ \mathcal{R} $ and the Haar measure $ \mathbb{H} $, finding agreement up to corrections of size $\mc{O}(N\chi^{-2})$ for local observables with non-zero trace. The RMPU ensemble therefore reflects the central role of locality in quantum thermalization (i.e., randomization), akin to ETH-abiding Hamiltonian systems. The scaling of the bond dimension required to suppress the error in Eq.~\eqref{eq:otoc} matches that needed to generate unitary designs, despite fundamentally different mechanisms. 
Our proofs rely on the non-crossing partition structure of free probability, going beyond the so-called \emph{diagonal approximation} in the algebraic structure of unitary ensembles which is at the core of recent unitary design results~\cite{schuster2024randomun,Lami2025}. 
In contrast, for traceless operators we show that reproducing the Haar value of OTOCs up to small multiplicative error requires exponentially large bond dimension $\chi=O(\exp(N))$ in the RMPU. This difference can be seen as stemming from the fact that Haar-random OTOCs are exponentially small for traceless observables, and so require a high-complexity unitary to reproduce. 
This highlights a key insight: any observable probing design properties beyond polynomial bond dimension necessarily also captures corrections beyond the {diagonal approximation}, thereby becoming sensitive to the refined structure of the unitary ensemble.

To probe the emergence of freeness beyond local observables, we also compute the \emph{frame potential} of the RMPU ensemble~\cite{Bisio2009,roberts2017chaos}. In contrast to other, inequivalent measures of approximate designs (reviewed in App.~\ref{ap:a}), the frame potential captures an average OTOC over all operators~\cite{roberts2017chaos}. This measure can therefore be interpreted as an average-case probe of freeness. Through exact computation up to the leading orders in large $\chi$, we find that the frame potential for $\mc{R}$ is approximately equal to the Haar value, also with an error of $\mathcal O(N\chi^{-2})$. This result therefore resolves an open question for whether frame potential designs can be reached in an equivalently `extremely' low depth compared to relative error designs, while simultaneously showing that RMPUs lead to approximate freeness not only in specific settings, but also typically. A summary of our main results is presented in Table~\ref{tab:example}, along with previous relevant results~\cite{schuster2024randomun,Lami2025}. 

In summary, our results show that when considering correlations that stem from both forwards and backwards evolution in a many-body system, higher-order characteristics of randomness emerge, encoded in Heisenberg operators and predicted by free probability. This perspective is orthogonal to traditional approaches in characterizing unitary randomness, which implicitly prioritize forward-in-time correlations and quantum states. We take the first steps toward integrating these approaches. Our results lead to natural questions regarding both the time scales of universality in the random matrix behavior of chaotic systems and the source of advantage in quantum algorithms. We discuss these points and other avenues in Sec.~\ref{sec:discussion}.

\section{From free probability to Weingarten Calculus and Back}
\label{sec:methods}
This section lays the groundwork for the analytical framework employed throughout this work. Namely, we first explain the precise meaning of free independence between distributions of observables, before introducing the Weingarten calculus for Haar integrals over the unitary group~\cite{collins_integration_2006,Mele_2024,th2025}. Relating these two fields, we discuss the asymptotic structure of the Weingarten calculus, where free probability governs the behavior of higher-order correlations, cf. Eq.~\eqref{eq:otoc}~\cite{nica2006lectures,SilviaDesigns}.

\subsection{Free Independence}
\label{sec:free}
\begin{figure}
    \centering
    \includegraphics[width=0.9\linewidth]{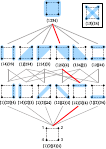}
    \caption{The non-crossing partition lattice (\emph{Hasse diagram}) for $k=4$. A diagrammatic representation of the integer partition is shown together with the corresponding element of the permutation group $S_k$ in cyclic notation. Solid lines connecting partitions indicate a unit distance on the lattice, i.e. permutations related via a single transposition, or equivalently integer partitions related by refinement. A multichain is defined as an ordered set of permutations lying on a path from the bottom to the top, satisfying the geodesic condition $\pi_1 \leq \sigma_1 \leq \dots \leq \gamma$ as detailed around Eq.~\eqref{eq:geodesic}. An example geodesic in shown in bold [red], where for instance, $\pi=(12)(3)(4)$ and $\sigma=(123)(4)$ satisfy the $2$-chain condition $\pi \leq \sigma \leq \gamma $. Inset: in the top right the one crossing partition of $k=4$ elements is given. Cumulants corresponding to this crossing partition do not appear in the leading order Haar-twirled OTOC [Eq.~\eqref{eq:loOTOC}]. }
    \label{fig:nc_lattice}
\end{figure}

To understand the concepts behind free probability, we must first explain some basics on the combinatorics of \emph{non-crossing partitions} of $k$ elements, denoted $NC(k)$. These are defined as the set of partitions whose blocks do not cross when drawn diagrammatically in a loop. For instance, the partitions 
\begin{equation}
(123) (45) ( 6)  =\includegraphics[scale=3.7, valign=c]{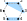}, \; \;  (13) (25) (4 ) (6)=\includegraphics[scale=3.7, valign=c]{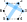} \nn
\end{equation}
are non-crossing and crossing, respectively. In Fig.~\ref{fig:nc_lattice} we show all the non-crossing partitions of \( k = 4 \), with the single crossing partition of $4$ elements given in the top right inset. Note that, as presented in this figure, the non-crossing partitions are isomorphic to a relevant set of permutations which appear throughout this work and which will be discussed in Sec.~\ref{sec:fp}. Non-crossing partitions enjoy elegant combinatorial properties~\cite{simion2000noncrossing}. For instance, they are enumerated by the Catalan numbers $C_k:=\frac{1}{k+1} \binom{2k}{k}$ and are characterized by partial ordering: two non-crossing partitions $\pi, \sigma \in NC(k)$ obey $\pi \leq \sigma$ if all the blocks of $\pi$ are contained in the blocks of $\sigma$. So, for example, $(12)(3)(4) \leq (123)(4)$. This leads to the lattice structure in Fig.~\ref{fig:nc_lattice}. Moreover, each partition \( \pi \in NC(k) \) admits a unique dual partition \( \pi^* \in NC(k) \), known as its \emph{Kreweras complement}. $\pi^*$ can be determined diagrammatically from $\pi$~\footnote{In this method, one first places new vertices at the center of each edge, and then maximal partitions are drawn without crossing any of the original partition boundaries. For instance, two examples of Kreweras complements are given diagrammatically to the right [in red] of their duals [in blue] in Fig.~\ref{fig:rmpu} (c). See e.g. Refs.~\cite{SilviaDesigns,fritzsch2025freeprob}.}, and we will see a different way to construct them in Sec.~\ref{sec:fp} when viewing $\pi$ as a permutation~\footnote{A partition can be mapped to a permutation through interpreting its elements as cycles, ordered according to the canonical cyclic permutation $\gamma = (12 3\dots k)$. So, for example, the partition $\pi=(12)(3)(4)$ corresponds to the permutation in four replica space which acts as a SWAP on the first two replicas and identity on the second two.}.

The above language allows us to compactly express the concept of free independence. Statistical independence between classical random variables allows one to compute mixed moments from the individual moments of each variable, stemming from simple factorization rules. Free probability extends this idea to non-commuting variables. In this setting, moments of distributions of matrices are defined by normalized expectation values, $\int \langle \cdot \rangle := \frac{1}{D}   \int \operatorname{tr}[\cdot]$, averaged over the relevant distribution(s); cf. Eq.~\eqref{eq:otoc}. Then, two distributions of operators $A$ and $B$ are said to be \emph{freely independent}, or simply \emph{free}, if their mixed moments satisfy the factorization rule~\footnote{At the lowest orders this reads $ \int \braket{A B} = \braket{A}\braket{B}$ and $\int  \braket{ABAB} =  \braket{A^2}\braket{B}^2 + \braket{A}^2\braket{B^2}  -\braket{A}^2\braket{B}^2$. Equivalently, two non-commutative random variables are free if their mixed free cumulants vanish (not defined here)~\cite{nica2006lectures,SilviaDesigns}.
}
\begin{equation}
\label{eq_free_def}
    \int \left\langle (AB)^k \right\rangle = \sum_{\pi \in NC(k)} 
    \langle A,\dots,A \rangle_{\pi^*} \, \kappa_\pi(B, \dots, B)\,
    =: \OTOC^{(k)}_{\mathrm{FP}},
\end{equation}
which defines the right-hand side of Eq.~\eqref{eq:freeOTOC}. In the above expression, $\langle \cdot \rangle_{\pi^*}$ are partitioned moments, defined on the partition's blocks 
\begin{equation}
    \langle A,\dots,A \rangle_{\pi^*} := \prod_{b\in \pi^*} \langle A^{|b|} \rangle, \label{eq:fact_moms}
\end{equation}
and \( \kappa_\pi \) are the \emph{free cumulants}. The latter are defined recursively through the moment-cumulant relation:
\begin{equation}
    \langle B^{ k} \rangle =: \sum_{\pi \in NC(k)} \kappa_\pi(B,\dots,B)\,,\label{eq:mom-cuma}
\end{equation}
which can be inverted using the corresponding M\"obius function (discussed in detail in Sec.~\ref{sec:fp}):
\begin{equation}
    \kappa_\pi(B, \dots, B) = \sum_{\sigma \le \pi}  \langle B,\dots,B \rangle_\sigma \mu(\sigma, \pi)\,. \label{eq:mom-cum}
\end{equation}
The above definitions readily generalize to arbitrary sets of distributions of operators, $\{A^{(i)}\} $ and $\{B^{(i)}\}$~\cite{nica2006lectures}, which is relevant to our results on the frame potential in Sec.~\ref{sec:frame}. As we generally consider the simple case of a single $A$ and $B$, we also adopt the shorthand $\langle A,\dots,A \rangle_{\pi^*} =: \langle A \rangle_{\pi^*}$. 

The definition of freeness may appear artificial, but is in fact a recurring feature of sufficiently large and random matrices.
For example, two independent distributions of Gaussian random matrices are free for $D\to \infty$, and this fact can be used to prove Wigner’s semicircle law~\cite{Voiculescu1991,nica2006lectures}. Most pertinent to our analysis, a Haar rotated distribution of matrices $\{ A_U \}_{U \sim \rm \mathbb{H}}$ is asymptotically free compared to \textit{any} constant set of matrices $B$, such that the left-hand side of Eq.~\eqref{eq_free_def} coincides with the average OTOC, Eq.~\eqref{eq:otoc}~\cite{Voiculescu1991}. In the next sections, we will review how Weingarten calculus can be used to arrive at this result.

\subsection{Weingarten Calculus}
Consider a Hilbert space $\mathcal{H}=\mathrm{span}\{|x\rangle\;|\;x=0,\dots,D-1\}$ of total dimension $D = d^N$. 
Denoting with $S_k$ the symmetric group on $k$ elements, we define $T_\pi|x_1,\dots,x_k\rangle=|x_{\pi^{-1}(1)},\dots,x_{\pi^{-1}(k)}\rangle$ the unitary representation of $\pi\in S_k$ acting on $k$ replicas of Hilbert space, $\mathcal{H}^{\otimes k}$. 
We define $e \in S_k$ as the identity element and $\gamma \in S_k$ to be the canonically ordered cyclic permutation.
In the following, we occasionally use the cyclic notation for permutations, with $e = (1)(2)\dots(k)$ and $\gamma = (1 2 \dots k)$. 
Relevant to the asymptotic analysis employed throughout this work, the symmetric group $S_k$ admits a natural metric, called the \textit{Cayley distance},
\begin{equation}
    \dist(\mu, \nu):= k-\#(\mu^{-1} \nu), \label{eq:cayley}
\end{equation}
where $\#(\pi)$ denotes the number of cycles of the permutation $\pi$, for instance: $\#((123)(4)) = 2$ and $\#(e) = k$. The Cayley distance quantifies the minimal number of transpositions (i.e., SWAP operations) needed to transform $\mu$ into $\nu$, and it satisfies standard metric properties such as the triangle inequality. 
An important subset of permutations, relevant also to our later analysis in Sec.~\ref{sec:fp}, are the ones that lie between the identity and the cyclic permutation, i.e. $\mu\in S_k$:
\begin{equation}
    \dist(e, \mu) + \dist(\mu, \gamma) = \dist(e, \gamma) \ .
\end{equation}
This subset is isomorphic to the non-crossing partitions \cite{nica2006lectures, Collins2015random}, and can be arranged in a lattice structure where each line corresponds to a unit step (single transposition) according to $\dist$, as illustrated in Fig.~\ref{fig:nc_lattice}.

To characterize the moments of an ensemble of unitaries $\mc{E}$, we define
the $k$-fold \textit{twirling channel}, 
\begin{equation}
    \Phi^{(k)}_{\mc{E}}(X) := \int_{U \sim \mc{E}} U^{\otimes k} X (U^\dagger)^{\otimes k} \, . \label{eq:twirl}
\end{equation}
Here and throughout this work, by $U \sim \mathcal{E}$ we mean that $U$ is sampled according to the distribution $\mathcal{E}$, and $\int_{U \sim \mc{E}}$ indicates averaging according to this sampling. A unitary ensemble $\mathcal{E}$ is said to form a unitary $k$-design if, for any operator $X$, its $k$-fold twirl is equal to that of the Haar ensemble $\mathbb{H}$,
\begin{equation}
    \Phi^{(k)}_{\mathcal{E}}(X) = \Phi^{(k)}_{\mathbb{H}}(X). \label{eq:deisgn_cond}
\end{equation}
When Eq.~\eqref{eq:deisgn_cond} is only satisfied up to some error, the notion of approximation becomes a subtle point: one must both specify a restricted set of input operators $X$ and a concrete method for distinguishing the resulting outputs. We review inequivalent measures of approximate unitary designs in App.~\ref{ap:a}.

For technical convenience, we occasionally work in the vectorized (or \textit{superoperator}) representation, induced by the (unnormalized) maximally entangled state across two copies of the relevant Hilbert space. In this representation, an operator $X$ on a $\mathfrak{D}$-dimensional space is mapped to a (unnormalized) state via $|X\rangle \! \rangle := (X \otimes \id) |\phi^+ \rangle$, where $|\phi^+\rangle = \sum_{x=0}^{\mathfrak{D}-1} |x x\rangle $ and the inner product becomes $\langle\!\langle X | Y \rangle \!  \rangle = \mathrm{tr}[Y^\dagger X]$. 
Unitary conjugation then acts as $(U \otimes U^\ast) |X\rangle \! \rangle = |UXU^\dagger\rangle \! \rangle$. 
Accordingly, the $k$-fold twirl channel can be represented as 
\begin{align}    \tilde{\Phi}^{(k)}_{\mc{E}}:=\int_{U\sim \mc{E}} (U \otimes U^*)^{\otimes k} &= \int_{U\sim \mc{E}} \includegraphics[scale=1.6, valign=c]{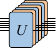}, \label{eq:moment}
\end{align}
where there are $2k$ alternating layers of $U$ and $U^*$ shown in blue and orange respectively, and where each wire represents a copy of the physical Hilbert space $\mc{H}$. In this representation, the twirl channel acts via left/right matrix multiplication on vectorized operators in $k$-replica space,
\begin{equation}
    \tilde{\Phi}^{(k)}_{\mc{E}}| A^{\otimes k} \rangle \! \rangle = | {\Phi}^{(k)}_{\mc{E}}(A^{\otimes k} )\rangle \! \rangle. \label{eq:choi}
\end{equation}
Through standard tensor-network replica manipulations, all quantities of interest considered in this work can be expressed as functions of $\tilde{\Phi}^{(k)}_{\mathcal{E}}$.

The twirl channel over the unitary Haar ensemble $\mathbb{H}$ can be evaluated using the Weingarten calculus~\cite{collins_integration_2006}. Namely, Eq.~\eqref{eq:twirl} reduces to a projection onto the symmetric group $S_k$ of permutations (i.e., the commutant of the unitary group) 
\begin{equation}
    \Phi^{(k)}_{\mathbb{H}}(A^{\otimes k} ) = \sum_{ \pi ,\sigma \in S_k} \mathrm{Wg}_{\pi ,\sigma }(D,k)  \tr[A^{\otimes k}  T_{\sigma^{-1}}] T_\pi.\label{eq:twirl1}
\end{equation}
Here, $\mathrm{Wg}_{\pi,\sigma}(D,k)$ denotes the Weingarten matrix, defined as the inverse of the Gram matrix~\footnote{We here assume the existence of the inverse, which is assured by taking $D>k$, covering all relevant regime considered in this work. } 
\begin{equation}
    G_{\pi,\mu}(D,k) :=\langle \! \langle T_\mu|T_\pi\rangle \! \rangle = D^{k-\dist(\mu,\pi)}\ ,
\end{equation} 
which quantifies how permutation operators are not orthonormal under the Hilbert-Schmidt inner product. For clarity of notation, we will omit the arguments of $\mathrm{Wg}$ and $G$ when clear from context. Moreover, the corresponding moment operator of $\mathbb{H}$ admits a compact form, for which we introduce additional graphical notation,
\begin{align}
   \tilde{\Phi}^{(k)}_{\mathbb{H}}&= \sum_{\pi,\sigma \in S_k} \mathrm{Wg}_{\pi,\sigma} |{T_\pi}\rangle \! \rangle \langle \! \langle {T_\sigma}| \\
   &=\includegraphics[scale=1.4, valign=c]{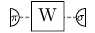}\;, \label{eq:transfer}
\end{align}
and the Gram matrix
\begin{equation}
     G_{\pi ,\mu} = \includegraphics[scale=1.4, valign=c]{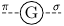}. \label{eq:gram}
\end{equation}
Here, we use dashed wires to denote vectors in the $k!$-dimensional permutation space (i.e., abstract elements of the group algebra $\mathbb{C}[S_k]$). 
In contrast, semicircles represent the corresponding permutation operators acting on the physical Hilbert space $\mathcal{H}^{\otimes k}$ via the representation defined in Eq.~\eqref{eq:moment}.  
For the more intricate expressions arising in the RMPU ensemble, we will omit the explicit $\mathrm{G}$ and $\mathrm{W}$ labels, implicitly assuming that: square boxes represent Weingarten matrices, circular nodes represent Gram matrices, and semicircles represent a projection onto a permutation matrix.

\subsection{Asymptotics of the Weingarten Symbols I: the Diagonal Approximation}\label{sec:diag}

The Weingarten matrix is generally a complicated object. However, since in the context of many-body physics we are primarily interested in the large dimensional limit, we can focus on its asymptotic expression
\begin{equation}
    \mathrm{Wg}_{\pi ,\sigma }(D,k) = \frac{\mu(\pi, \sigma  ) + \mc{O}(D^{-2})}{D^{k+\dist (\pi,\sigma)}} \label{eq:leading-wg}
\end{equation}
where $\mu(\pi,\sigma)$ is a constant independent from $D$ determined by the cycle structure of $\sigma^{-1}\pi$. 
Some of its properties useful for the following are: (i) it depends only on the product of its arguments, $\mu(\pi,\sigma) = \mu(\sigma^{-1}\pi)$; (ii) it factorizes over disjoint cycles, $\mu(\sigma) = \prod_{a \in \sigma} \mu(a)$~\footnote{For example, $\mu((1)(234)) = \mu((1))\mu((234))$}; (iii) for a $k$-cycle $\gamma_k$, one has $\mu(\gamma_k) = (-1)^{k-1} C_{k-1}$, where $C_k $ denotes the $k$-th Catalan number; and (iv) it satisfies the identity $\sum_{\sigma:\pi \leq \sigma \leq \mu} \mu(\pi,\sigma) = \delta_{\pi,\mu}$. When $\pi, \sigma$ correspond to non-crossing partitions---as depicted in Fig.~\ref{fig:nc_lattice} for $k=4$---then $\mu(\pi, \sigma)$ is the corresponding M\"obius function appearing in Eq.~\eqref{eq:mom-cum}~\cite{nica2006lectures}.

In many situations, it is not necessary to retain the full structure of Eq.~\eqref{eq:leading-wg}. As a matrix, the Weingarten function is {approximately diagonal} at leading order, $\mathrm{Wg}_{\pi ,\sigma } \propto \delta_{\pi, \sigma} D^{-k}$, since the dominant contributions in the $1/D$ expansion come from terms minimizing the Cayley distance in the denominator of Eq.~\eqref{eq:leading-wg}.
An important example when this approximation is relevant are \emph{state $k$-designs}. Considering the Haar twirl of a pure state $\psi := \ket{\psi}\! \bra{\psi}$, we have 
\begin{equation}
\begin{split}
\Phi^{(k)}_{\mathbb{H}}(\psi^{\otimes k}) &=  \sum_{ \pi ,\sigma \in S_k} \frac{\mu(\pi,\sigma  ) + \mc{O}(D^{-2})}{D^{k+\dist (\pi ,\sigma)}} T_\pi \tr[\psi^{\otimes k} T_{\sigma^{-1}}]  \\
     &= \sum_{ \pi ,\sigma \in S_k} \frac{\mu(\pi,\sigma  )+ \mc{O}(D^{-2}) }{D^{k+\dist (\pi ,\sigma)}} T_\pi \\
     &\approx \sum_{ \pi \in S_k} \frac{T_\pi }{D^k} \label{eq:lo}
     \end{split}
\end{equation}
where we used that $\tr[\psi^{\otimes k} T_{\sigma^{-1}}] = 1$ for all $\sigma\in S_k$, that the exponent is uniquely minimized for $\pi = \sigma$ since $\dist(\pi,\pi)=0$, and that $\mu(\pi,\pi) = 1$.
Throughout this work, we use ``$\approx$'' to denote the asymptotic leading-order expression according to the dominant dimensional variable: $D \to \infty$ here and $\chi \to \infty$ later for our calculations with respect to the RMPU ensemble. 

Thus, at least for the case of random quantum states, we can replace the twirl channel Eq.~\eqref{eq:twirl} with the following ``diagonal approximation''
\begin{equation}
    \Phi_{\mathrm{diag}}^{(k)}(X) = \sum_{\pi\in S_k} \frac{1}{D^k} T_{\pi^{-1}} \tr[X T_\pi]. \label{eq:diag} 
\end{equation}
where only terms with $\pi = \sigma$ in Eq.~\eqref{eq:twirl1} are retained. This diagonal approximation is not just an asymptotic convenience. It provides the exact $k$-twirling channel for ensembles of complex Gaussian random matrices with independent, identically distributed entries of zero mean and variance $1/\sqrt{D}$, even at finite $D$~\cite{deluca2024universalityclassespurificationnonunitary,christopoulos2025universaldistributionsoverlapsgeneric,lami2025quantumstatedesignemergent}. 
Moreover, the diagonal approximation to the twirl appears throughout different proofs of low-depth unitary designs~\cite{aharonov_quantum_2022,schuster2024randomun,Lami2025}; see Table~\ref{tab:example}. For instance, Eq.~\eqref{eq:lo} is a key step in the ``gluing lemma'' for relative error designs~\cite{schuster2024randomun}.

The validity of the diagonal approximation in the above example depended on the fact that the Haar twirl was applied to an input satisfying \(\tr[(\ket{\psi}\bra{\psi})^{\otimes k} T_{\sigma^{-1}}] = 1\), such as pure states or projectors. However, it fails when the input is a finite-rank operator with \(\tr(A^m) \propto D\), such as for OTOCs. This can be seen directly from Eq.~\eqref{eq:leading-wg}: if the Weingarten function is contracted with quantities that depend explicitly on the indices $\pi$ and $\sigma$, then the simplification of specifying $\sigma = \pi$ is no longer appropriate. As we will detail in the next section, in such settings free probability becomes relevant.

\subsection{Asymptotics of the Weingarten Symbols II: Free Probability} \label{sec:fp}
We will now review how the leading-order average of the Haar twirl leads to the free probability result for higher-order OTOCs~\cite{Voiculescu1991,nica2006lectures,SilviaDesigns}. Employing the replica method $\mathrm{tr}[A^k] = \mathrm{tr}[T_\gamma A^{\otimes k}]$, we can re-express the average OTOC Eq.~\eqref{eq:otoc} for any ensemble $\mathcal{E}$ in terms of the corresponding $k$-fold twirl, 
\begin{align}
    \OTOC^{(k)}_\mathcal{E}&= \frac{1}{D}\tr[\Phi^{(k)}_{\mc{E}}(A^{\otimes k}) B^{\otimes k} T_\gamma ] \nn \\
    &=\frac{1}{D}\llangle A^{\otimes k}| \tilde{\Phi}_{\mathcal{E}}^{(k)}|T_\gamma B^{\otimes k}\rrangle. \label{eq:otocreplica}
\end{align}
This expression can be readily evaluated for the Haar ensemble for large $D$, employing the asymptotic expression Eq.~\eqref{eq:leading-wg}. This leads to
\begin{equation}
\begin{split}
\OTOC_\mathbb{H}^{(k)}
    \approx& \sum_{\pi,\sigma} \frac{\mu(\pi,\sigma)\tr[A^{\otimes k} T_\pi]\tr[T_{\sigma^{-1}}B^{\otimes k}  T_\gamma ] }{D^{k+1+\dist (\pi, \sigma)}} \label{eq:derivation1}  \\
    =& \sum_{\pi,\sigma} \frac{\mu(\pi,\sigma)\braket{A }_\pi \braket{B }_{\sigma^{-1} \gamma} }{D^{1-k+\dist (\pi,\sigma) + \dist(e,\pi) +\dist(\sigma, \gamma)}} \ .
\end{split}
\end{equation}
To ease notation, we have introduced the normalized replica expectation value with respect to a permutation $\pi \in S_k$: $\braket{B}_{\pi} := D^{k-\dist(\pi,e)} \tr[B^{\otimes k} T_\pi]$, which agrees with the partitioned moment expression of Eq.~\eqref{eq:fact_moms}. 
Since $0 \leq |\braket{B}_{\pi}| \leq 1$ and the coefficients $\mu(\pi,\sigma)$ are independent of $D$, the leading-order behavior in the large-$D$ limit is determined by minimizing the exponent of $1/D$. Using the fact that the cyclic permutation $\gamma \in S_k$ consists of a single cycle, so that $\dist(e,\gamma) = k - \#(\gamma) = k - 1$, the exponent of $1/D$ in Eq.~\eqref{eq:derivation1} can be re-written as
\begin{equation}
    -\dist(e, \gamma)+\dist(e,\pi) +  \dist(\pi,\sigma)+\dist(\sigma,\gamma) \geq   0. \label{eq:2chain}
\end{equation}
where the inequality comes from the successive application of triangle inequalities. Therefore, the leading order to Eq.~\eqref{eq:derivation1} is given by a summation restricted to a subset of permutations $\pi, \sigma$ that saturate the inequality in Eq.~\eqref{eq:2chain}, i.e., that are ordered along the geodesic between $e$ and $\gamma$:
\begin{equation}
    \dist(e,\pi) +  \dist(\pi,\sigma)+\dist(\sigma,\gamma) = \dist(e, \gamma) \, . \label{eq:geodesic}
\end{equation}
For brevity, we denote the condition Eq.~\eqref{eq:geodesic} as $\pi \leq \sigma \leq \gamma$. As discussed above in Sec.~\ref{sec:free}, the permutations satisfying $\sigma \leq \gamma$ correspond to the {non-crossing partitions} $NC(k)$~\cite{nica2006lectures, Collins2015random}. When viewed as partitions, $\sigma^{-1} \gamma = \sigma^*$ is the Kreweras complement of $\sigma$, and the condition $\pi \leq \sigma \leq \gamma$ is a hierarchy of containment properties (i.e., partial orderings) between the partitions, called a $2$-chain on the non-crossing partition lattice. Building on these combinatorial insights, one arrives at the leading-order expression for the Haar-averaged OTOC
\begin{align}
    \OTOC_\mathbb{H}^{(k)}\approx \sum_{\pi\leq\sigma\leq \gamma} \mu(\pi,\sigma)\braket{A }_\pi \braket{B }_{\sigma^{-1} \gamma}=: \OTOC^{(k)}_{\mathrm{FP}},\label{eq:loOTOC}
\end{align}
which corresponds precisely with the condition of free independence between $A_U$ and $B$, Eq.~\eqref{eq_free_def}. 
Just as with the diagonal approximation, Eq.~\eqref{eq:diag}, we can define the restricted twirl channel according to the free probability approximation, 
\begin{equation}
    \Phi_{\mathrm{FP}}^{(k)}(X) := \sum_{\pi \leq \sigma \leq \gamma} \frac{\mu(\pi,\sigma  )}{D^{k+\dist (\pi ,\sigma)}} T_{\sigma^{-1}} \tr[X T_\pi].  \label{eq:fp_channel}
\end{equation}
Remarkably, solely from considerations of the Weingarten calculus, one arrives at the free independence of unitarily rotated operators, which itself is, a priori, independent of the concept of Haar-random unitaries~\cite{Voiculescu1991}.
While Eq.~\eqref{eq:loOTOC} becomes exact in the limit $D \to \infty$, it is natural to ask: what is the quantitative error for large but finite $D$?

There are two sources of subleading corrections to the Haar-averaged expression. The first arises from higher-order terms in the power series expansion of the Weingarten function. Namely~\cite{collins2017weinga,Collins2022-hr},
\begin{equation}
    \mathrm{Wg}_{\pi,\sigma}(D,k) = \sum_{g\geq0} \frac{ \mathrm{Wg}^{(g)}_{\pi,\sigma}(k)}{D^{k+\dist(\pi ,\sigma) +2g} }, \label{eq:collins}
\end{equation}
where $\mathrm{Wg}^{(0)}_{\pi,\sigma}(k) = \mu(\pi,\sigma)$ as in Eq.~\eqref{eq:leading-wg}. The subleading correction $\mathrm{Wg}^{(1)}_{\pi,\sigma}$ has no known closed form, but can be bounded by $\mathrm{Wg}^{(1)}_{\pi,\sigma}\le 6k^{7/2} \mu(\pi,\sigma)$~\cite{collins2017weinga}. 
Since this correction is proportional to the leading-order coefficient $\mu(\pi,\sigma)$, it contributes a relative error of $6k^{7/2} D^{-2}$ compared to the free probability approximation, Eq.~\eqref{eq:loOTOC}. 
This term also precisely accounts for the $\mathcal{O}(D^{-2})$ correction arising from the diagonal approximation in Eq.~\eqref{eq:diag}, and plays a central role in proving design properties of logarithmic-depth circuits~\cite{aharonov_quantum_2022,schuster2024randomun}. For the OTOC calculation, however, there exists an additional source of error, arising from deviations from the free probability approximation. Specifically, subleading contributions can appear from permutations that do not saturate the triangle inequality in Eq.~\eqref{eq:2chain}, and yet are exactly proportional to the leading order of the Weingarten matrix, $\mathrm{Wg}^{(0)}_{\pi,\sigma}(k) = \mu(\pi,\sigma)$. 
One can show that there exist permutations $\pi, \sigma$ such that the next-smallest value of the exponent in Eq.~\eqref{eq:derivation1} is exactly two, equal to the left-hand side of Eq.~\eqref{eq:2chain}~\footnote{This can be proven through identifying $S_k$ with an orientable surface, and applying Euler's formula. Namely, we have that the only allowable solutions to the triangle inequality here are $\dist(e,\pi)+\dist(\pi,\sigma) +\dist(\sigma,\gamma)=\dist(e,\gamma)+2g$, for integer $g\geq 0$~\cite{nica2006lectures}.}. These terms thus also contribute corrections of size $\mathcal{O}(1/D^2)$, and so must be taken into account alongside the Weingarten matrix corrections in Eq.~\eqref{eq:collins}. We return to this point in detail in Sec.~\ref{sec:subleading}.

\section{Random Matrix Product Unitaries} \label{sec:rmpu}
We now introduce the random matrix product unitary (RMPU) ensemble as depicted in Fig.~\ref{fig:rmpu} (a).  
Consider a system of $N$ qudits, each of local dimension $d$, and define $\chi =d^r$ for some $r\leq N-1$. An RMPU is constructed from a sequence of $n:=N - r$ independent unitary matrices $\{U_i\}_{i=1}^n$, each independently and identically drawn from the Haar measure on the $(\chi d)$-dimensional unitary group, and overlapping on an $r$-qudit space of dimension $\chi$. Viewing each $U_i$ as a rank-$4$ tensor of dimensions $d \times \chi \times d \times \chi$, the full RMPU is formed by contracting the $\chi$-dimensional virtual indices.  
Specifically, the fourth leg of $U_i$ is contracted with the second leg of $U_{i+1}$ for all $i = 1, \dots, n-1$. The total Hilbert space dimension according these various quantities is then: $D = d^N = \chi d^n = d^{r+n}$. In standard graphical tensor network notation, RMPUs are represented as a sequence of unitary boxes with thin (physical) and bold (bond) legs, arranged in a linear geometry, 
\begin{equation}
    \includegraphics[scale=1.6, valign=c]{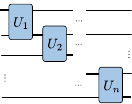} = \includegraphics[scale=1.6, valign=c]{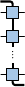}\; . \label{eq:mpo}
\end{equation}
Here, we have also written the staircase circuit as a linear tensor network, showing how the shared space of dimension $\chi$ between each brick leads to a matrix product operator (MPO) of this bond dimension~\cite{Styliaris2025matrixproduct}. 
Note that the Haar ensemble is recovered in the case $n = 1$, corresponding to the application of a single global gate across the entire system. 

Before proceeding to compute operator properties with respect to this ensemble, a few remarks are in order.

First, the staircase geometry considered here does not exhaust the class of matrix product unitaries (MPUs) in one dimension~\cite{Cirac_2017mpu,Styliaris2025matrixproduct}---that is, unitaries that admit an exact representation as an MPO with a fixed bond dimension. For instance, the staircase circuit exhibits a directional causal structure (a light cone) depending on whether the unitary sequence is ascending or descending, and the resulting output ensemble lacks translational invariance~\cite{leontica2025unbias}.
While various ensembles of MPUs can be defined, we focus on the ansatz shown in Fig.~\ref{fig:rmpu}(a) as a tractable and illustrative toy model. We briefly consider a so-called `two-floor staircase' later in Sec.~\ref{sec:twofloor}, while an alternative construction from Ref.~\cite{schuster2024randomun} employs a two-layer brickwork geometry. 
Nonetheless, we note that the key results obtained there---such as the applicability of the `gluing lemma' therein---also extends directly to the staircase architecture considered here for sufficiently large overlap $\chi$. In particular, this means that the staircase RMPU ensemble with polynomial bond dimension is a relative error design, as schematically depicted in Fig.~\ref{fig:rmpu}(b). 

Further, applying an RMPU from the ensemble to a product state yields a random matrix product state. This means that implying RMPUs with bond dimension polynomial in the system size $N$ exhibit anticoncentration~\cite{Lami2025} and universal correlation scaling~\cite{Haag2023}. 

Finally, we note that by construction, RMPUs have operator entanglement \cite{Zanardi2001} bounded by its bond dimension $\log(\chi)$. Consequently, RMPUs obeying an area law scaling $\chi = \mathrm{poly}(N)$ cannot reproduce certain global properties of random operators, such as the volume-law growth of local-operator entanglement typically observed in chaotic spin chains~\cite{Prosen2007,Kos2020} and (deep) random circuits~\cite{Jonay2018,Dowlin2024LOE-OSRE}.

Our goal is to assess how well the RMPU ensemble---denoted by $\mathcal{R}$ and induced by the Haar measure over the component unitaries $U_i$---can reproduce global Haar values of quantities that go beyond the regime of unitary designs, such as the free independence of Heisenberg operators. 
Recalling Eq.~\eqref{eq:moment} and using blue and orange to depict $U$ and $U^\ast$, respectively, the moment operator of the RMPU ensemble given by
\begin{equation}
\tilde{\Phi}^{(k)}_{\mc{R}}=\int_{U_1\in \mathbb{H}}\cdots \int_{U_n \in \mathbb{H}}\includegraphics[scale=1.6, valign=c]{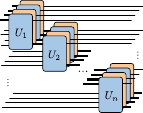} .
\end{equation}
Using the independence and identical distribution of the components $\{U_i\}$, we can factor the average into $n$ local Haar integrals, each of which can be written in terms of the transfer matrix Eq.~\eqref{eq:transfer}, to arrive at 
\begin{equation}
    \tilde{\Phi}^{(k)}_{\mc{R}}= \adjustbox{scale =\scalefactor}{\input{rmpu_mom.tikz}}\label{eq:momentrmpu}.
\end{equation}
We have used the replica graphical notation introduced in Eqs.~\eqref{eq:transfer}–\eqref{eq:gram}, additionally color-coding the Gram matrices to distinguish between contributions from bond versus physical replica Hilbert spaces (with single-replica dimensions of $\chi=d^r$ and $d$, respectively),
\begin{equation}
\begin{split}
     &\mathrm{Wg}(\chi d, k) =: \adjustbox{scale =\scalefactor}{\input{ident1a.tikz}}\;,\\
    &G(\chi,k) =: \adjustbox{scale =\scalefactor}{\input{gr.tikz}} \;,\label{eq:graphical}\\
    &G(d,k) =: \adjustbox{scale =\scalefactor}{\input{gd.tikz}}. 
\end{split}
\end{equation}
Here, we represent $G(\chi, k)$ with a thicker wire in the diagrams to emphasize its dependence on $\chi > d$. In contrast, $\mathrm{Wg}(\chi d, k)$ is depicted with two output lines, highlighting both its role as the inverse of $G(\chi d, k)$ and the factorized structure $G(\chi d, k)_{\pi, \sigma} = G(\chi, k)_{\pi, \sigma} G(d, k)_{\pi, \sigma}$. We use dashed wires to indicate contractions with respect to the permutation indices $\pi,\sigma$  (of dimension $k!$), to distinguish them from the fundamental representation (as in the graphical notation of Eq.~\eqref{eq:mpo}).

\subsection{Asymptotic Expression for OTOCs} \label{sec:loRMPU}
We now sketch the proof of the asymptotic expression for the average multipoint OTOC [Eq.~\eqref{eq:otoc}] over the RMPU ensemble $\mathcal{R}$, with a more detailed analysis provided in App.~\ref{ap:otocProofs}. 
In contrast to Eq.~\eqref{eq:derivation1}, which holds for arbitrary observables $A$ and $B$, let us start by restricting ourselves to $A$ and $B$ to be (quasi-)local operators. Importantly, {we require $A$ and $B$ to be within each other's light cone}; otherwise, $A_U$ and $B$ trivially commute for any bond dimension $\chi$. We ensure this by assuming without loss of generality that they have nontrivial support only on the first and last unitaries making up the RMPU, for $A$ and $B$ respectively. Namely, we generate a random Heisenberg operator $A_U$ through acting with the staircase RMPU on the initial operator $A$ on the first qudit (`bottom' of the staircase), and study its freeness properties with respect to any other local $B$. We will discuss at the end of this section how the results change for different choices of operators. We adopt a slight abuse of notation: $A \equiv A \otimes \id^{\otimes N - r -1}$ and $B \equiv \id^{\otimes N - r -1} \otimes B$. 
We further assume that both operators have non-vanishing normalized trace: $D^{-1} \tr[A] = a$ and $D^{-1} \tr[B] = b$, with $a, b \in \mathbb{R} \backslash \{0\}$. 
This condition ensures that the Haar averaged OTOC is non-vanishing as $D \to \infty$, cf. Eq.~\eqref{eq:loOTOC}. From the graphical notation of Eq.~\eqref{eq:momentrmpu} and using the replica trick Eq.~\eqref{eq:otocreplica}, the average OTOC over the RMPU ensemble reads
\begin{align}
    \OTOC^{(k)}_\mathcal{R} &=\frac{1}{D} \,\int_{U_i\in \mathbb{H}}\includegraphics[scale=1.6, valign=c]{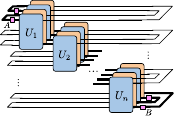} \label{eq:rmpumomentotoc} \\
    &=\frac{1}{D} \, \adjustbox{scale =\scalefactor}{\input{rmpu_2kotoc.tikz}}. \nn
\end{align}
Here, we denote the single-index vectors in replica space  
\begin{align}
    \langle \! \langle A^{\otimes k} &| T_\pi\rangle\!\rangle =:  \adjustbox{scale =\scalefactor}{\input{A_choi.tikz}},\quad  \langle \! \langle T_{\pi \gamma} | B^{\otimes k}\rangle\!\rangle =:     \adjustbox{scale =\scalefactor}{\input{B_choi.tikz}}, \quad \nn\text{and}\\ &G(d,k)_{\pi,\sigma}  =\llangle T_\pi| T_\sigma\rangle \! \rangle =:   \adjustbox{scale =\scalefactor}{\input{gen_sigma.tikz}}\, , 
\end{align}
for a given $\sigma\in S_k$.
To obtain the leading-order asymptotic expression for large $\chi$, we replace each Weingarten matrix in the RMPU moment operator with its leading term from Eq.~\eqref{eq:leading-wg}. For clarity, we present the analysis for a two-layer RMPU ($n = 2$), corresponding to a total Hilbert space dimension $D = \chi d^2$, with the steps of the proof extending directly to arbitrary $n$ (see App.~\ref{ap:otocProofs}). 
The resulting moment operator reads
\begin{align}
    \tilde{\Phi}^{(k)}_{\mc{R}}= \frac{1}{\chi d^2}&\sum_{\{\sigma_i,\pi_i \in S_k\}} \frac{\mu(\pi_1,\sigma_1)\mu(\pi_2,\sigma_2)}{(d\chi)^{(2k+\dist(\sigma_1,\pi_1)+\dist(\sigma_2,\pi_2))}\chi^{\dist(\sigma_1,\pi_2) -k}}\nn \\
    &\times \left(| T_{\pi_1}^{(1  \dots r+1)} T_{\pi_2}^{(r+2)}  \rangle \! \rangle  \! \langle \! \langle  T_{\sigma_1}^{(1)} T_{\sigma_2}^{(2  \dots r+2)} |\right),
\end{align}
where the sum runs over all permutations $\pi_i, \sigma_i \in S_k$ for $i \in \{1,2\}$, and superscripts indicate the specific $k$-replica qudits on which the corresponding permutation operators act. 
Substituting this expression into Eq.~\eqref{eq:rmpumomentotoc} and applying the leading-order form of the Weingarten function from Eq.~\eqref{eq:leading-wg}, we obtain
\begin{equation}
   \OTOC^{(k)}_\mathcal{R}\approx  \sum_{\{\sigma_i,\pi_i \in S_k\}} \frac{\left(\prod_{j=1}^2\mu(\pi_j,\sigma_j) \right)\braket{A}_{\pi_1} \braket{B}_{\sigma_2^{-1} \gamma}}{\chi^{f_2(\vec{\pi},\vec{\sigma}) } d^{g_2(\vec{\pi},\vec{\sigma})}}, \label{eq:rmpuotoc1}
\end{equation}
where we have introduced the exponent functions
\begin{align}
\begin{split}
    f_2(\vec{\pi},\vec{\sigma}) :=& 1-k+ \dist(e,\pi_1) + \dist(\pi_1,\sigma_1) + \dist(\sigma_1,\pi_2)  \\
    &+ \dist(\pi_2,\sigma_2) +\dist(\sigma_2,\gamma), \label{eq:f}
\end{split}\\
    g_2(\vec{\pi},\vec{\sigma}):= &\sum_{i=1}^2 [1-k+ \dist(e,\pi_i) + \dist(\pi_i,\sigma_i) + \dist(\sigma_i,\gamma)]. \label{eq:g}
\end{align}
To determine the leading-order contributions to the sum in Eq.~\eqref{eq:rmpuotoc1}, we characterize the set of permutations $\vec{\pi}=(\pi_1,\pi_2), \vec{\sigma}=(\sigma_1,\sigma_2)$ that simultaneously minimize the exponents $f_2(\vec{\pi}, \vec{\sigma})$ and $g_2(\vec{\pi}, \vec{\sigma})$. Focusing first on the $\chi$ exponent, from a sequence of triangle inequalities, we find,
\begin{equation}
    f_2(\vec{\pi},\vec{\sigma}) \geq 1-k + \dist(e,\gamma) =0.
\end{equation}
The permutations that saturate this bound form the sequence
\begin{equation}
    e \leq \pi_1 \leq \sigma_1 \leq \pi_2 \leq \sigma_2 \leq \gamma, \label{eq:multi}
\end{equation}
which defines a multichain in the poset of non-crossing partitions; see Fig.~\ref{fig:nc_lattice}. 
Turning now to Eq.~\eqref{eq:g}, again from triangle inequalities, we observe that $g_2(\vec{\pi}, \vec{\sigma}) \geq 0$ is saturated when both of the following independent conditions hold: $\pi_j \leq \sigma_j \leq \gamma$ for each $j \in \{1, 2\}$. These are clearly satisfied when Eq.~\eqref{eq:multi} holds.
Imposing \emph{non-crossing multichain condition} in Eq.~\eqref{eq:multi} therefore simultaneously minimizes both $f_2(\vec{\pi}, \vec{\sigma})$ and $g_2(\vec{\pi}, \vec{\sigma})$ to be zero. We note that this non-crossing multichain condition also arises in a computation of Refs.~\cite{chen2024freeindependen, fritzsch2025freeprob}. 
Asymptotically, in the large-$\chi$ limit, Eq.~\eqref{eq:rmpuotoc1} then reduces to
\begin{equation}
     \OTOC^{(k)}_\mathcal{R}\approx \!\sum_{\pi_1 \leq \sigma_1 \leq \pi_2 \leq \sigma_2 \leq \gamma}  \left( \prod_{j=1}^2\mu(\pi_j,\sigma_j) \right)\braket{A}_{\pi_1}  \braket{B}_{\sigma_2^{-1} \gamma}, \label{eq:multi1}
\end{equation}
where we neglect terms $\mc{O}(\chi^{-2})$. 
Finally, we apply a fundamental identity of M\"obius functions~\cite{nica2006lectures},
\begin{equation}
    \sum_{\mu:\nu \leq \mu \leq \alpha} \mu(\nu, \mu) = \delta_{\nu,\alpha}. \label{eq:mobius_ident}
\end{equation}
This identity allows us to cancel two of the sums in Eq.~\eqref{eq:multi1}, yielding
\begin{equation}
    \begin{split}
        \OTOC^{(k)}_\mathcal{R} \!&\approx \!\sum_{\pi_1 \leq \pi_2 \leq \sigma_2 \leq \gamma}  \delta_{\pi_1,\pi_2} \mu(\pi_2,\sigma_2) \braket{A}_{\pi_1} \! \braket{B}_{\sigma_2^{-1} \gamma}\\
        &= \!\sum_{\pi_2 \leq \sigma_2 \leq \gamma} \! \mu(\pi_2,\sigma_2) \braket{A}_{\pi_2} \! \braket{B}_{\sigma_2^{-1} }=\OTOC^{(k)}_{\mathrm{FP}},\label{eq:multi2}
    \end{split}
\end{equation}
which exactly matches the asymptotic global Haar-averaged OTOC expression in Eq.~\eqref{eq:loOTOC}.
We thus conclude that, for sufficiently large bond dimension $\chi$, rotating with the RMPU ensemble $U\sim \mc{R}$ causes $A_U$ and $B$ to become freely independent, reproducing the Haar value of higher-order OTOCs 
\begin{equation}
    \OTOC^{(k)}_\mathcal{R} = \OTOC^{(k)}_{\mathrm{FP}} + \mc{O}\left(n  {\chi^{-2}}\right) \approx \OTOC^{(k)}_{\mathbb{H}} +\mc{O}\left( N {\chi^{-2}}\right)\;. \label{eq:lormpu}
\end{equation}
The displayed subleading scaling is justified in Sec.~\ref{sec:subleading} and App.~\ref{ap:otocProofs}. In the final equality above we have assumed that $\chi \gg N \gg r$ such that $n = N-r \approx N$ (as $d^N=D=\chi d^n=d^{r+n}$), which is also the limit presented in Table~\ref{tab:example}.
Given that for finite trace operators $A$ and $B$, $\OTOC^{(k)}_{\mathrm{FP}}= \mathcal{O}(1)$ and $\chi = d^r$, Eq.~\eqref{eq:lormpu} shows that RMPUs with $r = \mathcal{O}(\log n)$ suffice to approximate the Haar value with polynomial precision, while $r = \mathrm{poly}(\log N)$ ensures errors smaller than any inverse polynomial in $N$.

Eq.~\eqref{eq:lormpu} constitutes one of the central results of this work. It is surprising: OTOCs lie strictly beyond the set of observables guaranteed to be close to their Haar value from the $k$-design property of the RMPU ensemble.
Yet, for physically relevant observables, we find that RMPUs with only polynomial bond dimension (i.e., shallow depth) already reproduce the Haar value of these higher-order diagnostics. 
In the thermalization of isolated many-body systems, higher order OTOCs encode the off-diagonal matrix elements in the full ETH~\cite{Foini2019,Pappalardi2022-vo}. 
At infinite temperature, free probability predicts that these cumulants take the form given by Eq.~\eqref{eq:loOTOC}, and the ETH asserts that such freeness structure governs the long-time behavior of chaotic, closed systems. 
{Our result, Eq.~\eqref{eq:lormpu}, shows that this thermal behavior can emerges at only logarithmic depth, even for moderately large moments $k$}; see Sec.~\ref{sec:subleading} below. It is worth noting that the claimed correspondence between logarithmic depth and polynomial bond dimension is not always valid. For instance, for a locally interacting dynamics (local Hamiltonian or a local circuit), operators exhibit a Lieb-Robinson light cone and so local $A$ and $B$ separated by a linear number of spins will be uncorrelated, leading to a trivial value for the OTOC (see Sec.~\ref{sec:twofloor}).  

Before turning to the error analysis of the asymptotic expression in Eq.~\eqref{eq:lormpu}, we first examine the large-$\chi$ behavior of RMPU averaged OTOCs for different classes of observables.
\subsubsection{Near Observables}
First, we consider local operators that are separated by $M < N$ sites, rather than the $N$-site separation considered above. That is, we take $A \equiv A \otimes \id^{N-1}$ as before, but $ B \equiv \id^{M-1} \otimes B \otimes \id^{N-M}$, such that the average OTOC is equal to 
\begin{equation}
     \OTOC^{(k)}_{\mathcal{R}(n)} =\frac{1}{D}\, \adjustbox{scale =\scalefactor}{\input{rmpu_2kotoc2.tikz} }.
\end{equation}
Here, we see an emergent light-cone property,
\begin{equation}
    \begin{split}
        \OTOC^{(k)}_{\mathcal{R}(n)} &=\frac{1}{D}\, \adjustbox{scale =\scalefactor}{\input{rmpu_2kotoc3.tikz} } \\
    &= \OTOC^{(k)}_{\mathcal{R}(n')}\;,
    \end{split}
\end{equation}
where we have reduced the quantity to a function of a smaller RMPU with $n'=\lceil \log_d( d^n - M/\chi ) \rceil$ layers, using the identity 
\begin{equation}
    \begin{split}
        G_{\alpha,\pi}(d) &G_{\beta,\pi}(\chi) \sum_\sigma \mathrm{Wg}_{\pi,\sigma} G_{\sigma,\mu}(\chi d)  = \adjustbox{scale =\scalefactor}{\input{identlightcone1.tikz} } \\ &= \adjustbox{scale =\scalefactor}{\input{identlightcone2.tikz} } = G_{\alpha,\mu}(d) G_{\beta,\mu}(\chi).
    \end{split}
\end{equation}
Therefore, our preceding analysis also holds in this case, but replacing $n$ with $n'$ and thus achieving a smaller subleading correction. When $M\leq \chi $, we clearly obtain the Haar value exactly, as both $A$ and $B$ are directly connected to the same Weingarten matrix. The case of $M=N$ therefore gives the worst-case approximation, and so we consider this case throughout most of this work. 

\subsubsection{Commuting observables: $[A_U,B]=0$} \label{sec:twofloor}
We next observe that the staircase geometry used to define the RMPU ensemble introduces a clear asymmetry. 
In particular, if the operators $A$ and $B$ lie outside each other’s effective light cone---such as in the case where $A \equiv \id^{\otimes N-1} \otimes A$ and $B \equiv B \otimes \id^{\otimes N-1} $---then the OTOC becomes trivial,
\begin{align}
    \OTOC^{(k)}_\mathcal{R} &=\frac{1}{D} \, \adjustbox{scale =\scalefactor}{\input{rmpu_2kotoc1.tikz} }\nn \\
    &=\frac{1}{D}\tr[A^k] \tr[B^k] \;.
\end{align}
To restore spatial symmetry, one can instead adopt a two-floor staircase geometry for the MPU~\cite{Styliaris2025matrixproduct},
\begin{equation}
    \includegraphics[scale=1.6, valign=c]{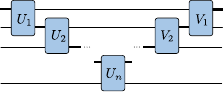}. \label{eq:two-floor}
\end{equation}
Here, independent Haar sampling of both $\{U_i\}_{i=1}^n$ and $\{V_i \}_{i=1}^{n-1}$ generates the corresponding RMPU ensemble. Compared to the asymmetric staircase in Fig.~\ref{fig:rmpu}, this symmetric variant therefore requires approximately twice as many gates to achieve the Haar-limit value $\OTOC^{(k)}_{\mathrm{FP}}$ (Eq.~\eqref{eq:loOTOC}) to the same precision $\mathcal{O}(\chi^{-2})$, while ensuring that Eq.~\eqref{eq:loOTOC} holds symmetrically for $A$ and $B$ acting on $N$-separated sites. Note that while the $\chi$ in this geometry corresponds to the size of the bonds within the two-dimensional tensor network of Eq.~\eqref{eq:two-floor}, unlike Eq.~\eqref{eq:mpo} it no longer corresponds to the bond dimension of the corresponding MPO. 

\subsubsection{Traceless Observables} \label{sec:traceless}
For traceless operators, $\tr[A]=\tr[B]=0$, the leading order of the Haar averaged OTOC in Eq.~\eqref{eq:loOTOC} vanishes, and so the final result is instead exponentially suppressed, i.e.
\begin{equation}
    \OTOC^{(k)}_{\mathrm{FP}} = 0 \implies 
    \OTOC^{(k)}_{\mathbb H}= \mathcal O(D^{-2})
    \ .
\end{equation} 
This follows from the fact that every term in the sum in $\OTOC^{(k)}_{\mathrm{FP}}$ is proportional to either $\tr[A]$ or $\tr[B]$, hence it vanishes. Specifically, the non-crossing partition chain condition from Eq.~\eqref{eq:geodesic} ensures that either $\pi$ or $\sigma^{-1}\gamma$ must contain a cycle of length one, also known as singleton~\cite{nica2006lectures}, leading to
\begin{equation}
    \pi \leq \sigma \leq \gamma \implies \braket{A}_\pi \braket{B}_{\sigma^{-1} \gamma} \propto \tr[A]= 0 \, {\rm or} \, \tr[B]=0  .\label{eq:singleton}
\end{equation} 
This follows from a property of the non-crossing partitions lattice \footnote{Recall that $\sigma^*= \sigma^{-1}\gamma$ is the Kreweras complement of $\sigma$. We note that for $\sigma\in NC(k)$ with $k>2$, either $\sigma^*$ or $\sigma$ must contain at least one singleton (i.e., a unit-length cycle). Eq.~\eqref{eq:singleton} can be proven directly from the identity~\cite{nica2006lectures}
\begin{equation}
    k-\dist(e,\sigma) +k-\dist(e,\sigma^*)=k+1= \#(\sigma)+\#(\sigma^*). \label{eq:KrewerasIdent}
\end{equation}
Recalling that $\#(\sigma)$ is the number of cycles of $\sigma$, then if neither $\sigma$ nor $\sigma^*$ contain a singleton, then the most cycles they could have corresponds to each of them being a product of $2$-cycles, meaning $\#(\sigma),\#(\sigma^*)  \leq k/2$. This clearly contradicts Eq.~\eqref{eq:KrewerasIdent} and so we know that one of $\sigma$ or $\sigma^*$ contains a singleton.
Now, the condition $\pi \leq \sigma \leq \gamma$ implies that both $\pi$ and $\sigma$ lie on the lattice (i.e., are non-crossing partitions), with $\sigma$ either equal to or strictly above $\pi$ along a geodesic. If $\pi$ contains a unit-length cycle, then $\braket{A}_\pi \propto \tr[A]$ and we are done. If $\pi$ does not contain a singleton, then neither does $\sigma$ by the partition refinement property $\pi \leq \sigma$. However, by the identity stated above, this means that $\sigma^{-1}\gamma$ must contain a singleton and so $\braket{B}_{\sigma^{-1} \gamma} \propto \tr[B]$. This can also be visualized diagrammatically in Fig.~\ref{fig:nc_lattice}: if $\pi$ has no singleton, then it lies in the top half of the lattice. Then $\pi \leq \sigma$ ensures that $\sigma$ lies above $\pi$ along a chain, and so $\sigma$ also contains no singleton. However, taking the Kreweras complement necessarily maps $\sigma$ to an element in the bottom half of the lattice~\cite{nica2006lectures}, in which case it must have at least one unit-length cycle. 
This confirms that every term in the sum of Eq.~\eqref{eq:loOTOC} is proportional to either $\tr[A]$ or $\tr[B]$.}. Consequently, the $\mathcal{O}(1)$ leading-order contribution present in the finite trace case vanishes when both operators are instead traceless. In the traceless case, the $\mathcal{O}(\chi^{-2})$ terms become the dominant contribution, leading to
\begin{equation}
    \OTOC^{(k)}_\mathcal{R} =  \mc{O}\left(n  {\chi^{-2}}\right)  \ .
\end{equation}
Furthermore, the coefficient agrees exactly with the Haar value for traceless $A$ and $B$, which we discuss further in Sec.~\ref{sec:subleading} and App.~\ref{ap:otocProofs}.

\subsubsection{Non-local Observables}
For non-local operators, the asymptotic expression for the RMPU average OTOC need not agree with the global Haar value, even at leading order. For simplicity, let us show how the discrepancy can be seen already in the $n = 2$ layer case for tensor product observables,  $A \equiv A \otimes \cdots \otimes A$ and $B\equiv B\otimes \cdots \otimes B$. In this case, we find that  
\begin{align}
    \OTOC^{(k)}_\mathcal{R} &=\frac{1}{D}\, \adjustbox{scale =\scalefactor}{\input{nonlocalops.tikz}}\nn\\
    &\!\approx \!\sum_{\pi_1 \leq \sigma_1 \leq \pi_2 \leq \sigma_2 \leq \gamma} \! \mu(\pi_1,\sigma_1)\mu(\pi_2,\sigma_2) \label{eq:nonlocal1}\\
    &\qquad \quad \times \braket{A}_{\pi_1} \! \braket{A}_{\pi_2} \! \braket{B}_{\sigma_1^{-1} \gamma} \! \braket{B}_{\sigma_2^{-1} \gamma}\nn.
\end{align}
Although the summation exhibits the structure of a non-crossing multichain, it cannot be simplified as in Eq.~\eqref{eq:multi2}, due to the separate dependence on the moments $\braket{A}_{\pi_2}$ and $\braket{B}_{\sigma_2^{-1} \gamma}$. 
We therefore find an $\mathcal{O}(1)$ discrepancy compared to the Haar case [Eq.~\eqref{eq:loOTOC}], justifying the final result for the OTOCs as depicted in the first row of Table~\ref{tab:example}. This argument directly generalizes to arbitrary non-local operators and $n$ layers.

\subsection{Subleading Corrections to the Free Approximation} \label{sec:subleading}
We are now in the position to analyze the subleading corrections to Eq.~\eqref{eq:lormpu}, specifically the $k$- and $n$-dependence of the terms proportional to $\chi^{-2}$. 
This turns out to be a challenging task. 
As discussed below Eq.~\eqref{eq:collins}, quantities whose leading-order terms follow the structure of non-crossing partitions exhibit subleading corrections that are significantly harder to characterize or bound tightly compared to the diagonal approximation. 
To gain insight into these terms---and to contrast them with quantities governed purely by the diagonal approximation---we first reexamine the global Haar case.

Recall that the error in the asymptotic expression for the Haar-averaged OTOC arises not only from subleading corrections to the Weingarten function (from Eq.~\eqref{eq:collins}), but also from permutations that violate the non-crossing partition geodesic condition (Eq.~\eqref{eq:geodesic}). In other words, by pairs of permutations satisfying
\begin{equation}
    \label{eq_geod_2}\dist(e,\pi)+\dist(\pi,\sigma)+\dist(\sigma,\gamma)=\dist(e,\gamma) + 2\ .
\end{equation} We can bound the resulting correction via a simple combinatorial counting argument (see App.~\ref{ap:otocProofs}),
\begin{equation}
\begin{split}
     |\OTOC^{(k)}_\mathbb{H}| &=|\OTOC^{(k)}_{\mathrm{FP}} +\frac{c_{k}(A,B)}{ D^{2}} + \mc{O}(D^{-4}) | \label{eq:haarsub}\\
     &\leq \left(1+\frac{6k^{7/2}}{D^2}\right) |\OTOC^{(k)}_{\mathrm{FP}}| + \frac{(k!)^2 C_k}{D^2} +\mc{O}(D^{-4}).
\end{split}
\end{equation}
Here, we recall that $\OTOC^{(k)}_{\mathrm{FP}} = \mathcal{O}(1)$ is the leading-order Haar value given in Eq.~\eqref{eq:loOTOC} and $C_k$ are the Catalan numbers, and we have assumed that $D>\sqrt{6}k^{7/4}$. The coefficient $c_{k}(A,B)$ depends only on $k$ and the (normalized) moments of $A$ and $B$, and is given exactly by the expression
\begin{align}
c_k(A,B) & :=\sum_{\pi\leq\sigma\leq \gamma} \mathrm{Wg}^{(1)}_{\pi,\sigma}(k) \braket{A }_\pi \braket{B }_{\sigma^{-1} \gamma}  \label{eq:ck_haar}
\\ &
+  \sum_{\dist(e,\pi)  \dist(\pi,\sigma)+\dist(\sigma,\gamma)= \ell(e, \gamma) + 2}  \mu(\pi,\sigma) \braket{A }_\pi \braket{B }_{\sigma^{-1} \gamma}\ ,
\nonumber
\end{align}
where we recall that $\mathrm{Wg}^{(1)}_{\pi,\sigma}(k)$ are the subleading coefficients for the asymptotic expansion of the Weingarten function from Eq.~\eqref{eq:collins}. These coefficients can be worked out explicitly for small $k$, see Eqs.~\eqref{ck_haar}-\eqref{eq:traceless_ck} in App.~\ref{ap:otocProofs}.

We emphasize that while the final line of Eq.~\eqref{eq:haarsub} is an exact relation, it provides only a loose upper bound. It could be refined, for instance, by counting the number of pairs of permutations $\{\pi,\sigma\}$ which satisfy Eq.\eqref{eq_geod_2}. However, to the best of our knowledge, no closed combinatorial expression for this count is currently known, although they are likely related to genus one permutations~\cite{walsh_counting_1972,sorbonne_universite_counting_2023}. In App.~\ref{ap:otocProofs} we explicitly calculate the number of such pairs of permutations for $k \leq 10$.

We now turn to the $\mathcal{O}(n \chi^{-2})$ correction for the RMPU calculation, Eq.~\eqref{eq:lormpu}. Specifically, our objective is to characterize the $n$- and $k$-dependence of the coefficient $\tilde{c}_{k,n}(A,B)$, for
\begin{equation}
    \OTOC^{(k)}_{\mathcal{R}}= \OTOC^{(k)}_{\mathrm{FP}} + \frac{\tilde{c}_{k,n}(A,B)}{\chi^2}+\mc{O}\left( {\chi^{-4}}\right). \label{eq:hahaha}
\end{equation}
Also in this case (cf. Eq.~\eqref{eq:ck_haar}), the corrections admit an explicit formula in terms of a permutation multi-chain condition, 
\begin{align}
   & \tilde{c}_{k,n}(A,B)=
n \sum_{\pi \leq \sigma \leq \gamma} {\rm Wg_{\sigma \pi}^{(1)}}(k) \braket{A}_{\pi} \braket{B}_{\sigma^{-1}\gamma}\nn
\\ & \, +
\sum_{ f_n(\vec \pi, \vec \sigma)=2 }
 \left( \prod_{i=1}^n \frac{\mu(\sigma_i^{-1} \pi_i) }{d^{  \dist(e,\pi_i)+\dist(\pi_i, \sigma_i)+\dist( \sigma_i,\gamma) }} \right)\braket{A}_{\pi_1} \braket{B}_{\sigma_n^{-1}\gamma}    \, , \label{eq:anal_ck_rmpu}
    \end{align}
where $f_n(\vec \pi, \vec \sigma):=\dist(e,\pi_1)+\dist(\pi_1,\sigma_1)+\dots+\dist(\sigma_n,\gamma)$ is defined analogously to Eq.~\eqref{eq:f}. This can be proven through repeated application of the formula for the subleading correction in the Haar case.
By explicit computation, we find a relation between the RMPU corrections and the Haar ones (see Eqs.~\eqref{eq_ck_rmpu} in App.~\ref{ap:otocProofs}). For $k>2$ this reads:
\begin{equation}
  \tilde{c}_{k,n}(A,B)=  \left(\frac n{d^2} - (n-1)\right) \, a_k(A,B) + \frac {b_k(A,B)}{d^{2n}}, \label{eq:rmpucoeff}
\end{equation}
where $a_k(A,B)$ and $b_k(A,B)$ are coefficients that depend only on $k$ and the separate moments of $A$ and $B$, and satisfy
\begin{equation}
    a_k(A,B)+b_k(A,B)= c_k(A,B), \label{eq:abc}
\end{equation}
with $c_k(A,B)$ the Haar correction coefficient defined in Eq.~\eqref{eq:haarsub}. The case $k = 2$ is qualitatively different, and we discuss this case at the end of this section. 
We note that Eq.~\eqref{eq:rmpucoeff} is a conjectured form, supported by exact symbolic computations for fixed $n$ and $k$. 
Characterizing $a_k(A,B)$ and $b_k(A,B)$ analytically remains an open problem. It would require significant progress on two fronts: (i) a better understanding of power series expansion of the Weingarten symbols, for which no closed-form expressions for the subleading coefficients are currently known~\cite{collins2017weinga}; and (ii) a combinatorial characterization of permutation chains beyond the non-crossing partition lattice~\cite{walsh_counting_1972,sorbonne_universite_counting_2023}. We discuss this in more detail in App.~\ref{ap:otocProofs}.

In the traceless case, $\tr[A]=\tr[B]=0$, we find that $a_k(A,B) = 0$ and so $c_k(A,B)=b_k(A,B)$. Recalling also that the $\mc{O}(1)$ terms from free probability vanish in this case (see Eq.~\eqref{eq:singleton}), the $\mathcal{O}(D^{-2})$ terms become leading order,
\begin{equation}
    \OTOC^{(k)}_{\mathbb{H}}\vert_{\tr[A]=\tr[B]=0} = \frac{b_k(A,B)}{D^2} + \mc{O}(D^{-4}). \label{eq:lotraceless}
\end{equation}
For the RMPU ensemble, where the total Hilbert space dimension is $D=\chi d^{2n}$ (see Fig.~\ref{fig:rmpu}), we similarly obtain 
\begin{align}
    \OTOC_\mathcal{R}^{(k)} \vert_{\tr[A]=\tr[B]=0}&= \frac {b_k(A,B)}{d^{2n} \chi^2} + \mc{O}(\chi^{-4}) \\
    &= \frac{1}{D^2} \lim_{D \to \infty} \left(D^2  \OTOC^{(k)}_\mathbb{H}\right) + \mc{O}(\chi^{-4}). \label{eq:tracelesstohaar}
\end{align}
Using this, we can compute the relative multiplicative error in approximating the Haar value of the OTOC,
\begin{align}
    \Delta \OTOC^{(k)}_{\mc{R}}\vert_{\tr[A]=\tr[B]=0} &:=  \left|\frac{\OTOC^{(k)}_\mathcal{R}-\OTOC^{(k)}_\mathbb{H}}{ \OTOC^{(k)}_\mathbb{H}} \right| \label{eq:relative}= \mc{O}\left({D^{2}}{\chi^{-4}}\right) .
\end{align}
This result contrasts with the non-zero trace case, where the Haar value of the OTOC remains $\mathcal{O}(1)$. For traceless observables, the Haar value is exponentially small in system size and so accurately reproducing it with RMPUs demands a volume-law operator entangled unitary. 
In other words, while a polynomial bond dimension $\chi=\mathrm{poly}(N)$ suffices for finite trace observables, capturing the correct scaling of OTOCs of traceless observables requires $\chi=\mathrm{exp}(N)$, highlighting a qualitative separation in complexity between the two cases.

There is an important caveat for $k = 2$. In this case, the OTOC reduces to the standard $4$-point version which has been widely studied in the literature~\cite{Shenker_Stanford_2014,Maldacena2016-rl,Swingle2018,dowling2023scrambling,west2025graphth}:
\begin{equation}
    \OTOC^{(2)}_{\mc{E}}= \int_{U \sim \mc{E}}\frac{1}{D}\tr[A_U B A_U B].
\end{equation}
In contrast to Eq.~\eqref{eq:rmpucoeff}, we find that 
\begin{align}
\tilde{c}_{2,n}(A,B) &= \left(\frac n{d^2} -(n-1)\right) c_2(A, B) \\
    &
     = -\left(\frac n{d^2} -(n-1)\right)
(\braket{A^2}-\braket{A}^2)(\braket{B^2}-\braket{B}^2).\nn
\end{align}
This results from the fact that $b_2(A,B) \equiv 0$ in Eq.~\eqref{eq:abc}, and so this $\mc{O}(\chi^{-2})$ RMPU coefficient for the $4$-OTOC is the same for both traceless and finite trace observables. 
It is worth noting that $k=2$ is a special case also for designs: an $N$-qudit unitary $2$-design can be produced exactly using only Clifford gates~\cite{zhu2016cliffordgroupfailsgracefully}.

\subsection{Frame potential} \label{sec:frame}
We will now contrast the above results for OTOCs for the RMPU ensemble with a common probe of unitary design: the unitary frame potential. This measure is related to the average value of higher-order OTOCs on different observables~\cite{roberts2017chaos}, thus measuring average-case freeness. 
In particular, the frame potential quantifies the closeness of an ensemble $\mc{E}$ to a unitary $k$-design via a $2$-norm distance with respect to the moment operators defined in Eq.~\eqref{eq:moment} (see App.~\ref{ap:a}),
\begin{equation}
\begin{split}
    \mathcal{F}^{(k)}_{\mc{E}} &:= \int_{U,V \sim \mc{E}} |\tr[U V^\dagger ]|^{2k} \\&= \int_{U,V \sim \mc{E}}\mathrm{tr}\left[(U\otimes U^\ast)^{\otimes k}(V^\dagger \otimes V^T)^{\otimes k}\right] \\
    &= \tr[\tilde{\Phi}_{\mc{E}}^{(k)} (\tilde{\Phi}_{\mc{E}}^{(k)})^\dagger]\;,\label{eq:framepotMain}
\end{split}
\end{equation}
where in the second step we used the replica method and $\mathrm{tr}(A^T)=\mathrm{tr}(A)$. Relevant to our setting, we recall a key result from Ref.~\cite{roberts2017chaos} which relates the frame potential to a certain average of higher-order OTOCs
\begin{equation}
    \overline{|\OTOC^{(k)}_{\mc{E}}|^2}\! :=\!\sum_{A^{(i)},B^{(i)} \in \mathcal{P}_N} \!   \left[\frac{{\OTOC^{(k)}_{\mc{E}}(\vec{A}_U,\vec{B}) }}{D^{2k}}  \right]^2 = \frac{\mathcal{F}^{(k)}_{\mc{E}}}{D^{2(k+1)}} \;. \label{eq:otocvsfp}
\end{equation}
On the left-hand side, we average the square ensemble-averaged $2k$-point OTOC over all possible $N$-qudit Pauli strings $\{A^{(i)} \}_{i=1}^{k},\{B^{(i)} \}_{i=1}^{k} \in \mc{P}_N$, using the more general definition of the OTOC which allows for different operators for its arguments, $\vec{A}_U:= (A^{(1)}_U,\dots, A^{(k)}_U )$ and $\vec{B}_U:= (B^{(1)},\dots, B^{(k)} )$. Then Eqs.~\eqref{eq:otoc1}-\eqref{eq:otoc} generalize to: 
\begin{equation}
    \OTOC^{(k)}_{\mc{E}}(\vec{A}_U,\vec{B}) := \int_{U \sim \mc{E}} \left\langle A^{(1)}_U B^{(1)} \dots A^{(k)}_U B^{(k)} \right\rangle.
\end{equation}
Note that the concept of free independence extends directly to this setting: see e.g. Ref.~\cite{nica2006lectures,SilviaDesigns}. In the context of the free probability framework, Eq.~\eqref{eq:otocvsfp} then supplies a new meaning for the frame potential of $\mc{E}$: \textit{on average across all Pauli observables, how close to being free is a distribution of $k$ observables rotated by $U\sim \mc{E}$, compared to another constant set of $k$ observables? }

It is instructive to recall the frame potential for the Haar ensemble, before presenting our results for RMPUs. Using the Haar moment operator [Eq.~\eqref{eq:transfer}], one obtains immediately that
\begin{equation}
\begin{split}
\mathcal{F}^{(k)}_{\mathbb{H}}=&\sum_{{\mu, \nu, \pi, \sigma} \in S_k} \mathrm{Wg}_{\pi, \sigma} G_{\sigma,\mu}\mathrm{Wg}_{\mu, \nu} G_{\nu,\pi}\\&= \sum_{\pi,\sigma\in S_k}\delta_{\pi,\sigma}= k!\;,\label{eq:framepotHaar}
\end{split}
\end{equation}
where we have used that $\mathrm{Wg}(D,k)$ is the pseudoinverse of the Gram matrix $G_{\pi,\sigma}(D,k)=D^{k-\dist(\pi,\sigma)}$. 

In App.~\ref{ap:framepot} we compute the frame potential for the RMPU ensemble exactly to the first two leading orders for large $\chi$, which we shortly review the steps of here.
The first step in this derivation is to substitute the moment operator expression for the RMPU ensemble, Eq.~\eqref{eq:momentrmpu}, into Eq.~\eqref{eq:framepotMain} to obtain
\begin{align}
\mathcal{F}^{(k)}_{\mc{R}}&=\tr[\tilde{\Phi}^{(k)}_{\mc{R}} (\tilde{\Phi}^{(k)}_{\mc{R}})^\dagger ] \label{eq:fp_rmpu_graph} \\
     &= \, \input{rmpu1.tikz}\nn \; .
\end{align}
Recalling the notation introduced in Eq.~\eqref{eq:graphical}, the expression above can be derived via graphical manipulations and by applying again the inverse relation between the Weingarten and Gram matrices to eliminate two pairs of each. 
The task is then to carry out an asymptotic expansion of the matrix components appearing in Eq.~\eqref{eq:fp_rmpu_graph}.
Unlike the OTOC [Eq.~\eqref{eq:rmpumomentotoc}], Eq.~\eqref{eq:fp_rmpu_graph} does not feature the boundary conditions associated with $T_e$ and $T_\gamma$. As a result, non-crossing partitions do not appear in our asymptotic analysis.

Eq.~\eqref{eq:fp_rmpu_graph} admits a natural interpretation as the partition function of a classical statistical mechanics model: links carry permutation-valued “spins” $\pi \in S_k$, while circles and squares correspond to fixed two- and four-body interaction vertices that encode the underlying tensor network structure. 
The bond dimension $\chi$ plays the role of an inverse temperature or, equivalently, a strong ferromagnetic field—favoring configurations that minimize permutation distances $\dist(\pi,\sigma)$, cf. Sec.~\ref{sec:diag}.
This perspective enables a systematic ``low-temperature expansion'' of the frame potential, which can be resummed exactly at first subleading order in $1/\chi$. We find the frame potential of an $n$-layer RMPU with bond dimension $\chi$ to be
\begin{equation}
\begin{split}
    \mathcal{F}^{(k)}_{\mc{R}} &= \mathcal{F}^{(k)}_{\mathbb{H}}\left[ 1 + \frac{k(k-1)}{2 \chi^2}\left( n-1 -\frac{n}{d^2} 
    \right)\right] + \mc{O}({\chi^{-3}})\;,\label{eq:framepotcalc}
\end{split}
\end{equation}
constituting one of our main analytical results. Interestingly, the coefficients of the $\chi^{-2}$ scaling are similar to those numerically found in Ref.~\cite{Lami2025} for the state frame potential, signaling that the equivalence between the two measure of design goes beyond the pure Haar limit. In the neglected terms $\mc{O}({\chi^{-3}})$, we have implicitly assumed that $D^{-2} \leq\chi^{-3}$, which covers the most relevant cases, e.g. when $\chi \sim \mathrm{poly(N)}$. In contrast to the more intricate considerations for the subleading expressions in the previous section, this expression is rigorously proven. Notably, the derivation requires only the diagonal approximation [Eq.~\eqref{eq:diag}] and its subleading corrections, agreeing with the general intuition that such an approximation is sufficient to prove (state) design properties of a unitary ensemble. 

Otherwise, from Eq.~\eqref{eq:otocvsfp} we know that the frame potential also captures an average-case behavior of the ensemble-averaged OTOCs. Since the overall scaling prefactor $1/D^{2k+2}$ on the right-hand-side does not affect relative differences, from Eq.~\eqref{eq:framepotcalc} we find 
\begin{equation}
    \Delta \overline{|\OTOC^{(k)}_{\mc{R}}|^2} = \frac{k(k-1)}{2 \chi^2}\left( n-1 -\frac{n}{d^2} +\frac{1}{d^{2n}}\right) + \mc{O}\left(\frac{1}{\chi^3}\right), \label{eq:finalfp}
\end{equation}
where, as throughout this work, ``$\Delta$'' denotes the relative error in the quantity between the RMPU and Haar ensembles (cf. Eq.~\eqref{eq:relative}). 
Hence, for $\chi = \mathrm{poly}(N)$, the average value of the square OTOC over all Pauli observables becomes indistinguishable from the Haar value. In other words: \textit{on average for global observables, rotating one $k$-length set of operators by the RMPU ensemble causes it to be freely independent with respect to any other}.

We also point out another interpretation of the frame potential result, related to insights from state design literature~\cite{Lami2025,DeLuca2024Frame}. Eq.~\eqref{eq:framepotcalc} directly determines the probability distribution
\begin{equation}
    \mathcal{P}(w) = \int_{U,V\sim \mathcal{R}} \delta(w - |\mathrm{tr}(U^\dagger V)|^2),
\end{equation}  
for the random variable $w = |\mathrm{tr}(U^\dagger V)|^2$, with $U, V$ sampled independently from the ensemble $\mathcal{R}$. The frame potential then corresponds to the moments of this distribution $\mathcal{F}^{(k)}_{\mathcal{R}} = \int dw\, w^k \mathcal{P}(w)$. Consider the many-layer limit, $n := N-r = \mc{O}({N})$. Then we can introduce the scaling variable $x = \frac{N}{\chi^2}\left(1 - \frac{1}{d^2}\right)$, 
assume that both $\chi$ and $N$ are large, and then rewrite Eq.~\eqref{eq:framepotcalc} as $\mathcal{F}^{(k)}_{\mathcal{R}} =\mathcal{F}^{(k)}_{\mathbb{H}} \exp\left[\frac{k(k-1)}{2} x\right]$, up to subleading corrections $\chi^{-3}$. 
This expression underscores that the moments of $\mathcal{P}(w)$ are the product of the Haar moments $\mathcal{F}^{(k)}_{\mathbb{H}}=k!$ with those of the lognormal distribution $\mathcal{P}_\mathrm{LN}(w)=\exp\left[{-\frac{(\log(w)+x/2)^2}{x}}\right]/\sqrt{\pi x}$. 
Hence, $\mathcal{P}(w)$ is the convolution,
\begin{equation}
    \mathcal{P}(w) = [\mathcal{P}_\mathrm{Haar} \star \mathcal{P}_\mathrm{LN}](w),
\end{equation}
between $\mathcal{P}_\mathrm{Haar}(w) = e^{-w} $ and $\mathcal{P}_\mathrm{LN}$.

We conclude this section by emphasizing that the frame potential is a priori inequivalent to the definitions of unitary design considered in previous work on low-depth circuits, such as relative error designs~\cite{schuster2024randomun,laracuente_approximate_2024}; see App.~\ref{ap:a}. Therefore, our result in Eq.~\eqref{eq:framepotcalc} has two main implications:
(i) it adds to the growing body of evidence that a polynomial bond dimension suffices to produce approximate unitary designs
; and (ii) that on-average, freeness emerges on average over global observables, despite the frame potential depending solely on the diagonal approximation.

\section{Discussion} 
\label{sec:discussion}
In this work, we introduced and analytically studied the ensemble of \emph{random matrix product unitaries} (RMPUs) as a minimal and tractable model for probing the emergence of randomness encoded in local operators in quantum many-body systems. Namely, we investigated how well this ensemble produces observables which are \textit{freely independent} compared to another set of constant observables. This is a hallmark property of sufficiently large and random matrix ensembles.
We analytically demonstrated that RMPUs with only polynomial bond dimension $\chi$ replicate Haar-random values of higher-point out-of-time-ordered correlators for local, finite trace observables, while in contrast, an exponentially large bond dimension is necessary to produce the Haar value of traceless observables for replicas $k\geq 3$. This marks an intriguingly sharp boundary between what features of random unitaries are ``easy'' and what are ``hard'' to reproduce. Our derivations relied on a large-$\chi$ analysis that goes beyond the local perturbative diagonal approximation for the Weingarten calculus, harnessing the random matrix theory toolbox of free probability. We further calculated the frame potential of the RMPU ensemble exactly to second order, verifying that it converges to the Haar value with deviations $\mc{O}(N \chi^{-2})$, 
widening the set of unitary design properties emerging already at polynomial bond dimension. At the same time, from the result of Ref.~\cite{roberts2017chaos}, our frame potential result informed us about the average-case emergence of freeness for the RMPU ensemble: that typically, global and traceless observables are approximately free for polynomial bond dimension. \\

Our results have broad implications and raise pressing questions across both many-body physics and quantum information. \\

First, the RMPU ensemble captures key signatures of thermalization and quantum chaos, providing a \emph{toy model for more realistic dynamics from locally-interacting Hamiltonian or Floquet structures}. Many-body systems satisfying the ETH have also been argued to lead to freeness for local observables~\cite{Foini2019,Pappalardi2022-vo}, whereby individual high-energy eigenstates of chaotic Hamiltonians locally mimic the features of a thermal (Gibbs) state~\cite{Deutsch1991,Srednicki}. Therefore, the recovery of Haar values of higher-order OTOCs demonstrates on firm ground the fast emergence of the predictions of the ETH~\cite{SilviaDesigns}. Our results are therefore complementary to the recent toy model of freeness discussed in Ref.~\cite{fritzsch2025freeprob}, where the authors prove an exponentially fast approach to freeness for global observables. In contrast, our results explicitly depend on the locality of the observables, and can be readily compared to previous ans\"atze for anticoncentration of the wavefunction~\cite{Lami2025,magni2025anticliff,sauliere2025unianticoncen} and emergence of unitary design~\cite{schuster2024randomun,west2025nogot,grevink2025glueshortdepthdesignsunitary}. \\

Further, the dependence of our results on the bond dimension $\chi$ represents \emph{a proxy for time-dependence at which freeness is reached}: namely, for non-localized dynamics, the operator entanglement of $U$ tends to grow linearly with time $t$, implying $\chi \sim d^t$~\cite{Zhou2017}. For circuit implementations, this translates into corrections $\mc{O}(\exp[{-2v_E t}])$, with $v_E$ the entanglement velocity~\cite{Nahum2017}. Thus, our findings translate into a dynamical statement: approximate freeness, and thus thermal behavior for OTOCs, can arise in times 
\[t\sim \log (\chi) \sim \log (N) \ , \] i.e. at logarithmic depth. Moreover, in contrast to the model of Ref.~\cite{fritzsch2025freeprob}, the time scale of the emergence of freeness is $k$-dependent. This suggests that {entanglement spreading at early times is sufficient to wash out higher-point correlations between local operators}, at least provided they have non-zero trace. This motivates the search for more physical systems for which the predictions of free probability emerge at short time scales, such as local random brickwork circuits~\cite{Nahum2017,Nahum2018}.\\

Next, we foresee direct implications regarding \emph{quantum advantage in the generation of random unitaries}. Beyond unitary design properties being reachable with polynomial bond dimension~\cite{schuster2024randomun,cui2025unitarydesignsnearlyoptimal}, our results indicate that also free independence of relevant observables can be replicated with similar cost. This prompts the question of which quantities can witness the genuine complexity of deep, random quantum circuits? OTOCs of traceless or global observables seem to be examples. Our results therefore motivate the development of more refined randomness diagnostics and distance measures that capture the emergence of freeness in multitime correlations. Relating freeness to the simulability of Heisenberg operators~\cite{Prosen2007,dowling2025} is an interesting research direction along these lines: How does the generation of non-classical resources impact the statistics of multitime correlations~\cite{dowling2023scrambling,Dowlin2024LOE-OSRE}?\\

From the viewpoint of quantum information, our work provides \emph{a precise characterization of the statistical properties of the RMPU in terms of the free independence}. A natural question is how these results would read for generic restricted gate sets, such as Clifford circuits. While Haar-random Cliffords and unitaries coincide up to the third moment in qubit systems~\cite{zhu2016cliffordgroupfailsgracefully}, it remains an open question whether the commutant of the full Clifford group exhibits any free probability structure asymptotically~\cite{bittel2025completethe}. 
Moreover, we should compare this setting to recent no-go results for sub-linear depth relative-error designs over other Haar ensembles, including the Clifford group~\cite{west2025nogot,grevink2025glueshortdepthdesignsunitary}. 
Probing freeness in these groups will further inform us of the distinction and similarities between unitary designs and the emergence of free independence. Some notion of {partial freeness} for specific classes of observables (e.g., diagonal or stabilizer-preserving operators) could still be viable within such restricted ensembles. Results in this direction would be consistent with the fast emergence of a modified notion of anticoncentration for different group ensembles~\cite{magni2025anticliff,sauliere2025unianticoncen}. \\

From a \emph{mathematical perspective}, we arrived at pressing questions regarding the approximation to free independence at finite dimension, even for the simpler case of globally Haar-random matrices. Refs.~\cite{collins2017weinga,aharonov_quantum_2022} provide results for the asymptotics of the Weingarten functions which lay the foundation for seminal results on short-depth unitary design. As detailed in Sec.~\ref{sec:methods}, the asymptotics of the Weingarten function is insufficient to compute the corrections to freeness at finite $D$. Indeed, more work needs to be done to characterize the permutations that satisfy the modified multichain condition (cf. the discussion around Eq.~\eqref{eq:anal_ck_rmpu}), which contribute on equal footing with the subleading corrections to the Weingarten functions. These appear to be related to the class of genus $1$ permutations~\cite{walsh_counting_1972,sorbonne_universite_counting_2023}. Moreover, the concept of second-order freeness may offer insight into formal quantifiers of the emergence of free probability from random circuit ensembles~\cite{collins_second_2007}. \\

Finally, in terms of \emph{experiments}, our findings suggest that approximate freeness may be observable at modest circuit depths. 
Since OTOCs are experimentally accessible via interferometric~\cite{garttner_measuring_2017,PhysRevX.7.031011,abanin2025const} or randomized measurement protocols \cite{PhysRevLett.124.240505}, this raises the exciting possibility that freeness could serve as an operational benchmark of quantum chaos in near-term devices. Our results thus invite experimental efforts to measure the onset of freeness in programmable quantum systems, and to explore how it correlates with conventional benchmarks like entanglement growth or fidelity decay.


\begin{acknowledgments}
    We are thankful to Pieter Claeys and Benoit Collins for useful discussions and to Guglielmo Lami for close collaborations on related topics. We thank Thomas Schuster for helpful discussion on unitary designs, and Gregory White for useful discussions and comments on the manuscript. N.D., M.H., S.P., X.T. acknowledge funding by the Deutsche Forschungsgemeinschaft (DFG, German Research Foundation) under Germany’s Excellence Strategy - Cluster of Excellence Matter and Light for Quantum Computing (ML4Q) EXC 2004/1 - 390534769. S.P. and X.T. acknowledge DFG Collaborative Research Center (CRC) 183 Project No. 277101999 - project B02 and B01. M. H. acknowledges funding by the Deutsche Forschungsgemeinschaft (DFG, German Research Foundation) - 547595784.
    J.D.N. is funded by the ERC Starting Grant 101042293 (HEPIQ) and the ANR-22-CPJ1-0021-01.
\end{acknowledgments}


%


\onecolumngrid
\appendix
\newpage

\section*{Appendices}


\section{Approximate Unitary Designs do not Imply Freeness}\label{ap:a}
Here we overview the basics of unitary designs, loosely following Refs.~\cite{low2010pseudorando,Mele_2024}. In particular, we will define relevant definitions of unitary designs used in the literature and relations between them. Important for the results of this work, we explain why the commonly used definitions of designs do not imply free independence of higher-order correlations: namely, that $2k$-point OTOCs [Eq.~\eqref{eq:otoc}] twirled over approximate designs do not necessarily attain the Haar value; cf. Table~\ref{tab:example}.

Suppose we wish to approximately generate a $k$-design, such that Eq.~\eqref{eq:deisgn_cond} is satisfied up to some allowable error $\varepsilon$. There are inequivalent ways to measure this error, and to understand these (and their connections) we first need to define distance measures on quantum channels. The natural class of distance measures between two quantum channels is defined in terms of the superoperator norm of a map $\mc{X}:\mc{H}_A \to \mc{H}_B$, which depends on the matrix norm of the input versus output spaces,
\begin{equation}
    \| \mc{X} \|_{p \to q} := \sup_{Y \neq 0} \frac{\| \mc{X}(Y) \|_q}{\|Y \|_p}. \label{eq:channelnorm}
\end{equation}
Here, $\|Y \|_p$ is the Schatten $p-$norm of the matrix $Y$. Different choices of $p$ and $q$ define different measures~\cite{wilde_2017}. If one is interested in channels on quantum states, a natural choice is $p=q=1$ as the one-norm defines the canonical distance between states~\cite{wilde_2017}. The diamond norm distance is then defined between the two channels $\mc{X}$ and $\mc{Y}$ acting on the Hilbert space $\mc{H}$ as
\begin{align}
    \| \mc{X}- \mc{Y} \|_{\diamond} &:= \sup_{\mathfrak{D}} \| (\mc{X} - \mc{Y})\otimes \mc{I}_{\mathfrak{D}} \|_{1 \to 1} \label{eq:diamond1}\\
    &=\sup_{\ket{\psi} \in \mc{H}^{\otimes 2}} \| (\mc{X}-\mc{Y})\otimes \mc{I} (\ket{\psi}\!\bra{\psi}) \|_1 \label{eq:diamond2}
\end{align}
where $\mc{I}_{\mathfrak{D}}$ is the identity channel on a replica space of some dimension $\mathfrak{D}$. The second line here is a non-trivial result (see e.g. Ref.~\cite{wilde2020}) which tells us that it is equivalent to optimize the suprema of Eqs.~\eqref{eq:channelnorm} and \eqref{eq:diamond1} by instead only optimizing with respect to pure states on a doubled Hilbert space. Note that allowing access to an ancilla space provides a better ability to distinguish quantum channels in general. An $\varepsilon$-approximate \emph{additive error $k$-design} is then defined as an ensemble $\mc{E}$ such that
\begin{equation}
    \| \Phi_{\mathcal{E}}^{(k)}- \Phi_{\mathbb{H}}^{(k)}\|_{\diamond} \leq \varepsilon. \label{eq:additive}
\end{equation}
A careful reader will notice that being an additive error, Eq.~\eqref{eq:additive} indicates that if a quantity of interest is exponentially small in all relevant cases, then one would require an exponentially small $\varepsilon$ for the design to faithfully replicate the Haar value. For instance, the purity of a random state is $\mc{O}(\exp(-N))$, and so for a design to reproduce the Page curve, one would also require $\varepsilon = \mc{O}(\exp(-N))$ in Eq.~\eqref{eq:additive}. This motivates the introduction of an $\varepsilon$-approximate \emph{relative error $k$-design} $\mc{E}$, defined via the condition~\cite{Brand_o_2016}
\begin{equation}
    (1-\varepsilon) \Phi_{\mathbb{H}}^{(k)} \preceq \Phi_{\mathcal{E}}^{(k)}  \preceq (1+\varepsilon) \Phi_{\mathbb{H}}^{(k)}. \label{eq:relError}
\end{equation}
Here, `$\Phi_A \preceq \Phi_B$' is the complete semidefinite ordering of channels, meaning that $\Phi_B - \Phi_A$ is a completely positive map (i.e., it maps quantum states to possibly unnormalized quantum states~\cite{wilde_2017}). To make sense of this, we note that an immediate corollary of Eq.~\eqref{eq:relError}, presented in the final row of Table~\ref{tab:example}: for any positive matrices $X,Y \geq 0$, an $\varepsilon$-approximate relative error design $\mc{E}$ satisfies 
\begin{equation}
    \Delta \tr[\Phi_{\mc{E}}^{(k)}(X)Y ] := \left| \frac{\tr[\Phi_{\mc{E}}^{(k)}(X)Y ] -\tr[\Phi_{\mathbb{H}}^{(k)}(X)Y ] }{\tr[\Phi_{\mathbb{H}}^{(k)}(X)Y ]}\right|\leq \varepsilon.
\end{equation}
For example, for a relative error design, the higher moments of a projective measurement outcome of a random quantum state are assured to be relatively $\varepsilon$-close to the Haar prediction. We stress that, unlike the additive error definition which bounds the one-norm distance for \textit{any} observable (Eqs.~\eqref{eq:channelnorm}-\eqref{eq:diamond2}), relative error designs explicitly depend on the positivity of both the input $X$ and measurement $Y$.

\begin{figure}
    \centering
    \includegraphics[width=0.4\linewidth]{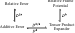}
   \caption{Known relations between unitary design definitions as discussed in this work. An arrow between definitions, $A \rightarrow B$, possibly accompanied with a dimensional factor $D^a$ indicates that an $\varepsilon$-approximate design according to definition $A$ is necessarily a $(D^a\varepsilon)$-approximate design according to definition $B$. No factor means there is only a constant cost going from one to the other. The relations between relative error [Eq.~\eqref{eq:relError}] and additive error [Eq.~\eqref{eq:additive}] designs is proven in Ref.~\cite{Brand_o_2016}, between additive error and tensor product expander [Eq.~\eqref{eq:tpe}] is proven in Ref.~\cite{low2010pseudorando}, and the relation between tensor product expander and relative frame potential [Eq.~\eqref{eq:relframedesign}] designs is immediate from first noticing that $\| {\Phi}_\mathcal{E}^{(k)} -{\Phi}_{\mathbb{H}}^{(k)} \|_{2  \gets 2} = \| \tilde{\Phi}_\mathcal{E}^{(k)} -\tilde{\Phi}_{\mathbb{H}}^{(k)} \|_{\infty}$ by definition, and then using the inequalities between matrix Schatten norms: for $0 < r < p$, $\|X \|_p \leq \|X \|_r \leq d^{1/r - 1/p} \|X\|_p$.}
    \label{fig:designs}
\end{figure}

One can also define a design in terms of different distant metrics. An $\varepsilon$-approximate \textit{relative frame potential} design satisfies [Eq.~\eqref{eq:framepotMain}]
\begin{equation}
    \|\tilde{\Phi}_\mathcal{E}^{(k)} -\tilde{\Phi}_{\mathbb{H}}^{(k)} \|_2 = \sqrt{\mc{F}^{(k)}_{\mathcal{E}} - \mc{F}_{\mathbb{H}}^{(k)} } \leq \varepsilon \label{eq:relframedesign}
\end{equation}
which we notice is equivalent to a $2$-norm distance between moment operators [Eq.~\eqref{eq:moment}].
Finally, a $\varepsilon$-approximate \textit{tensor product expander} design (TPE) is defined in terms of the $2 \to 2$ channel norm distance between twirls,   
\begin{equation}
    \| {\Phi}_\mathcal{E}^{(k)} -{\Phi}_{\mathbb{H}}^{(k)} \|_{2  \gets 2} \leq \varepsilon. \label{eq:tpe}
\end{equation}
Notably, the above discussed measures of approximate unitary designs are generally equivalent to each other only up to exponential factors; see Fig.~\ref{fig:designs}. Therefore, proving that an ensemble forms a design according to one definition, for polynomial bond dimension (or in logarithmic depth) does not necessarily imply that it is an efficient design according to another. It is an interesting open question what further relations can be derived between the discussed metrics, particularly if those relations come with a sub-exponential cost to the error.

So why do the relative and additive error unitary design not ensure the attaining of the approximate Haar value of OTOCs [Eq.~\eqref{eq:otoc}]? First, for relative error designs, the positive semi-definite ordering of condition Eq.~\eqref{eq:relError} means the error $\varepsilon$ only bounds (in a relative sense) quantities with positive inputs (e.g., replicas of quantum states) and positive observables (e.g., projective measurements). Clearly, in the $k$-replica representation of the OTOC,
\begin{equation}
    F=\frac{1}{D}\tr[\Phi(A^{\otimes k}) B^{\otimes k} T_{\gamma} ]
\end{equation}
neither the input $X\equiv A^{\otimes k}$ nor the `measurement' $Y \equiv B^{\otimes k} T_{\gamma} $ are generally positive, and so the relative error design condition does not encompass OTOCs. 
Additive error designs do not have the same constraint. Recalling the channel norm definition of Eq.~\eqref{eq:channelnorm} together with Eq.~\eqref{eq:diamond1}, and assuming that the unitary ensemble $\mc{E}$ forms an additive error design with error $\varepsilon$, we have 
\begin{equation}
     \sup_{X \neq 0} \frac{\| (\Phi_{\mc{E}}- \Phi_{\mathbb{H}})[X] \|_1}{\|X \|_1} \leq \| \Phi_{\mc{E}} - \Phi_{\mathbb{H}} \|_{\diamond} \leq \varepsilon. \label{eq:bounddia}
\end{equation}
That is, the additive error bounds the distance for any operator input to the twirl channels. From the definition of the ensemble-averaged OTOCs [Eq.~\eqref{eq:otoc}], we can only bound its relative value as
\begin{equation}
    |\OTOC^{(k)}_{\mc{R}}- \OTOC^{(k)}_{\mathbb{H}}| \leq \frac{1}{D}|\tr[(\Phi_{\mc{E}}- \Phi_{\mathbb{H}})[A^{\otimes k}] B^{\otimes k} T_{\gamma} ] |\leq \frac{1}{D} \|(\Phi_{\mc{E}}- \Phi_{\mathbb{H}})[A^{\otimes k}] \|_1 \| B^{\otimes k} T_{\gamma} \|_\infty \leq \frac{1}{D}\|(\Phi_{\mc{E}}- \Phi_{\mathbb{H}})[A^{\otimes k}] \|_1, \label{eq:useholders}
\end{equation}
where we have first applied H\"older's inequality and then used the fact that $T_\gamma $ is unitary together with the assumption that the operators $A,B$ have bounded operator norm, $\| A\|_\infty ,\| B\|_\infty \leq 1$. Then the normalization of \eqref{eq:bounddia} means that 
\begin{equation}
    |\OTOC^{(k)}_{\mc{R}}- \OTOC^{(k)}_{\mathbb{H}}|  \leq \frac{1}{D}\frac{\|(\Phi_{\mc{E}}- \Phi_{\mathbb{H}})[A^{\otimes k}] \|_1}{\|A^{\otimes k} \|_1} \|A^{\otimes k} \|_1 \leq \frac{\varepsilon}{D} \|A^{\otimes k} \|_1  \leq D^{k-1} \varepsilon .
\end{equation}
where we have taken $\| A \|_1 = \tr[ |A| ] \leq D$. Note that this final inequality is saturated for $A,B$ unitary, where $\| A \|_1 =\| B \|_1 = \tr[\id]= D$. It is an open question whether the above inequalities can be saturated in typical scenarios, or whether improved bounds can be derived (particularly with respect to the application of H\"older's inequality in Eq.~\eqref{eq:useholders}). The above discussion nonetheless strongly suggests that an additive error design need not lead to freeness, unless $\varepsilon$ is already exponentially small. This should be compared to Eq.~\eqref{eq:otocvsfp} from the main text, which presents how a certain averaged OTOC is proportional to the frame potential~\cite{roberts2017chaos}.

\section{Calculation of OTOCs for the RMPU ensemble }\label{ap:otocProofs}
We now provide the details underlying the results of Secs~\ref{sec:loRMPU} and \ref{sec:subleading}. 

\subsection{Leading Order OTOC calculation for arbitrary number of layers and replicas}
We begin by extending the leading-order calculation of the twirled OTOC for the $2$-layer RMPU [Eq.~\eqref{eq:rmpuotoc1}] to arbitrary number of layers $n$. For simplicity, we consider observables $A$ and $B$ with finite trace. As in the main text (cf. Eq.~\eqref{eq:rmpumomentotoc}), from the linearity of the OTOC with respect to the twirl channel and the independence of these component Haar integrals allow us to directly replace each local ($2k$-replica) unitary block with its corresponding moment operator, yielding 
\begin{equation}
   \OTOC^{(k)}_\mathcal{R}= \sum_{\sigma_i,\pi_i \in S_k} \frac{\left(\prod_{j=1}^n\mu(\pi_j,\sigma_j)\right) \braket{A}_{\pi_1} \braket{B}_{\sigma_n^{-1} \gamma}+\mc{O}(\chi^{-2})}{\chi^{f_n(\vec{\pi},\vec{\sigma}) } d^{g_n(\vec{\pi},\vec{\sigma})}}, \label{eq:rmpuotoc1_app}
\end{equation}
with the generalized exponent functions,
\begin{equation}
    \begin{split}
        f_n(\vec{\pi},\vec{\sigma}) &:= 1-k+ \dist(e,\pi_1) + \dist(\pi_1,\sigma_1) + \dist(\sigma_1,\pi_2) + \dist(\pi_2,\sigma_2) +\dots + \dist(\sigma_{n-1},\pi_n)+ \dist(\pi_n,\sigma_n)+\dist(\sigma_n,\gamma), \\
    g_n(\vec{\pi},\vec{\sigma})&:= \sum_{i=1}^n [1-k+ \dist(e,\pi_i) + \dist(\pi_i,\sigma_i) + \dist(\sigma_i,\gamma)]. \label{eq:fgn}
    \end{split}
\end{equation}
Similar to the $n=2$ case, we can minimize $f_n(\vec{\pi},\vec{\sigma})$ through choosing permutations in the sum that satisfy the multichain condition, 
\begin{equation}
    \pi_1 \leq \sigma_1 \leq \pi_2 \leq \sigma_2 \leq \dots \leq \sigma_n \leq \gamma. \label{eq:multichain}
\end{equation}
This can be proven through saturating successive triangle inequalities: $\dist(e,\pi_1) + \dist(\pi_1,\sigma_1) \leq \dist(e,\sigma_1)$, then $\dist(e,\sigma_1) + \dist(\sigma_1,\pi_2) \leq \dist(e,\pi_2) $, and so on, leading to the ordering of permutations on a geodesic between $e$ and $\gamma$ shown in Eq.~\eqref{eq:multichain}. For $k=4$, this condition is illustrated in Fig.~\ref{fig:nc_lattice}, where permutations are arranged along wires (geodesics) from bottom to top in the non-crossing partition lattice; one example is highlighted in bold [red]. Under the same constraint, the cost function $g_n(\vec{\pi}, \vec{\sigma})$ is also minimized, as Eq.~\eqref{eq:multichain} enforces the geodesic condition
\begin{equation}
    \pi_i \leq \sigma_i \leq \gamma,
\end{equation}
for all $1 \leq i \leq n$. 
This yields the minimal value $g_n(\vec{\pi}, \vec{\sigma}) = f_n(\vec{\pi}, \vec{\sigma}) = 0$ for all permutations satisfying Eq.~\eqref{eq:multichain}. 
Consequently, the leading-order contribution to the RMPU-averaged OTOC becomes
\begin{equation}
    \OTOC^{(k)}_\mathcal{R}\approx \sum_{ \pi_1 \leq \sigma_1 \leq \dots \leq \sigma_n \leq \gamma  } \left(\prod_{j=1}^n\mu(\pi_j,\sigma_j)\right) \braket{A}_{\pi_1} \braket{B}_{\sigma_n^{-1} \gamma}.
\end{equation}
Applying the M\"obius identity [Eq.~\eqref{eq:mobius_ident}] to $n-1$ of the functions—e.g., all but $\mu(\pi_1,\sigma_1)$—and resolving the resulting delta functions, we obtain 
\begin{equation}
    \OTOC^{(k)}_\mathcal{R}\approx \sum_{\pi_1 \leq \sigma_1 \leq \gamma} \mu(\pi_1,\sigma_1) \braket{A}_{\pi_1} \braket{B}_{\sigma^{-1}_1 \gamma} = \OTOC^{(k)}_\mathrm{FP} + \mc{O}(\chi^{-2}). 
\end{equation}
Recalling that $\OTOC^{(k)}_\mathbb{H} = \OTOC^{(k)}_\mathrm{FP} + \mathcal{O}(D^{-2})$ [Eq.~\eqref{eq:loOTOC}], and taking the limit $\chi \ll D$, we find that for local operators $A$ and $B$ with finite trace, the RMPU ensemble reproduces the Haar value of the OTOC up to subleading corrections up to error scaling with the inverse bond dimension, 
\begin{equation}
    \OTOC^{(k)}_\mathcal{R} = \OTOC^{(k)}_\mathbb{H} + \mc{O}(\chi^{-2}).
\end{equation}
\subsection{Subleading Correction to the Haar Averaged OTOC}
We will now elaborate on coefficients of the subleading corrections which scale as $\mc{O}(\chi^{-2})$, discussed in Sec.~\ref{sec:subleading}. We begin by revisiting the global Haar case. While subleading deviations from the diagonal approximation can be bound in a relatively straightforward way (see Sec.~\ref{sec:fp}), the analysis becomes more complex for quantities whose leading-order behavior depends on free probability considerations. In this setting, there are two distinct sources of subleading corrections to freeness for Haar-random matrices at finite Hilbert space dimension $D$
\begin{equation}
    \begin{split}
        \OTOC_{\mathbb{H}}^{(k)} &= \sum_{\pi,\sigma} \frac{\left[\mu(\pi,\sigma)  + D^{-2} \mathrm{Wg}^{(1)}_{\pi,\sigma}(k) + \mc{O}(D^{-4})\right]\braket{A }_\pi \braket{B }_{\sigma^{-1} \gamma} }{D^{1-k+\dist (\pi,\sigma) + \dist(e,\pi) +\dist(\sigma, \gamma)}}   \\
    &=\sum_{\pi\leq\sigma\leq \gamma} \left[\mu(\pi,\sigma) + \frac{\mathrm{Wg}^{(1)}_{\pi,\sigma}(k) }{D^{2 }} \right]\braket{A }_\pi \braket{B }_{\sigma^{-1} \gamma} +  \frac{1}{D^2}\sum_{1-k+\dist(e,\pi) +  \dist(\pi,\sigma)+\dist(\sigma,\gamma)=2}  \mu(\pi,\sigma) \braket{A }_\pi \braket{B }_{\sigma^{-1} \gamma} + \mc{O}(D^{-4})\\
    &=\OTOC_{\mathrm{FP}}^{(k)} +\frac{1}{D^2} \left(\sum_{\pi\leq\sigma\leq \gamma} \mathrm{Wg}^{(1)}_{\pi,\sigma}(k) +  \sum_{1-k+\dist(e,\pi) +  \dist(\pi,\sigma)+\dist(\sigma,\gamma)=2}  \mu(\pi,\sigma) \right)  \braket{A }_\pi \braket{B }_{\sigma^{-1} \gamma} + \mc{O}(D^{-4})\\
    &=\OTOC_{\mathrm{FP}}^{(k)} +\frac{c_k(A,B)}{D^2} + \mc{O}(D^{-4})\;,\label{eq:globalHaarOTOCapp}
    \end{split}
\end{equation}
as reported in Eq.~\eqref{eq:nonlocal1}.
In the expression above, we have used both the power-series expansion of the Weingarten function [Eq.~\eqref{eq:collins}] and the fact that the next-smallest value of the cost function in Eq.~\eqref{eq:2chain} is $2$. First, the subleading terms $\mathrm{Wg}^{(g)}_{\pi,\sigma}(k)$ for $g \geq 1$ generally lack closed-form expressions, though they can be systematically characterized and bound via a combinatorial correspondence~\cite{collins2017weinga}. 
Similarly, the permutations satisfying the constraint $1-k+\dist(e,\pi) +  \dist(\pi,\sigma)+\dist(\sigma,\gamma)=2$ lack an apparent combinatorial structure analogous to the non-crossing partition multichains that characterize the leading-order contribution (cf. Fig.~\ref{fig:nc_lattice}).
We note that the related condition for a single permutation, $1-k+\dist(e,\pi) +\dist(\pi,\gamma)=2$, characterizes the set of genus-one permutations. As such, the constraint appearing in Eq.~\eqref{eq:globalHaarOTOCapp} can be viewed as a natural generalization of non-crossing partition multichains to a higher-genus setting. This observation, combined with the fact that the coefficients $c_k(A, B)$ depend explicitly on the (potentially negative) moments of $A$ and $B$, highlights the difficulty in determining subleading corrections to freeness—--even in the global Haar case.  
We therefore resort to somewhat loose bounds, before performing a similar analysis for the RMPU average OTOC. In particular, to prove Eq.~\eqref{eq:haarsub}, we first recall the result of Ref.~\cite{collins2017weinga}: for any $\pi,\sigma \in S_k$ and $D>\sqrt{6}k^{7/4}$,
\begin{equation}
    \frac{1}{1-\frac{k-1}{D^2}} \leq \frac{D^{k+\dist(\pi,\sigma)}\mathrm{Wg}_{\pi,\sigma}(D,k)}{\mu(\pi,\sigma)} \leq \frac{1}{1-\frac{6 k^{7/2}}{D^2}}.
\end{equation}
This bound is often used in unitary design literature, because it is the only correction that appears for quantities that asymptotically depend on the diagonal approximation; see Sec.~\ref{sec:diag}. Expanding the series of Eq.~\eqref{eq:collins}, we can deduce that 
\begin{equation}
    \mathrm{Wg}_{\pi,\sigma}-\mathrm{Wg}^{(0)}_{\pi,\sigma} \leq \frac{6k^{7/2} \mu({\pi,\sigma})}{D^2} + \mc{O}(D^{-4}).
\end{equation}
Otherwise, we can also bound the other contribution to $c_k(A,B)$ through a simple counting argument, 
\begin{equation}
     |\sum_{1-k+\dist(e,\pi) +  \dist(\pi,\sigma)+\dist(\sigma,\gamma)=2}  \mu(\pi,\sigma)  \braket{A }_\pi \braket{B }_{\sigma^{-1} \gamma}|\leq C_k \sum_{1-k+\dist(e,\pi) +  \dist(\pi,\sigma)+\dist(\sigma,\gamma)=2}  \leq C_k (k!)^2 \label{eq:looseboundapp}
\end{equation}
where we have used that the moments are normalized, $|\braket{A }_\pi|,|\braket{B }_{\sigma^{-1} \gamma}| \leq 1$ together with properties of the M\"obius functions as described below Eq.~\eqref{eq:leading-wg}: that $|\mu(\pi,\sigma)|= \prod_{a \in \sigma^{-1}\pi} C_a \leq C_k$, with $C_k$ the $k$-th Catalan number. Note that in the final line, we have upper-bound the number of terms in the sum simply by the number of pairs of permutations, essentially ignoring the constraint, $1-k+\dist(e,\pi) +  \dist(\pi,\sigma)+\dist(\sigma,\gamma)=2$. We compute the number of terms contributing to this sum via exact enumeration for $k \leq 10$, with results summarized in Table~\ref{tab:genus1}. We notice that the naive upper bound of $(k!)^2$ vastly overestimates the actual count, and the number of leading-order terms (satisfying $\pi \leq \sigma \leq \gamma$) is significantly smaller. Characterizing the set $\{\pi,\sigma \in S_k:1-k+\dist(e,\pi) +  \dist(\pi,\sigma)+\dist(\sigma,\gamma)=2\}$ remains an interesting open problem. A deeper understanding of this structure could shed light on the precise nature of subleading corrections to freeness in Haar-random unitaries, and clarify the error incurred when approximating the freeness of $A_U$ with respect to a fixed matrix $B$ at finite, though large, $D$.

\begin{table}
    \begin{center}
       \begin{tabular}{c|c|c|c}
    $k$ & $\# \mathrm{NC}_2^{g=0}$ & $\# \mathrm{NC}_2^{g=1}$ & $(k!)^2$\\
    \hline
1 & 1 & 0 & 1\\
2 & 3 & 1 & 4\\
3 & 12 & 21 & 36\\
4 & 55 & 270 & 576\\
5 & 273 & 2860 & 14400\\
6 & 1428 & 27300 & 518400\\
7 & 7752 & 244188 & 25401600\\
8 & 43263 & 2089164 & 1625702400\\
9 & 246675 & 17305200 & 131681894400\\
10 & 1430715 & 139864725 & 13168189440000
    \end{tabular}
    \end{center}
    \caption{The number of pairs of permutations under relevant constraints. We define the set of $2$-chains on non-crossing partitions as $\mathrm{NC}_2^{g=0}:= \{\pi,\sigma \in S_k: 1-k+\dist(e,\pi) +  \dist(\pi,\sigma)+\dist(\sigma,\gamma)=0 \}$, as detailed in Fig.~\ref{fig:nc_lattice} and appearing in the leading order expression to the Haar-averaged OTOC, $\OTOC^{(k)}_{\mathrm{FP}}$. The number of elements of this set is given by the Fuss-Catalan numbers $ \# \mathrm{NC}_2^{g=0}= (2k+1)^{-1}\binom{3k}{k}$~\cite{nica2006lectures}. We define the generalization of this condition to genus one permutations, $\mathrm{NC}_2^{g=1}:=\{\pi,\sigma \in S_k: 1-k+\dist(e,\pi) +  \dist(\pi,\sigma)+\dist(\sigma,\gamma)=2 \}$, which are the permutations appearing in the subleading corrections to the Haar averaged OTOC, Eq.~\eqref{eq:globalHaarOTOCapp}. These do not admit any apparent closed form expression, and we find the first $10$ of them through brute force search. The final column denotes the number of pairs of permutations without constraint, which we use in the bound Eq.~\eqref{eq:looseboundapp}. Note that the sum of the second and third columns must be less than the fourth, as the union of all $\mathrm{NC}_2^{g}$ for non-negative integer $g$ is equal to the set of all pairs of permutations.   } 
    \label{tab:genus1}
\end{table}

Having established the general arguments above, we now turn to an exact evaluation of the subleading correction for small values of $k$. 
To ease notation in the following, we introduce a shorthand notation for the free cumulants of the replica operators $A$ and $B$ (cf. Eqs.~\eqref{eq:mom-cuma}-\eqref{eq:mom-cum}),
\begin{equation}
    \kappa_m(A) = \sum_{\pi \in S_m: \pi\leq \gamma}\mu(\pi,\gamma) \langle A \rangle_\pi \ ,
\end{equation}
where we also recall the notation of normalized moments, $\langle A \rangle_\pi = D^{m-\dist(e,\pi)}\tr[A^{\otimes m} T_{\pi} ] $ for $\pi \in S_m$. We find explicit expressions symbolically for the coefficient $c_k(A,B)$ appearing in Eq.~\eqref{eq:globalHaarOTOCapp} for small $k$, which we present explicitly up to $k=4$:
\begin{equation}
    \begin{split}
        \label{ck_haar}
c_1(A, B) & = 0\\
c_2(A, B) & = - \kappa_2(A) \kappa_2(B)\\
c_3(A, B) & =\kappa_3(A) \left(-3 \kappa_1(B) \kappa_2(B) + \kappa_3(B)\right)
- 3 \kappa_1(A) \kappa_2(A) \left(2 \kappa_1(B) \kappa_2(B) + \kappa_3(B)\right) \\
c_4(A, B) & = 
\kappa_4(A) \left(-6\, \kappa_1(B)^2 \kappa_2(B) + \kappa_2(B)^2 + 4\, \kappa_1(B) \kappa_3(B)\right)  \\
&+ \kappa_2(A)^2 \left(-10\, \kappa_1(B)^2 \kappa_2(B) + \kappa_4(B)\right)  \\
&+ 4\, \kappa_1(A) \kappa_3(A) \left(-5\, \kappa_1(B)^2 \kappa_2(B) + \kappa_4(B)\right)  \\
&- 2\, \kappa_1(A)^2 \kappa_2(A) \left(5 \left(2\, \kappa_1(B)^2 \kappa_2(B) + \kappa_2(B)^2 + 2\, \kappa_1(B) \kappa_3(B)\right) + 3\, \kappa_4(B)\right)\ .
    \end{split}
\end{equation}
In the case of traceless operators, $\tr[A]=\tr[B]=0$, the above coefficients of Eq.~\eqref{ck_haar} simplify to:
\begin{equation}
    \begin{split}
         c_2(A, B) & = - \kappa_2(A) \kappa_2(B)\\
        c_3(A, B) & = \kappa_3(A) \kappa_3( B)\\
        c_4(A, B) & = \kappa_4(A) \kappa_2(B)^2 + \kappa_4(B) \kappa_2(A)^2\ . \label{eq:traceless_ck}
    \end{split}
\end{equation}

\subsection{Subleading Correction to the RMPU Averaged OTOC}
Turning to the RMPU ensemble, we can also determine an exact form for the correction coefficient $\tilde{c}_{k,n}$ in Eq.~\eqref{eq:hahaha}. Starting from Eq.~\eqref{eq:rmpuotoc1_app},
we both expand the component Weingarten functions and break the multichain condition Eq.~\eqref{eq:multichain} (cf. Eq.~\eqref{eq:globalHaarOTOCapp})
\begin{equation}
    \begin{split}
        \OTOC_{\mc{R}}^{(k)} =& \OTOC_{\mathrm{FP}}^{(k)} + \frac{1}{\chi^2} \Bigg[\sum_{f_n(\vec{\pi},\vec{\sigma})=0 } { \sum_{p=1}^n\left[\left(\prod_{j\in[1,n]\backslash \{ p \}}\mu(\pi_j,\sigma_j)\right) \mathrm{Wg}^{(1)}_{\pi_p,\sigma_p}(k) \right] \braket{A }_{\pi_1} \braket{B }_{\sigma^{-1}_n \gamma} }  \\
         &+\sum_{ f_n(\vec \pi, \vec \sigma)=2 }
     \left( \prod_{i=1}^n \frac{\mu(\sigma_i^{-1} \pi_i) }{d^{  \dist(e,\pi_i)+\dist(\pi_i, \sigma_i)+\dist( \sigma_i,\gamma) }} \right)\braket{A}_{\pi_1} \braket{B}_{\sigma_n^{-1}\gamma}  \Bigg] +\mc{O}(\chi^{-4}).
    \end{split}
\end{equation}
The first line comes from expanding each Weingarten matrix in the staircase to subleading order (using Eq.~\eqref{eq:collins}), while the second line comes from a single breaking of the geodesic condition such that $f_n(\vec \pi, \vec \sigma)=2$ (recall Eq.~\eqref{eq:fgn}). Now, as the first sum involves the multichain condition of non-crossing partitions, $f_n(\vec \pi, \vec \sigma)=0$, we can recursively apply the identity of M\"obius functions, $\sum_{\sigma:\pi \leq \sigma \leq \mu} \mu(\pi,\sigma) = \delta_{\pi,\mu}$, to simplify the first line above to arrive at Eq.~\eqref{eq:anal_ck_rmpu} in the main text. To determine a scaling with $n$ and to compare with the global Haar case, we symbolically calculate this exact expression for small $k$ and increasing $n$, 
\begin{equation}
\label{eq_ck_rmpu}
    \begin{split}
        \tilde c_2(A, B, n) & = \left(-\frac n{d^2} + (n-1)\right) \kappa_2(A) \, \kappa_2(B)
    \\ 
    \tilde c_3(A, B, n) & =  \left(-\frac n{d^2} + (n-1)\right) \left [3  \kappa_1(A) \kappa_2(A) \left(2 \kappa_1(B) \kappa_2(B) + \kappa_3(B)\right) 
+ 3 \kappa_3(A) \kappa_1(B) \kappa_2(B)\right ]
 \\ & \quad +\frac{\kappa_3(A) \kappa_3(B)}{d^{2n}}       \\
    \tilde c_4(A, B, n) & =  \left(-\frac n{d^2} + (n-1)\right) \Big [
    6 \kappa_4(A) \kappa_1(B)^2 \kappa_2(B)
+ 2 \kappa_2(A)^2 \kappa_1(B) \left(5 \kappa_1(B) \kappa_2(B) + 2 \kappa_3(B)\right)
  \\ & \quad \quad \quad 
 + 4 \kappa_1(A) \kappa_3(A) \left( \kappa_2(B) \left(5 \kappa_1(B)^2 + \kappa_2(B) \right)
 + 2 \kappa_1(B) \kappa_3(B) \right)
 \\ & \quad \quad \quad 
 + 2 \kappa_1(A)^2 \kappa_2(A) \left(
 10 \kappa_1(B)^2 \kappa_2(B) + 5 \kappa_2(B)^2 + 10 \kappa_1(B) \kappa_3(B) + 3 \kappa_4(B)\right )
    \Big ]
\\ & \quad +\frac{1}{d^{2n}} \Big [\kappa_4(A) \left( \kappa_2(B)^2 + 4 \kappa_1(B) \kappa_3(B) \right)
+ 4 \kappa_1(A) \kappa_3(A) \left( \kappa_2(B)^2 
 + 2 \kappa_1(B) \kappa_3(B) + \kappa_4(B) \right)  \\ & 
 \quad \quad
+ \kappa_2(A)^2 \left( 4 \kappa_1(B) \kappa_3(B) + \kappa_4(B) \right ) \Big] 
    \end{split}
\end{equation}
Comparing with the Haar case, we find that the coefficients have the following scaling form, as reported in Eq.~\eqref{eq:rmpucoeff},
\begin{equation}
    {\tilde c_{k,n}(A, B) = 
    \left(\frac n{d^2} - (n-1)\right) \, a_k(A, B) + \frac {b_k(A, B)}{d^{2n}} \ .}
\end{equation}
The functions $a_k(A, B)$ and $b_k(A, B)$ satisfy $a_k + b_k = c_k$, reproducing the full Haar result for $n=1$, Eq.~\eqref{ck_haar}. For traceless operators, we find that
\begin{equation}
 \tilde c_{2,n}(A, B) = \begin{cases}
     \left(\frac n{d^2} -(n-1)\right) c_2(A, B) &\quad{ \rm for } \quad k=2 \\
     \tilde c_{k,n}(A, B) = \frac{ c_k(A, B)}{d^{2n}}& \quad{ \rm for } \quad k>2
 \end{cases} \;.
\end{equation}

\section{Calculation of the Frame Potential for the RMPU ensemble }\label{ap:framepot}
In this section, we analytically compute the leading and subleading terms of the frame potential for the RMPU ensemble in the regime $D \gg \chi \gg 1$.


Throughout this section, we adopt the graphical notation for Weingarten and Gram matrices introduced in Eq.~\eqref{eq:graphical}, which streamlines the computation by providing efficient combinatorial bookkeeping. For example, the identity $\mathrm{Wg}(d\chi,k) = [G(d\chi,k)]^{-1}$ can be expressed diagrammatically as
\begin{equation}
\adjustbox{scale = \scalefactor}{\input{ident1.tikz}} = \adjustbox{scale = \scalefactor}{\input{ident1a.tikz}} \;,\label{eq:ident1}
\end{equation}
where we note that Gram matrices factorize: $G(\chi d,k)_{\pi \sigma} = G(\chi ,k)_{\pi \sigma} G(d,k)_{\pi \sigma}$.
A crucial point is that \textit{all wires (thin or thick) connected to the same side of a given box must carry the same index}; otherwise, the contraction becomes inconsistent. 
For instance, if the thick and thin wires are contracted with different permutations, the product becomes
\begin{equation}
    G(\chi ,k)_{\pi \sigma} G(d,k)_{\pi \mu},
\end{equation}
which no longer corresponds to the structured inverse relation of Eq.~\eqref{eq:ident1}. In this graphical formalism, {wires emerging from a single side of a box do not represent separate tensor factors (as in standard tensor network diagrams), but instead share a common label in the permutation basis}.

Using the described graphical formalism, it is insightful to rederive the Haar frame potential in Eq.~\eqref{eq:framepotHaar} 
\begin{equation}
    \mathcal{F}^{(k)}_{\mathbb{H}} = \input{fpwg1.tikz} = \input{fpwg2.tikz} =\input{fpwg3.tikz} = \sum_{\pi \in S_k} = k!\;.
\end{equation}
where we have recalled the frame potential definition in terms of moment operators in Eq.~\eqref{eq:framepotMain}.
The above computation follows from the fact that the Weingarten and Gram matrices involved are mutual inverses, with only diagonal terms surviving. 
We now proceed to prove the main result for the RMPU ensemble, Eq.~\eqref{eq:framepotcalc} from the Main Text, recalled here for convenience
\begin{equation}
    \mathcal{F}_{\mc{R}}^{(k)}= k!\left( 1 + \frac{k(k-1)}{2 \chi^2}( n-1 -\frac{n}{d^2} +\frac{1}{d^{2n}}) + \mathcal O(\chi^{-3})\right).
\end{equation}
Graphically, the frame potential of the RMPU ensemble takes the exact form
\begin{align}
    \mathcal{F}^{(k)}_{\mc{R}(n)}&= \tr\left[ \input{rmpu0.tikz} \right]\\
    &= \input{rmpu1.tikz}, \label{eq:rmpufp}
\end{align}
where in the second line we have applied the identity~\eqref{eq:ident1} twice: once to eliminate the central Weingarten matrix and once at the boundary.

Importantly, Eq.~\eqref{eq:rmpufp} can now be interpreted as the partition function of a classical spin model on a 1D chain of $n$ sites, with spin degrees of freedom given by permutations $\pi \in S_k$, and interactions governed by the contraction structure of Weingarten and Gram matrices. This classical mapping allows us to treat the frame potential using methods of statistical mechanics, in particular a low-temperature expansion in the inverse bond dimension $\chi^{-1}$, which plays the role of a small parameter. 
To proceed, we recall the asymptotic expansions of the Gram and Weingarten matrices
\begin{equation}
    \begin{split}
        G(\chi,k)_{\sigma \pi}  &= \chi^{k} \left[ \delta_{\sigma \pi}  + \frac{1}{\chi} \ad_{\sigma \pi}^{(1)} + \mathcal{O}\left(\frac{1}{\chi^2}\right) \right]\;,\\
    \mathrm{Wg}(d\chi,k)_{\sigma \pi} &= \frac{1}{(d\chi)^{k}} \left[ \delta_{\sigma \pi}  - \frac{1}{d\chi} \ad_{\sigma \pi}^{(1)} - \mathcal{O}\left(\frac{1}{(d\chi)^2}\right) \right]\;, 
    \end{split}
\end{equation}
where we denoted as $\ad^{(\alpha)}_{\sigma \pi} = \delta_{\dist(\sigma, \pi), \alpha}$ the $\alpha$ adjacency matrix. That is, $\ad^{(1)}_{\sigma \pi}$ is equal to one only if $\sigma $ and $\pi$ are related via a single transposition.
With this in hand, our task is to perform an asymptotic expansion of Eq.~\eqref{eq:rmpufp} for large $\chi$, namely
\begin{equation}
    \mathcal{F}_\mathcal{R}^{(k)}=\sum_{\alpha=0}^\infty \frac{F^{(k)}_\alpha(n,d)}{\chi^\alpha}. 
    \label{eq:partfun}
\end{equation}  
We now proceed with the low-temperature expansion of the partition function Eq.~\eqref{eq:partfun}. 
For this purpose, we define graphical notation for the leading order elements of the asymptotic expansion
\begin{equation}
    \frac{\delta_{\pi, \sigma} }{d^{k}}=: \input{wglo.tikz}, \qquad \text{and}\qquad 
    \frac{ \ad^{(1)}_{\pi, \sigma} (1-d^{-2})}{ d^{k} \chi}  =: \input{wgslo.tikz}\;, \label{eq:graphiddel}
\end{equation}
such that the local asymptotic expansion of the Weingarten matrix together with the Gram, can be written as
\begin{equation}
    \mathrm{Wg}(d \chi)_{\pi , \mu} G_{\mu, \sigma }(\chi) = \input{lo_wgG.tikz} \approx \input{wglo.tikz} + \input{wgslo.tikz} + \mc{O}(\chi^{-2}).
\end{equation}
The task is now to compute the leading-order contribution in the limit $\chi \to \infty$ explicitly. At leading order, we retain only the dominant terms in the expansions of the Gram and Weingarten matrices, namely
\begin{equation}
    F_0^{(k)}(n,d):=\lim_{\chi\to\infty} \mathcal{F}_\mathcal{R}^{(k)}=  \input{gesu.tikz} = 
    \frac{\sum_{\pi\in S_k} G(d,k)_{\pi,\pi}^{2(n-1)}}{d^{k(2n-2)}}= \sum_{\pi\in S_k} 1= k!\;. \label{eq:loframe}
\end{equation}
Here, recalling the bookkeeping rules, the multiple sums over permutations implicit in Eq.~\eqref{eq:rmpufp} reduce to a single sum over $\pi$, as each Kronecker delta $\delta_{\pi, \sigma}$ enforces index identifications across the diagram.
Thus, at leading order, the RMPU frame potential recovers that of the Haar ensemble.

We now compute the subleading correction exactly. 
Consider expanding a single Gram or Weingarten matrix in Eq.~\eqref{eq:rmpufp} to subleading order, while keeping all others at leading order (i.e., proportional to $\delta_{\pi,\sigma}$). For instance, one such term is
\begin{align}
      \input{rmpu2.tikz} = \input{chi1.tikz}\propto  \sum_{\pi \in S_k} \ad_{\pi,\pi}^{{(1)}} = \sum_{\pi \in S_k} \delta_{\dist(\pi,\pi),1} = 0\, \label{eq:rmpufp1}
\end{align}
where the loop of delta functions enforces equality of all permutation indices. Due to the symmetry of the problem, the above holds for any choice of expansion of a single Weingarten matrix to subleading order. Since the identity is never a transposition, the summand vanishes, and thus the $\mc{O}(1/\chi)$ correction is zero,
\begin{equation}
    F_1^{(k)}(n,d):=\lim_{\chi \to \infty} \left[ \chi \left(\mathcal{F}^{(k)}_{\mc{R}} - F_0(n,d)\right) \right]= \lim_{\chi \to \infty} \left[ \chi \left(\mathcal{F}^{(k)}_{\mc{R}} - k!\right) \right]=0\;.
\end{equation}
We must therefore proceed to next-to-leading order to capture the deviation from the Haar value: $F_2^{(k)}(n,d):= \lim_{\chi\to\infty} [ \chi^2(\mathcal{F}_\mathcal{R}^{(k)}-F_0(n,d))]$. 

Next, consider the case where a single Weingarten or Gram matrix in Eq.~\eqref{eq:rmpufp} is expanded to sub-subleading order, i.e., $\mathcal{O}(1/\chi^2)$, while all others remain proportional to $\delta_{\pi,\sigma}$. In this case, by an analogous arugment the correction is proportional to $\ad^{(2)}_{\pi,\pi} = \delta_{\dist(\pi,\pi),2} = 0$, and hence also vanishes. We deduce that the leading nonvanishing correction arises from the simultaneous expansion of two distinct Weingarten or Gram matrices to first subleading order, $\ad^{(1)}_{\pi,\sigma}$, while all remaining matrices contribute only at leading order, i.e., $\propto \delta_{\pi,\sigma}$. 
We explicitly compute an identity for non-negative integer $j$, which will recursively appear in our following calculation
\begin{equation}
    \begin{split}
         \input{rmpu3.tikz} &=   \frac{1}{d^{2j k}} \, \input{rmpu4.tikz} \\
    &= k!+ \sum_{\pi,\sigma \in S_k } \frac{G(d,k)_{\pi,\sigma }^{2j}   (1 - \frac{1}{d^2})^2   \ad^{(1)}_{\pi,\sigma}}{\chi^2 d^{2j k }} + \mc{O}(1/\chi^3) \\
    &= k!+ \sum_{\pi,\sigma \in S_k: \pi^{-1}\sigma =\tau } \frac{d^{2j(k-1) }  (1 - \frac{1}{d^2})^2  }{\chi^2 d^{2j k }} + \mc{O}(1/\chi^3) \\
    &= k!\left(1 + \frac{k(k-1) (1 - \frac{1}{d^2})^2}{2 \chi^2 d^{2j}}\right)+ \mc{O}(1/\chi^3) \\
    &=: k!\left(1 + \frac{\Xi}{\chi^2 d^{2j}}\right)+ \mc{O}(1/\chi^3), \label{eq:keyIdent}
    \end{split}
\end{equation}
where we have introduced the constant $\Xi:=k(k-1) (1 - {d^{-2}})^2/2$ for ease of reference. 
We explain in details the above steps. (i) In the first line, the factor $1/d^{2j k}$ arises because the leading-order Weingarten matrices scale as $\propto (d\chi)^{-k}$, and hence they do not completely cancel the leading-order contribution from the Gram matrices on the thick wire, which scale as $(\chi)^{k}$.
(ii) The factor $(1 - \frac{1}{d^2})^2$ in the second line comes from two applications of the subleading expansion of the matrix
    \begin{equation}
    \begin{split}
        \input{rmpu5.tikz} &= \sum_{\sigma \in S_k} \mathrm{Wg}(\chi d ,k)_{\pi \sigma} G(d ,k)_{\pi \sigma}G(\chi,k)_{ \sigma \mu} \\
        &=\delta_{\pi,\mu} + \ad_{\mu,\pi}^{(1)}\left((\chi d)^{-k} d^k  \chi^{k-1} - (\chi d)^{-k-1}d^{k-1} \chi^{k} \right)\\
        &=\delta_{\pi,\mu} + \frac{\delta_{\dist(\pi,\mu),1}}{\chi}\left(1- \frac{1}{d^2} \right). \label{eq:seconddelta}
    \end{split}
\end{equation}
(iii) In the final line we no longer have any dependence on $\pi,\sigma$ except in the double sum. We count the terms: $\pi \in S_k$ is freely summed over the symmetric group, while $\sigma \in S_k: \dist(\pi,\sigma) =1$ sums over all possible single transpositions in the symmetric group, of which there are $\binom{k}{2}$. 

Returning to the computation of the subleading correction to the frame potential for the RMPU ensemble, Eq.~\eqref{eq:rmpufp}, we will first expand only the center Weingarten/Gram matrix (indicated by the red box),
\begin{equation}
    \input{rmpu6.tikz}. \label{eq:circled}
\end{equation}
For its leading order expression, we find that 
\begin{align}
    \input{rmpu7.tikz}=\input{rmpu8.tikz} = \mathcal{F}^{(k)}_\mathcal{R}(n-1,d) \label{eq:1}
\end{align}
Here, we have applied the identity Eq.~\eqref{eq:ident1} once to cancel a single  Weingarten and Gram matrix in the center. 
In the complementary case, the circled matrices in Eq.~\eqref{eq:circled} are given by the subleading correction from Eq.~\eqref{eq:graphiddel}. If the bottom matrix is also taken to be subleading, we apply Eq.~\eqref{eq:keyIdent} to arrive at 
\begin{equation}
    \input{rmpu9.tikz}=\frac{1}{\chi^2}\frac{k!  \Xi}{d^{2(n-2))}}. \label{eq:firstsub}
\end{equation}
Alternatively, if this bottom matrix is instead leading order ($\propto \delta_{\pi,\sigma}$), we find 
\begin{align}
    \input{rmpu10.tikz}&=\input{rmpu11.tikz}\\
    &=\input{rmpu12.tikz} \, , \label{eq:recursive}
\end{align}
where in the final line we have applied the identity Eq.~\eqref{eq:ident1}. Now, the new bottom matrix in Eq.~\eqref{eq:recursive} is either subleading, in which case we have exactly Eq.~\eqref{eq:firstsub} but with the label $n-3$, or otherwise it is leading, in which case we get back Eq.~\eqref{eq:recursive} recursively but for label $n-4$. Combining Eqs.~\eqref{eq:1}, \eqref{eq:firstsub}, and the relation of Eq.~\eqref{eq:recursive}, we find the following relation for the subleading correction to the frame potential,
\begin{equation}
    \mathcal{F}^{(k)}_\mathcal{R}(n,d) = \mathcal{F}^{(k)}_\mathcal{R}(n-1,d)  + k!  \frac{\Xi}{\chi^2} \sum_{j=0}^{n-2} \frac{1}{d^{2j}} + \mc{O}\left(\frac{1}{\chi^3}\right).
\end{equation}
Solving this recursion relation using standard techniques, and recalling the leading order expression Eq.~\eqref{eq:loframe}, we have for $n\geq 3$ layers, that 
\begin{align}
    \mathcal{F}^{(k)}_\mathcal{R}(n,d) &= k!+k!  \frac{\Xi}{\chi^2} \left[ 1 + \sum_{m=0}^{n-3} \sum_{i=0}^{m+1} \frac{1}{d^{2i}} \right]+\mc{O}\left(\frac{1}{\chi^3}\right)\\
    &=k!\left( 1 + \frac{k(k-1)}{2 \chi^2}\left( n-1 -\frac{n}{d^2} +\frac{1}{d^{2n}}\right) + \mc{O}(\chi^{-3}) \right), \label{eq:final}
\end{align}
where we have solved twice a finite geometric series and recalled that $\Xi=k(k-1) (1 - \frac{1}{d^2})^2 /2$. Note Eq.~\eqref{eq:final} is valid also for $n<3$: for a single layer $n=1$, we have exactly the global Haar case, while $n=2$ is given by \eqref{eq:keyIdent} by setting $j=0$. Recalling also that we generally take $\chi^{-3} \gg D^{-2} =\chi^{-2} d^{-2n} $, we can safely neglect this term, thereby completing the proof of Eq.~\eqref{eq:framepotcalc}.



\end{document}

%% file: zh.tikzstyles
\tikzstyle{dot}=[inner sep=0.3mm, minimum width=2mm, minimum height=2mm, draw, shape=circle, font={\footnotesize}, tikzit fill=magenta]
\tikzstyle{white dot}=[dot, fill=white, text depth=-0.2mm, tikzit category=ZH-pf, draw=black]
\tikzstyle{white phase dot}=[minimum size=5mm, font={\footnotesize\boldmath}, shape=rectangle, rounded corners=2mm, inner sep=0.2mm, outer sep=-2mm, scale=0.8, tikzit shape=circle, draw=black, fill=white, tikzit category=ZH-pf, tikzit fill=white, tikzit draw=blue]
\tikzstyle{gray dot}=[dot, fill={rgb,255: red,180; green,180; blue,180}, text depth=-0.2mm, tikzit category=ZH-pf]
\tikzstyle{gray phase dot}=[white phase dot, tikzit shape=circle, tikzit draw=blue, fill={rgb,255: red,180; green,180; blue,180}, font={\footnotesize\boldmath}]
\tikzstyle{hadamard}=[fill=white, draw, inner sep=0.6mm, minimum height=1.5mm, minimum width=1.5mm, shape=rectangle, tikzit shape=rectangle, tikzit category=ZH-pf]
\tikzstyle{small hadamard}=[hadamard]
\tikzstyle{lambda}=[hadamard, fill={rgb,255: red,180; green,180; blue,180}, tikzit shape=rectangle]
\tikzstyle{halfscalar}=[star, fill=black, draw=black, minimum size=8pt, inner sep=0pt]
\tikzstyle{box}=[shape=rectangle, text height=1.5ex, text depth=0.25ex, yshift=0.2mm, fill=zx_yellow,tikzit fill=yellow, draw=black, minimum height=3mm, minimum width=5mm, font={\small}]
\tikzstyle{Z dot}=[inner sep=0mm, minimum size=2mm, shape=circle, draw=black, fill={zx_green}, tikzit fill=green]
\tikzstyle{Zhalf R}=[
  inner sep=0mm,
  minimum size=1mm,
  shape=semicircle,
  shape border rotate=90, 
  draw=black,
  fill={zx_green}, tikzit fill=green
]
\tikzstyle{Zhalf L}=[
  inner sep=0mm,
  minimum size=1mm,
  shape=semicircle,
  shape border rotate=270, 
  draw=black,
  fill={zx_green}, tikzit fill=green
]
\tikzstyle{Z phase dot}=[minimum size=5mm, font={\footnotesize\boldmath}, shape=rectangle, rounded corners=2mm, inner sep=0.2mm, outer sep=-2mm, scale=0.8, tikzit shape=circle, draw=black, fill={zx_green}, tikzit draw=blue, tikzit fill=green]
\tikzstyle{X dot}=[Z dot, shape=circle, draw=black, fill={zx_red}, tikzit fill=red]
\tikzstyle{Xhalf R}=[
  inner sep=0mm,
  minimum size=1mm,
  shape=semicircle,
  shape border rotate=90, 
  draw=black,
  fill={zx_red}, tikzit fill=red
]
\tikzstyle{Xhalf L}=[
  inner sep=0mm,
  minimum size=1mm,
  shape=semicircle,
  shape border rotate=270, 
  draw=black,
  fill={zx_red}, tikzit fill=red
]
\tikzstyle{X phase dot}=[Z phase dot, tikzit shape=circle, tikzit draw=blue, fill={zx_red}, font={\footnotesize\color{black}\boldmath}, tikzit fill=red]
\tikzstyle{H box}=[hadamard]
\tikzstyle{st}=[star, star points=5, fill=white, draw=black, inner sep=1.2pt, line width=1.2pt, tikzit fill=blue, tikzit draw=red, tikzit category=ZH-pf]
\tikzstyle{triangle}=[regular polygon, regular polygon sides=3, fill=white, draw=black, inner sep=0pt, minimum width=1em, tikzit draw=blue, tikzit category=ZH-pf, tikzit fill=cyan]
\tikzstyle{not}=[fill={rgb,255: red,180; green,180; blue,180}, draw=black, shape=circle, font={$\neg$}, dot]
\tikzstyle{vertex}=[inner sep=0mm, minimum size=1mm, shape=circle, draw=black, fill=black]
\tikzstyle{vertex set}=[inner sep=0mm, minimum size=1mm, shape=circle, draw=black, fill=white, font={\footnotesize\boldmath}]
\tikzstyle{wide point}=[fill=white, draw, shape=isosceles triangle, shape border rotate=-90, isosceles triangle stretches=true, inner sep=0pt, minimum width=1.5cm, minimum height=6.12mm, yshift=-0.0mm]
\tikzstyle{medium gray box}=[semilarge box, fill={rgb,255: red,180; green,180; blue,180}]
\tikzstyle{small box}=[rectangle, inline text, fill=white, draw, minimum height=5mm, yshift=-0.5mm, minimum width=5mm, font={\small}]
\tikzstyle{small gray box}=[small box, fill={rgb,255: red,180; green,180; blue,180}]
\tikzstyle{medium box}=[rectangle, inline text, fill=white, draw, minimum height=5mm, yshift=-0.5mm, minimum width=8mm, font={\small}]
\tikzstyle{ddot}=[line width=1.0pt, inner sep=0mm, minimum width=2.5mm, minimum height=2.5mm, draw, shape=square]
\tikzstyle{ddotR}=[inner sep=0mm, minimum width=2.5mm, minimum height=2.5mm, draw, shape=rectangle,fill={zx_red}]
\tikzstyle{ddotG}=[inner sep=0mm, minimum width=2.5mm, minimum height=2.5mm, draw, shape=rectangle,fill={zx_green}]
\tikzstyle{dd white}=[ddot, fill=white, tikzit draw=green]
\tikzstyle{dd white phase}=[white phase dot, line width=1.6pt, tikzit draw=yellow]
\tikzstyle{dd gray}=[ddot, fill={rgb,255: red,180; green,180; blue,180}, tikzit draw=green]
\tikzstyle{dd gray phase}=[gray phase dot, line width=1.6pt, tikzit draw=yellow]

\tikzstyle{simple}=[-, dashed, dash pattern=on 2pt off 1pt, thick, draw=gray]
\tikzstyle{hadamard edge}=[-, dashed, dash pattern=on 2pt off 1pt, thick, draw=gray]
\tikzstyle{gray}=[-, draw={blue!60!white}, tikzit draw=blue]
\tikzstyle{blue}=[-, draw={blue!60!white}, tikzit draw=blue]
\tikzstyle{brace edge}=[-, tikzit draw=blue, decorate, decoration={brace,amplitude=1mm,raise=-1mm}]
\tikzstyle{diredge}=[->]
\tikzstyle{not edge}=[-, thick, draw={rgb,255: red,255; green,68; blue,68}]
\tikzstyle{double edge}=[-, double, shorten <=-1mm, shorten >=-1mm, double distance=2pt]
\tikzstyle{boldedge}=[-, dashed, dash pattern=on 2pt off 1pt, line width=1.6pt, shorten <=-0.17mm, shorten >=-0.17mm, tikzit draw=gray,draw=gray]

%% file: figures/rmpu_mom.tikz
\begin{tikzpicture}
	\begin{pgfonlayer}{nodelayer}
		\node [style=box] (1) at (-2.75, 1.25) {};
		\node [style=box] (3) at (-0.75, 0.75) {};
		\node [style=box] (5) at (1.25, 0.25) {};
		\node [style=none] (10) at (-0.75, 0.5) {};
		\node [style=none] (11) at (1, 0.5) {};
		\node [style=none] (15) at (-2.75, 1.5) {};
		\node [style=none] (16) at (-2.75, 1) {};
		\node [style=none] (17) at (-0.75, 1) {};
		\node [style=none] (19) at (1, 0) {};
		\node [style=none] (23) at (2.75, 0) {};
		\node [style=X dot] (30) at (2.25, 0) {};
		\node [style=X dot] (31) at (0.25, 0.5) {};
		\node [style=X dot] (32) at (-1.75, 1) {};
		\node [style=none] (36) at (3.25, -0.5) {\dots};
		\node [style=none, rotate=90] (48) at (-3.75, -0.5) {...};
		\node [style=none, rotate=90] (50) at (5, -0.5) {...};
		\node [style=box] (51) at (4.25, -1.5) {};
		\node [style=none] (52) at (4.25, -1.25) {};
		\node [style=none] (53) at (3.5, -1.25) {};
		\node [style=none] (55) at (4.25, -1.75) {};
		\node [style=none] (59) at (-2.25, -1.25) {};
		\node [style=none] (60) at (0.75, -1.25) {\dots};
		\node [style=Zhalf R] (68) at (5.25, 1.5) {};
		\node [style=Zhalf R] (69) at (5.25, 1) {};
		\node [style=Zhalf R] (70) at (5.25, 0.5) {};
		\node [style=Zhalf R] (72) at (5.25, -1.25) {};
		\node [style=Xhalf R] (73) at (5.25, -1.75) {};
		\node [style=Zhalf L] (75) at (-3.75, 1) {};
		\node [style=Zhalf L] (76) at (-3.75, 0.5) {};
		\node [style=Zhalf L] (77) at (-3.75, 0) {};
		\node [style=Zhalf L] (78) at (-3.75, -1.25) {};
		\node [style=Zhalf L] (79) at (-3.75, -1.75) {};
		\node [style=Xhalf L] (80) at (-3.75, 1.5) {};
	\end{pgfonlayer}
	\begin{pgfonlayer}{edgelayer}
		\draw [style=boldedge] (10.center) to (11.center);
		\draw [style=boldedge] (16.center) to (17.center);
		\draw [style=boldedge] (23.center) to (19.center);
		\draw [style=boldedge] (52.center) to (53.center);
		\draw [style=hadamard edge] (17.center) to (69);
		\draw [style=hadamard edge] (11.center) to (70);
		\draw [style=hadamard edge] (15.center) to (68);
		\draw [style=hadamard edge] (52.center) to (72);
		\draw [style=boldedge] (55.center) to (73);
		\draw [style=hadamard edge] (78) to (59.center);
		\draw [style=hadamard edge] (79) to (55.center);
		\draw [style=hadamard edge] (19.center) to (77);
		\draw [style=hadamard edge] (76) to (10.center);
		\draw [style=hadamard edge] (75) to (16.center);
		\draw [style=boldedge] (80) to (15.center);
	\end{pgfonlayer}
\end{tikzpicture}

%% file: figures/ident1a.tikz
\begin{tikzpicture}
	\begin{pgfonlayer}{nodelayer}
		\node [style=none] (4) at (1, 0.25) {};
		\node [style=none] (5) at (-1, 0.25) {};
		\node [style=none] (6) at (-1, -0.25) {};
		\node [style=none] (8) at (1, -0.25) {};
		\node [style=box] (11) at (0, 0) {};
	\end{pgfonlayer}
	\begin{pgfonlayer}{edgelayer}
		\draw [style=simple] (4.center) to (5.center);
		\draw [style=boldedge] (6.center) to (8.center);
	\end{pgfonlayer}
\end{tikzpicture}

%% file: figures/gr.tikz
\begin{tikzpicture}
	\begin{pgfonlayer}{nodelayer}
		\node [style=X dot] (0) at (0, 0) {};
		\node [style=none] (1) at (-0.75, 0) {};
		\node [style=none] (2) at (0.75, 0) {};
	\end{pgfonlayer}
	\begin{pgfonlayer}{edgelayer}
		\draw [style=boldedge] (1.center) to (2.center);
	\end{pgfonlayer}
\end{tikzpicture}

%% file: figures/gd.tikz
\begin{tikzpicture}
	\begin{pgfonlayer}{nodelayer}
		\node [style=none] (1) at (-0.75, 0) {};
		\node [style=none] (2) at (0.75, 0) {};
		\node [style=Z dot] (3) at (0, 0) {};
	\end{pgfonlayer}
	\begin{pgfonlayer}{edgelayer}
		\draw [style=simple] (1.center) to (2.center);
	\end{pgfonlayer}
\end{tikzpicture}

%% file: figures/rmpu_2kotoc.tikz
\begin{tikzpicture}
	\begin{pgfonlayer}{nodelayer}
		\node [style=box] (0) at (-3.25, 1.25) {};
		\node [style=box] (1) at (-1.25, 0.75) {};
		\node [style=box] (2) at (0.75, 0.25) {};
		\node [style=none] (3) at (-1.25, 0.5) {};
		\node [style=none] (4) at (0.5, 0.5) {};
		\node [style=none] (5) at (-3.25, 1.6) {};
		\node [style=none] (6) at (-3.25, 1) {};
		\node [style=none] (7) at (-1.25, 1) {};
		\node [style=none] (8) at (0.5, 0) {};
		\node [style=none] (9) at (2.25, 0) {};
		\node [style=X dot] (10) at (1.75, 0) {};
		\node [style=X dot] (11) at (-0.25, 0.5) {};
		\node [style=X dot] (12) at (-2.25, 1) {};
		\node [style=none] (13) at (2.75, -0.5) {\dots};
		\node [style=none, rotate=90] (14) at (-4.25, -0.5) {...};
		\node [style=none, rotate=90] (15) at (4.5, -0.5) {...};
		\node [style=box] (16) at (3.75, -1.5) {};
		\node [style=none] (17) at (3.75, -1.25) {};
		\node [style=none] (18) at (3, -1.25) {};
		\node [style=none] (19) at (3.75, -1.85) {};
		\node [style=none] (20) at (-2.75, -1.25) {};
		\node [style=none] (21) at (0.25, -1.25) {\dots};
		\node [style=Z dot] (36) at (4.75, 0.5) {\tiny $\gamma$};
		\node [style=Z dot] (37) at (4.75, 1) {\tiny $\gamma$};
		\node [style=Z dot] (38) at (4.75, 1.6) {\tiny $\gamma$};
		\node [style=Z dot] (39) at (4.75, -1.25) {\tiny $\gamma$};
		\node [style=ddotR] (40) at (4.75, -1.85) {\tiny $B$};
		\node [style=ddotR] (41) at (-4.5, 1.6) {\tiny $A$};
		\node [style=Z dot] (42) at (-4.5, 1) {\tiny $e$};
		\node [style=Z dot] (43) at (-4.5, 0.5) {\tiny $e$};
		\node [style=Z dot] (44) at (-4.5, 0) {\tiny $e$};
		\node [style=Z dot] (45) at (-4.5, -1.25) {\tiny $e$};
		\node [style=Z dot] (46) at (-4.5, -1.85) {\tiny $e$};
		\node [style=X dot] (47) at (5.5, -1.85) {\tiny $\gamma$};
	\end{pgfonlayer}
	\begin{pgfonlayer}{edgelayer}
		\draw [style=boldedge] (3.center) to (4.center);
		\draw [style=boldedge] (6.center) to (7.center);
		\draw [style=boldedge] (9.center) to (8.center);
		\draw [style=boldedge] (17.center) to (18.center);
		\draw [style=hadamard edge] (5.center) to (38);
		\draw [style=hadamard edge] (7.center) to (37);
		\draw [style=hadamard edge] (4.center) to (36);
		\draw [style=hadamard edge] (17.center) to (39);
		\draw [style=boldedge] (19.center) to (40);
		\draw [style=boldedge] (5.center) to (41);
		\draw [style=hadamard edge] (19.center) to (46);
		\draw [style=hadamard edge] (8.center) to (44);
		\draw [style=hadamard edge] (43) to (3.center);
		\draw [style=hadamard edge] (42) to (6.center);
		\draw [style=hadamard edge] (20.center) to (45);
		\draw [style=boldedge] (47) to (40);
	\end{pgfonlayer}
\end{tikzpicture}

%% file: figures/A_choi.tikz
\begin{tikzpicture}
	\begin{pgfonlayer}{nodelayer}
		\node [style=none] (1) at (0.25, 0) {};
		\node [style=ddotR] (2) at (-0.5, 0) {\tiny $A$};
		\node [style=none] (3) at (0.25, 0.25) {\tiny  $ \pi$};
	\end{pgfonlayer}
	\begin{pgfonlayer}{edgelayer}
		\draw [style=boldedge] (1.center) to (2);
	\end{pgfonlayer}
\end{tikzpicture}

%% file: figures/B_choi.tikz
\begin{tikzpicture}
	\begin{pgfonlayer}{nodelayer}
		\node [style=none] (2) at (-0.25, 0.25) {\tiny  $\pi$};
		\node [style=none] (3) at (-0.25, 0) {};
		\node [style=X dot] (4) at (1.25, 0) {\tiny $\gamma$};
		\node [style=ddotR] (5) at (0.5, 0) {\tiny $B$};
	\end{pgfonlayer}
	\begin{pgfonlayer}{edgelayer}
		\draw [style=boldedge] (4) to (3.center);
	\end{pgfonlayer}
\end{tikzpicture}

%% file: figures/gen_sigma.tikz
\begin{tikzpicture}
	\begin{pgfonlayer}{nodelayer}
		\node [style=none] (0) at (-0.25, 0) {};
		\node [style=Z dot] (1) at (0.5, 0) {\tiny $\sigma$};
		\node [style=none] (2) at (-0.25, 0.25) {\tiny  $\pi$};
	\end{pgfonlayer}
	\begin{pgfonlayer}{edgelayer}
		\draw [style=hadamard edge] (0.center) to (1);
	\end{pgfonlayer}
\end{tikzpicture}

%% file: figures/rmpu_2kotoc2.tikz
\begin{tikzpicture}
	\begin{pgfonlayer}{nodelayer}
		\node [style=box] (0) at (-3.25, 1.25) {};
		\node [style=box] (1) at (-1.25, 0.75) {};
		\node [style=box] (2) at (0.75, 0.25) {};
		\node [style=none] (3) at (-1.25, 0.4) {};
		\node [style=none] (4) at (0.5, 0.4) {};
		\node [style=none] (5) at (-3.25, 1.6) {};
		\node [style=none] (6) at (-3.25, 1) {};
		\node [style=none] (7) at (-1.25, 1) {};
		\node [style=none] (8) at (0.5, 0) {};
		\node [style=none] (9) at (2.25, 0) {};
		\node [style=X dot] (10) at (1.75, 0) {};
		\node [style=X dot] (11) at (-0.25, 0.4) {};
		\node [style=X dot] (12) at (-2.25, 1) {};
		\node [style=none] (13) at (2.75, -0.5) {\dots};
		\node [style=none, rotate=90] (14) at (-4.25, -0.5) {...};
		\node [style=none, rotate=90] (15) at (4.5, -0.5) {...};
		\node [style=box] (16) at (3.75, -1.5) {};
		\node [style=none] (17) at (3.75, -1.25) {};
		\node [style=none] (18) at (3, -1.25) {};
		\node [style=none] (19) at (3.75, -1.75) {};
		\node [style=none] (20) at (-2.75, -1.25) {};
		\node [style=none] (21) at (0.25, -1.25) {\dots};
		\node [style=Z dot] (36) at (5.5, 0.4) {\tiny $\gamma$};
		\node [style=Z dot] (37) at (4.75, 1) {\tiny $\gamma$};
		\node [style=Z dot] (39) at (4.75, -1.25) {\tiny $\gamma$};
		\node [style=Z dot] (42) at (-4.5, 1) {\tiny $e$};
		\node [style=Z dot] (43) at (-4.5, 0.4) {\tiny $e$};
		\node [style=Z dot] (44) at (-4.5, 0) {\tiny $e$};
		\node [style=Z dot] (45) at (-4.5, -1.25) {\tiny $e$};
		\node [style=Z dot] (53) at (-4.5, -1.75) {\tiny $e$ };
		\node [style=X dot] (54) at (4.75, -1.75) {\tiny $\gamma$};
		\node [style=ddotR] (55) at (-4.5, 1.6) {\tiny $A$};
		\node [style=Z dot] (56) at (4.75, 1.6) {\tiny $\gamma$};
		\node [style=ddotG] (57) at (4.75, 0.4) {\tiny $B$};
	\end{pgfonlayer}
	\begin{pgfonlayer}{edgelayer}
		\draw [style=boldedge] (3.center) to (4.center);
		\draw [style=boldedge] (6.center) to (7.center);
		\draw [style=boldedge] (9.center) to (8.center);
		\draw [style=boldedge] (17.center) to (18.center);
		\draw [style=hadamard edge] (7.center) to (37);
		\draw [style=hadamard edge] (4.center) to (36);
		\draw [style=hadamard edge] (17.center) to (39);
		\draw [style=hadamard edge] (8.center) to (44);
		\draw [style=hadamard edge] (43) to (3.center);
		\draw [style=hadamard edge] (42) to (6.center);
		\draw [style=hadamard edge] (20.center) to (45);
		\draw [style=hadamard edge] (19.center) to (53);
		\draw [style=boldedge] (54) to (19.center);
		\draw [style=boldedge] (55) to (5.center);
		\draw [style=hadamard edge] (56) to (5.center);
	\end{pgfonlayer}
\end{tikzpicture}

%% file: figures/rmpu_2kotoc3.tikz
\begin{tikzpicture}
	\begin{pgfonlayer}{nodelayer}
		\node [style=box] (0) at (-3, 0.5) {};
		\node [style=box] (1) at (-1, 0) {};
		\node [style=box] (2) at (1, -0.5) {};
		\node [style=none] (3) at (-1, -0.3) {};
		\node [style=none] (4) at (0.75, -0.3) {};
		\node [style=none] (5) at (-3, 0.85) {};
		\node [style=none] (6) at (-3, 0.25) {};
		\node [style=none] (7) at (-1, 0.25) {};
		\node [style=none] (8) at (0.75, -0.85) {};
		\node [style=none] (9) at (2, -0.85) {};
		\node [style=X dot] (10) at (2, -0.85) {\tiny $\gamma$};
		\node [style=X dot] (11) at (0, -0.3) {};
		\node [style=X dot] (12) at (-2, 0.25) {};
		\node [style=Z dot] (22) at (2.75, -0.3) {\tiny $\gamma$};
		\node [style=Z dot] (23) at (2, 0.25) {\tiny $\gamma$};
		\node [style=Z dot] (25) at (-4.25, 0.25) {\tiny $e$};
		\node [style=Z dot] (26) at (-4.25, -0.3) {\tiny $e$};
		\node [style=Z dot] (27) at (-4.25, -0.85) {\tiny $e$};
		\node [style=ddotR] (31) at (-4.25, 0.85) {\tiny $A$};
		\node [style=Z dot] (32) at (2, 0.85) {\tiny $\gamma$};
		\node [style=ddotG] (33) at (2, -0.3) {\tiny $B$};
	\end{pgfonlayer}
	\begin{pgfonlayer}{edgelayer}
		\draw [style=boldedge] (3.center) to (4.center);
		\draw [style=boldedge] (6.center) to (7.center);
		\draw [style=boldedge] (9.center) to (8.center);
		\draw [style=hadamard edge] (7.center) to (23);
		\draw [style=hadamard edge] (4.center) to (22);
		\draw [style=hadamard edge] (8.center) to (27);
		\draw [style=hadamard edge] (26) to (3.center);
		\draw [style=hadamard edge] (25) to (6.center);
		\draw [style=boldedge] (31) to (5.center);
		\draw [style=hadamard edge] (32) to (5.center);
	\end{pgfonlayer}
\end{tikzpicture}

%% file: figures/identlightcone1.tikz
\begin{tikzpicture}
	\begin{pgfonlayer}{nodelayer}
		\node [style=box] (0) at (0, 0) {};
		\node [style=none] (1) at (-0.25, 0.25) {};
		\node [style=none] (2) at (-0.25, -0.25) {};
		\node [style=none] (3) at (1, -0.25) {};
		\node [style=X dot] (4) at (1, -0.25) {\tiny $\mu$};
		\node [style=Z dot] (6) at (1, 0.25) {\tiny $\mu$};
		\node [style=Z dot] (8) at (-1, -0.25) {};
		\node [style=X dot] (9) at (-1, 0.25) {};
		\node [style=none] (10) at (-1.5, 0.25) {};
		\node [style=none] (11) at (-1.5, -0.25) {};
	\end{pgfonlayer}
	\begin{pgfonlayer}{edgelayer}
		\draw [style=boldedge] (3.center) to (2.center);
		\draw [style=hadamard edge] (1.center) to (6);
		\draw [style=hadamard edge] (2.center) to (11.center);
		\draw [style=boldedge] (1.center) to (10.center);
	\end{pgfonlayer}
\end{tikzpicture}

%% file: figures/identlightcone2.tikz
\begin{tikzpicture}
	\begin{pgfonlayer}{nodelayer}
		\node [style=X dot] (4) at (0.5, 0.25) {\tiny $\mu$};
		\node [style=Z dot] (5) at (0.5, -0.25) {\tiny $\mu$};
		\node [style=none] (6) at (-0.5, 0.25) {};
		\node [style=none] (7) at (-0.5, -0.25) {};
	\end{pgfonlayer}
	\begin{pgfonlayer}{edgelayer}
		\draw [style=boldedge] (4) to (6.center);
		\draw [style=hadamard edge] (5) to (7.center);
	\end{pgfonlayer}
\end{tikzpicture}

%% file: figures/rmpu_2kotoc1.tikz
\begin{tikzpicture}
	\begin{pgfonlayer}{nodelayer}
		\node [style=box] (0) at (-3.25, 1.25) {};
		\node [style=box] (1) at (-1.25, 0.75) {};
		\node [style=box] (2) at (0.75, 0.25) {};
		\node [style=none] (3) at (-1.25, 0.5) {};
		\node [style=none] (4) at (0.5, 0.5) {};
		\node [style=none] (5) at (-3.25, 1.6) {};
		\node [style=none] (6) at (-3.25, 1) {};
		\node [style=none] (7) at (-1.25, 1) {};
		\node [style=none] (8) at (0.5, 0) {};
		\node [style=none] (9) at (2.25, 0) {};
		\node [style=X dot] (10) at (1.75, 0) {};
		\node [style=X dot] (11) at (-0.25, 0.5) {};
		\node [style=X dot] (12) at (-2.25, 1) {};
		\node [style=none] (13) at (2.75, -0.5) {\dots};
		\node [style=none, rotate=90] (14) at (-4.25, -0.5) {...};
		\node [style=none, rotate=90] (15) at (4.5, -0.5) {...};
		\node [style=box] (16) at (3.75, -1.5) {};
		\node [style=none] (17) at (3.75, -1.25) {};
		\node [style=none] (18) at (3, -1.25) {};
		\node [style=none] (19) at (3.75, -1.85) {};
		\node [style=none] (20) at (-2.75, -1.25) {};
		\node [style=none] (21) at (0.25, -1.25) {\dots};
		\node [style=Z dot] (36) at (4.75, 0.5) {\tiny $\gamma$};
		\node [style=Z dot] (37) at (4.75, 1) {\tiny $\gamma$};
		\node [style=Z dot] (39) at (4.75, -1.25) {\tiny $\gamma$};
		\node [style=Z dot] (42) at (-4.5, 1) {\tiny $e$};
		\node [style=Z dot] (43) at (-4.5, 0.5) {\tiny $e$};
		\node [style=Z dot] (44) at (-4.5, 0) {\tiny $e$};
		\node [style=Z dot] (45) at (-4.5, -1.25) {\tiny $e$};
		\node [style=ddotG] (53) at (-4.5, -1.85) {\tiny $A$};
		\node [style=X dot] (54) at (4.75, -1.85) {\tiny $\gamma$};
		\node [style=X dot] (55) at (-4.5, 1.6) {\tiny $e$};
		\node [style=ddotG] (56) at (4.75, 1.6) {\tiny $B$};
		\node [style=Z dot] (57) at (5.5, 1.6) {\tiny $\gamma$};
	\end{pgfonlayer}
	\begin{pgfonlayer}{edgelayer}
		\draw [style=boldedge] (3.center) to (4.center);
		\draw [style=boldedge] (6.center) to (7.center);
		\draw [style=boldedge] (9.center) to (8.center);
		\draw [style=boldedge] (17.center) to (18.center);
		\draw [style=hadamard edge] (7.center) to (37);
		\draw [style=hadamard edge] (4.center) to (36);
		\draw [style=hadamard edge] (17.center) to (39);
		\draw [style=hadamard edge] (8.center) to (44);
		\draw [style=hadamard edge] (43) to (3.center);
		\draw [style=hadamard edge] (42) to (6.center);
		\draw [style=hadamard edge] (20.center) to (45);
		\draw [style=hadamard edge] (19.center) to (53);
		\draw [style=boldedge] (54) to (19.center);
		\draw [style=boldedge] (55) to (5.center);
		\draw [style=hadamard edge] (5.center) to (57);
	\end{pgfonlayer}
\end{tikzpicture}

%% file: figures/nonlocalops.tikz
\begin{tikzpicture}
	\begin{pgfonlayer}{nodelayer}
		\node [style=box] (0) at (-1, 0.25) {};
		\node [style=box] (1) at (1, -0.25) {};
		\node [style=none] (2) at (1, -0.6) {};
		\node [style=none] (4) at (-1, 0.6) {};
		\node [style=none] (5) at (-1, 0) {};
		\node [style=none] (6) at (1, 0) {};
		\node [style=X dot] (7) at (2.75, -0.6) {\tiny $\gamma$};
		\node [style=X dot] (8) at (0, 0) {};
		\node [style=Z dot] (10) at (2.75, 0) {\tiny $\gamma$};
		\node [style=Z dot] (11) at (2.75, 0.6) {\tiny $\gamma$};
		\node [style=ddotR] (12) at (-2.25, 0.6) {\tiny $A$};
		\node [style=ddotG] (13) at (-2.25, 0) {\tiny $A$};
		\node [style=ddotG] (14) at (-2.25, -0.6) {\tiny $A$};
		\node [style=ddotG] (15) at (2, 0.6) {\tiny $B$};
		\node [style=ddotR] (17) at (2, -0.6) {\tiny $B$};
		\node [style=ddotG] (18) at (2, 0) {\tiny $B$};
	\end{pgfonlayer}
	\begin{pgfonlayer}{edgelayer}
		\draw [style=boldedge] (5.center) to (6.center);
		\draw [style=hadamard edge] (4.center) to (11);
		\draw [style=hadamard edge] (6.center) to (10);
		\draw [style=boldedge] (4.center) to (12);
		\draw [style=hadamard edge] (14) to (2.center);
		\draw [style=hadamard edge] (13) to (5.center);
		\draw [style=boldedge] (2.center) to (7);
	\end{pgfonlayer}
\end{tikzpicture}

%% file: figures/rmpu1.tikz
\begin{tikzpicture}
	\begin{pgfonlayer}{nodelayer}
		\node [style=box] (2) at (-2.25, 0) {};
		\node [style=none] (4) at (-3.75, 0) {\dots};
		\node [style=none] (5) at (-3, 0.25) {};
		\node [style=none] (6) at (-3, 2) {};
		\node [style=none] (7) at (3, 2) {};
		\node [style=none] (9) at (-0.25, -0.25) {};
		\node [style=none] (10) at (-3.25, -0.25) {};
		\node [style=none] (11) at (-2.5, 0.25) {};
		\node [style=box] (12) at (0, 0) {};
		\node [style=none] (13) at (-0.25, 0.25) {};
		\node [style=none] (14) at (-1, 0.25) {};
		\node [style=none] (15) at (-1, 1) {};
		\node [style=none] (16) at (1, 1) {};
		\node [style=none] (17) at (1, 0.25) {};
		\node [style=none] (18) at (0.25, 0.25) {};
		\node [style=box] (19) at (2.25, 0) {};
		\node [style=none] (20) at (4, 0) {\dots};
		\node [style=none] (21) at (2.5, 0.25) {};
		\node [style=none] (22) at (3.25, -0.25) {};
		\node [style=none] (24) at (3, 0.25) {};
		\node [style=none] (26) at (0.25, -0.25) {};
		\node [style=Z dot] (27) at (0, 1) {};
		\node [style=X dot] (28) at (1, -0.25) {};
		\node [style=X dot] (29) at (-1.25, -0.25) {};
		\node [style=Z dot] (30) at (0, 2) {};
		\node [style=none] (32) at (-1.5, 0.25) {};
		\node [style=none] (33) at (-1.5, 1.5) {};
		\node [style=none] (34) at (1.5, 1.5) {};
		\node [style=none] (35) at (1.5, 0.25) {};
		\node [style=none] (36) at (2, 0.25) {};
		\node [style=none] (37) at (-2, 0.25) {};
		\node [style=Z dot] (38) at (0, 1.5) {};
		\node [style=none] (39) at (-0.25, -2.25) {};
		\node [style=box] (40) at (0, -2) {};
		\node [style=none] (41) at (-0.25, -1.75) {};
		\node [style=none] (42) at (-1, -1.75) {};
		\node [style=none] (43) at (-1, -1) {};
		\node [style=none] (44) at (1, -1) {};
		\node [style=none] (45) at (1, -1.75) {};
		\node [style=none] (46) at (0.25, -1.75) {};
		\node [style=none] (47) at (0.25, -2.25) {};
		\node [style=Z dot] (48) at (0, -1) {};
		\node [style=X dot] (49) at (1, -2.25) {};
		\node [style=X dot] (50) at (-1.25, -2.25) {};
		\node [style=box] (51) at (-5.25, 0) {};
		\node [style=none] (52) at (-6.5, 0.25) {};
		\node [style=none] (53) at (-6.5, 4) {};
		\node [style=none] (54) at (6.5, 4) {};
		\node [style=none] (55) at (-5.5, 0.25) {};
		\node [style=box] (56) at (5.25, 0) {};
		\node [style=none] (57) at (5.5, 0.25) {};
		\node [style=none] (58) at (6.5, 0.25) {};
		\node [style=Z dot] (59) at (0, 4) {};
		\node [style=none] (60) at (-4.5, 0.25) {};
		\node [style=none] (61) at (-4.5, 3.5) {};
		\node [style=none] (62) at (4.5, 3.5) {};
		\node [style=none] (63) at (4.5, 0.25) {};
		\node [style=none] (64) at (5, 0.25) {};
		\node [style=none] (65) at (-5, 0.25) {};
		\node [style=Z dot] (66) at (0, 3.5) {};
		\node [style=none, rotate=90] (67) at (0, 2.75) {...};
		\node [style=none] (68) at (4.5, -0.25) {};
		\node [style=none] (69) at (6.5, -0.25) {};
		\node [style=none] (70) at (6.5, -2.25) {};
		\node [style=none] (71) at (-6.5, -0.25) {};
		\node [style=none] (72) at (-4.5, -0.25) {};
		\node [style=none] (73) at (-6.5, -2.25) {};
		\node [style=X dot] (74) at (3, -0.25) {};
		\node [style=X dot] (75) at (-3, -0.25) {};
		\node [style=none] (76) at (1.5, -1) {};
		\node [style=none] (77) at (6, -1) {};
		\node [style=none] (78) at (3.75, -1.5) {n-2};
	\end{pgfonlayer}
	\begin{pgfonlayer}{edgelayer}
		\draw [style=simple] (5.center) to (6.center);
		\draw [style=simple] (6.center) to (7.center);
		\draw [style=simple] (5.center) to (11.center);
		\draw [style=simple] (13.center) to (14.center);
		\draw [style=simple] (14.center) to (15.center);
		\draw [style=simple] (15.center) to (16.center);
		\draw [style=simple] (16.center) to (17.center);
		\draw [style=simple] (17.center) to (18.center);
		\draw [style=simple] (21.center) to (24.center);
		\draw [style=simple] (7.center) to (24.center);
		\draw [style=boldedge] (26.center) to (22.center);
		\draw [style=boldedge] (9.center) to (10.center);
		\draw [style=simple](32.center) to (33.center);
		\draw [style=simple](33.center) to (34.center);
		\draw [style=simple] (34.center) to (35.center);
		\draw [style=simple] (35.center) to (36.center);
		\draw [style=simple] (37.center) to (32.center);
		\draw [style=simple] (41.center) to (42.center);
		\draw [style=simple] (42.center) to (43.center);
		\draw [style=simple] (43.center) to (44.center);
		\draw [style=simple] (44.center) to (45.center);
		\draw [style=simple] (45.center) to (46.center);
		\draw [style=simple] (52.center) to (53.center);
		\draw [style=simple] (53.center) to (54.center);
		\draw [style=simple] (52.center) to (55.center);
		\draw [style=simple] (57.center) to (58.center);
		\draw [style=simple] (54.center) to (58.center);
		\draw [style=simple] (60.center) to (61.center);
		\draw [style=simple] (61.center) to (62.center);
		\draw [style=simple] (62.center) to (63.center);
		\draw [style=simple] (63.center) to (64.center);
		\draw [style=simple] (65.center) to (60.center);
		\draw [style=boldedge] (47.center) to (70.center);
		\draw [style=boldedge] (70.center) to (69.center);
		\draw [style=boldedge] (68.center) to (69.center);
		\draw [style=boldedge] (71.center) to (72.center);
		\draw [style=boldedge] (71.center) to (73.center);
		\draw [style=boldedge] (73.center) to (39.center);
		\draw [style=brace edge] (77.center) to (76.center);
	\end{pgfonlayer}
\end{tikzpicture}

%% file: figures/ident1.tikz
\begin{tikzpicture}
	\begin{pgfonlayer}{nodelayer}
		\node [style=none] (0) at (-1, -0.25) {};
		\node [style=box] (1) at (-0.75, 0) {};
		\node [style=none] (2) at (-1, 0.25) {};
		\node [style=none] (3) at (-1.75, 0.25) {};
		\node [style=none] (4) at (2, 0.25) {};
		\node [style=none] (5) at (-0.5, 0.25) {};
		\node [style=none] (6) at (-0.5, -0.25) {};
		\node [style=X dot] (7) at (0.25, -0.25) {};
		\node [style=none] (8) at (2, -0.25) {};
		\node [style=none] (9) at (-1.75, -0.25) {};
		\node [style=Z dot] (10) at (0.25, 0.25) {};
		\node [style=box] (11) at (1.25, 0) {};
	\end{pgfonlayer}
	\begin{pgfonlayer}{edgelayer}
		\draw [style=simple] (2.center) to (3.center);
		\draw [style=simple] (4.center) to (5.center);
		\draw [style=boldedge] (6.center) to (8.center);
		\draw [style=boldedge] (0.center) to (9.center);
	\end{pgfonlayer}
\end{tikzpicture}

%% file: figures/fpwg1.tikz
\begin{tikzpicture}
	\begin{pgfonlayer}{nodelayer}
		\node [style=none] (0) at (3, 0.25) {};
		\node [style=none] (1) at (-2.5, 0.25) {};
		\node [style=none] (2) at (-2.5, -0.25) {};
		\node [style=none] (3) at (3, -0.25) {};
		\node [style=box] (4) at (-1.5, 0) {};
		\node [style=box] (5) at (1.5, 0) {};
		\node [style=Z dot] (6) at (0, 0.25) {};
		\node [style=X dot] (7) at (0, -0.25) {};
		\node [style=Z dot] (8) at (2.5, 0.25) {};
		\node [style=X dot] (9) at (2.5, -0.25) {};
		\node [style=none] (10) at (-2.5, 1) {};
		\node [style=none] (11) at (3, 1) {};
		\node [style=none] (12) at (-2.5, -1) {};
		\node [style=none] (13) at (3, -1) {};
	\end{pgfonlayer}
	\begin{pgfonlayer}{edgelayer}
		\draw [style=simple] (0.center) to (1.center);
		\draw [style=boldedge] (2.center) to (3.center);
		\draw [style=boldedge] (3.center) to (13.center);
		\draw [style=boldedge] (13.center) to (12.center);
		\draw [style=boldedge] (12.center) to (2.center);
		\draw [style=simple] (1.center) to (10.center);
		\draw [style=simple] (10.center) to (11.center);
		\draw [style=simple] (11.center) to (0.center);
	\end{pgfonlayer}
\end{tikzpicture}

%% file: figures/fpwg2.tikz
\begin{tikzpicture}
	\begin{pgfonlayer}{nodelayer}
		\node [style=none] (0) at (1.5, 0.25) {};
		\node [style=none] (1) at (-1, 0.25) {};
		\node [style=none] (2) at (-1, -0.25) {};
		\node [style=none] (3) at (1.5, -0.25) {};
		\node [style=box] (5) at (0, 0) {};
		\node [style=Z dot] (8) at (1, 0.25) {};
		\node [style=X dot] (9) at (1, -0.25) {};
		\node [style=none] (10) at (-1, 1) {};
		\node [style=none] (11) at (1.5, 1) {};
		\node [style=none] (12) at (-1, -1) {};
		\node [style=none] (13) at (1.5, -1) {};
	\end{pgfonlayer}
	\begin{pgfonlayer}{edgelayer}
		\draw [style=simple] (0.center) to (1.center);
		\draw [style=boldedge] (2.center) to (3.center);
		\draw [style=boldedge] (3.center) to (13.center);
		\draw [style=boldedge] (13.center) to (12.center);
		\draw [style=boldedge] (12.center) to (2.center);
		\draw [style=simple] (1.center) to (10.center);
		\draw [style=simple] (10.center) to (11.center);
		\draw [style=simple] (11.center) to (0.center);
	\end{pgfonlayer}
\end{tikzpicture}

%% file: figures/fpwg3.tikz
\begin{tikzpicture}
	\begin{pgfonlayer}{nodelayer}
		\node [style=none] (0) at (0.5, 0.25) {};
		\node [style=none] (1) at (-1, 0.25) {};
		\node [style=none] (2) at (-1, -0.25) {};
		\node [style=none] (3) at (0.5, -0.25) {};
		\node [style=none] (10) at (-1, 1) {};
		\node [style=none] (11) at (0.5, 1) {};
		\node [style=none] (12) at (-1, -1) {};
		\node [style=none] (13) at (0.5, -1) {};
	\end{pgfonlayer}
	\begin{pgfonlayer}{edgelayer}
		\draw [style=simple] (0.center) to (1.center);
		\draw [style=boldedge] (2.center) to (3.center);
		\draw [style=boldedge] (3.center) to (13.center);
		\draw [style=boldedge] (13.center) to (12.center);
		\draw [style=boldedge] (12.center) to (2.center);
		\draw [style=simple] (1.center) to (10.center);
		\draw [style=simple] (10.center) to (11.center);
		\draw [style=simple] (11.center) to (0.center);
	\end{pgfonlayer}
\end{tikzpicture}

%% file: figures/rmpu0.tikz
\begin{tikzpicture}
	\begin{pgfonlayer}{nodelayer}
		\node [style=box] (0) at (-1, 1.5) {};
		\node [style=box] (1) at (1, 1.5) {};
		\node [style=box] (2) at (-3, 1) {};
		\node [style=box] (3) at (3, 1) {};
		\node [style=box] (4) at (-5, 0.5) {};
		\node [style=box] (5) at (5, 0.5) {};
		\node [style=none] (6) at (-11, 1.75) {};
		\node [style=none] (7) at (-1, 1.75) {};
		\node [style=none] (8) at (-3, 1.25) {};
		\node [style=none] (9) at (-1, 1.25) {};
		\node [style=none] (10) at (3, 0.75) {};
		\node [style=none] (11) at (4.75, 0.75) {};
		\node [style=none] (12) at (-3, 0.75) {};
		\node [style=none] (13) at (-4.75, 0.75) {};
		\node [style=none] (14) at (9, 1.75) {};
		\node [style=none] (15) at (1, 1.75) {};
		\node [style=none] (16) at (1, 1.25) {};
		\node [style=none] (17) at (3, 1.25) {};
		\node [style=none] (18) at (-4.75, 0.25) {};
		\node [style=none] (19) at (4.75, 0.25) {};
		\node [style=none] (20) at (-11, 1.25) {};
		\node [style=none] (21) at (-11, 0.75) {};
		\node [style=none] (22) at (-6.5, 0.25) {};
		\node [style=none] (23) at (6.5, 0.25) {};
		\node [style=none] (24) at (9, 1.25) {};
		\node [style=none] (25) at (9, 0.75) {};
		\node [style=X dot] (26) at (0, 1.75) {};
		\node [style=X dot] (27) at (-2, 1.25) {};
		\node [style=X dot] (28) at (-4, 0.75) {};
		\node [style=X dot] (29) at (-6, 0.25) {};
		\node [style=X dot] (30) at (6, 0.25) {};
		\node [style=X dot] (31) at (4, 0.75) {};
		\node [style=X dot] (32) at (2, 1.25) {};
		\node [style=Z dot] (33) at (6.75, 1.75) {};
		\node [style=Z dot] (34) at (6.75, 1.25) {};
		\node [style=Z dot] (35) at (6.75, 0.75) {};
		\node [style=none] (36) at (7.25, 0.25) {\dots};
		\node [style=none] (37) at (-7.25, 0.25) {\dots};
		\node [style=box] (38) at (-10, -1.25) {};
		\node [style=none] (39) at (-10, -1) {};
		\node [style=none] (40) at (-9.25, -1) {};
		\node [style=none] (41) at (-11, -1) {};
		\node [style=none] (42) at (-10, -1.5) {};
		\node [style=none] (43) at (-11, -1.5) {};
		\node [style=Z dot] (44) at (0, 1.25) {};
		\node [style=Z dot] (45) at (0, 0.75) {};
		\node [style=Z dot] (46) at (0, 0.25) {};
		\node [style=none] (47) at (-8.5, -1) {\dots};
		\node [style=none, rotate=90] (48) at (-10.75, -0.25) {...};
		\node [style=none, rotate=90] (49) at (0, -0.5) {...};
		\node [style=none, rotate=90] (51) at (8.75, -0.25) {...};
		\node [style=box] (52) at (8, -1.25) {};
		\node [style=none] (53) at (8, -1) {};
		\node [style=none] (54) at (7.25, -1) {};
		\node [style=none] (55) at (9, -1) {};
		\node [style=none] (56) at (8, -1.5) {};
		\node [style=none] (57) at (9, -1.5) {};
		\node [style=Z dot] (58) at (8.75, -1) {};
		\node [style=X dot] (59) at (8.75, -1.5) {};
	\end{pgfonlayer}
	\begin{pgfonlayer}{edgelayer}
		\draw [style=simple] (7.center) to (6.center);
		\draw [style=boldedge] (9.center) to (8.center);
		\draw [style=boldedge] (10.center) to (11.center);
		\draw [style=boldedge] (13.center) to (12.center);
		\draw [style=simple] (14.center) to (15.center);
		\draw [style=boldedge] (7.center) to (15.center);
		\draw [style=boldedge] (16.center) to (17.center);
		\draw [style=simple] (9.center) to (16.center);
		\draw [style=simple] (12.center) to (10.center);
		\draw [style=simple] (19.center) to (18.center);
		\draw [style=simple] (20.center) to (8.center);
		\draw [style=simple] (21.center) to (13.center);
		\draw [style=boldedge] (22.center) to (18.center);
		\draw [style=boldedge] (23.center) to (19.center);
		\draw [style=simple] (17.center) to (24.center);
		\draw [style=simple] (25.center) to (11.center);
		\draw [style=boldedge] (39.center) to (40.center);
		\draw [style=simple] (41.center) to (39.center);
		\draw [style=boldedge] (43.center) to (42.center);
		\draw [style=boldedge] (53.center) to (54.center);
		\draw [style=simple] (55.center) to (53.center);
		\draw [style=boldedge] (57.center) to (56.center);
		\draw [style=simple] (42.center) to (56.center);
	\end{pgfonlayer}
\end{tikzpicture}

%% file: figures/wglo.tikz
\begin{tikzpicture}
	\begin{pgfonlayer}{nodelayer}
		\node [style=box] (0) at (0, 0) {$\mathbb{I}$};
		\node [style=none] (1) at (1, -0.25) {};
		\node [style=none] (2) at (-1, -0.25) {};
		\node [style=none] (3) at (-1, 0.25) {};
		\node [style=none] (4) at (1, 0.25) {};
	\end{pgfonlayer}
	\begin{pgfonlayer}{edgelayer}
		\draw [style=boldedge] (1.center) to (2.center);
		\draw [style=hadamard edge] (3.center) to (4.center);
	\end{pgfonlayer}
\end{tikzpicture}

%% file: figures/wgslo.tikz
\begin{tikzpicture}
	\begin{pgfonlayer}{nodelayer}
		\node [style=box] (0) at (0, 0) {$\delta$};
		\node [style=none] (3) at (1, -0.25) {};
		\node [style=none] (4) at (-1, -0.25) {};
		\node [style=none] (5) at (-1, 0.25) {};
		\node [style=none] (6) at (1, 0.25) {};
	\end{pgfonlayer}
	\begin{pgfonlayer}{edgelayer}
		\draw [style=boldedge] (3.center) to (4.center);
		\draw [style=hadamard edge] (5.center) to (6.center);
	\end{pgfonlayer}
\end{tikzpicture}

%% file: figures/lo_wgG.tikz
\begin{tikzpicture}
	\begin{pgfonlayer}{nodelayer}
		\node [style=none] (0) at (1.5, -0.25) {};
		\node [style=box] (1) at (0, 0) {};
		\node [style=none] (2) at (1.5, 0.25) {};
		\node [style=none] (3) at (-1, 0.25) {};
		\node [style=none] (4) at (0.25, 0.25) {};
		\node [style=X dot] (6) at (1, -0.25) {};
		\node [style=none] (7) at (-1, -0.25) {};
	\end{pgfonlayer}
	\begin{pgfonlayer}{edgelayer}
		\draw [style=simple] (2.center) to (3.center);
		\draw [style=boldedge] (0.center) to (7.center);
	\end{pgfonlayer}
\end{tikzpicture}

%% file: figures/gesu.tikz
\begin{tikzpicture}
\begin{pgfonlayer}{nodelayer}
		\node [style=box] (2) at (-2.25, 0) {$\mathbb{I}$};
		\node [style=none] (4) at (-3.75, 0) {\dots};
		\node [style=none] (5) at (-3, 0.25) {};
		\node [style=none] (6) at (-3, 2) {};
		\node [style=none] (7) at (3, 2) {};
		\node [style=none] (9) at (-0.25, -0.25) {};
		\node [style=none] (10) at (-3.25, -0.25) {};
		\node [style=none] (11) at (-2.5, 0.25) {};
		\node [style=box] (12) at (0, 0) {$\mathbb{I}$};
		\node [style=none] (13) at (-0.25, 0.25) {};
		\node [style=none] (14) at (-1, 0.25) {};
		\node [style=none] (15) at (-1, 1) {};
		\node [style=none] (16) at (1, 1) {};
		\node [style=none] (17) at (1, 0.25) {};
		\node [style=none] (18) at (0.25, 0.25) {};
		\node [style=box] (19) at (2.25, 0) {$\mathbb{I}$};
		\node [style=none] (20) at (4, 0) {\dots};
		\node [style=none] (21) at (2.5, 0.25) {};
		\node [style=none] (22) at (3.25, -0.25) {};
		\node [style=none] (24) at (3, 0.25) {};
		\node [style=none] (26) at (0.25, -0.25) {};
		\node [style=Z dot] (27) at (0, 1) {};
		\node [style=Z dot] (30) at (0, 2) {};
		\node [style=none] (32) at (-1.5, 0.25) {};
		\node [style=none] (33) at (-1.5, 1.5) {};
		\node [style=none] (34) at (1.5, 1.5) {};
		\node [style=none] (35) at (1.5, 0.25) {};
		\node [style=none] (36) at (2, 0.25) {};
		\node [style=none] (37) at (-2, 0.25) {};
		\node [style=Z dot] (38) at (0, 1.5) {};
		\node [style=none] (39) at (-0.25, -2.25) {};
		\node [style=box] (40) at (0, -2) {$\mathbb{I}$};
		\node [style=none] (41) at (-0.25, -1.75) {};
		\node [style=none] (42) at (-1, -1.75) {};
		\node [style=none] (43) at (-1, -1) {};
		\node [style=none] (44) at (1, -1) {};
		\node [style=none] (45) at (1, -1.75) {};
		\node [style=none] (46) at (0.25, -1.75) {};
		\node [style=none] (47) at (0.25, -2.25) {};
		\node [style=Z dot] (48) at (0, -1) {};
		\node [style=box] (51) at (-5.25, 0) {$\mathbb{I}$};
		\node [style=none] (52) at (-6.5, 0.25) {};
		\node [style=none] (53) at (-6.5, 4) {};
		\node [style=none] (54) at (6.5, 4) {};
		\node [style=none] (55) at (-5.5, 0.25) {};
		\node [style=box] (56) at (5.25, 0) {$\mathbb{I}$};
		\node [style=none] (57) at (5.5, 0.25) {};
		\node [style=none] (58) at (6.5, 0.25) {};
		\node [style=Z dot] (59) at (0, 4) {};
		\node [style=none] (60) at (-4.5, 0.25) {};
		\node [style=none] (61) at (-4.5, 3.5) {};
		\node [style=none] (62) at (4.5, 3.5) {};
		\node [style=none] (63) at (4.5, 0.25) {};
		\node [style=none] (64) at (5, 0.25) {};
		\node [style=none] (65) at (-5, 0.25) {};
		\node [style=Z dot] (66) at (0, 3.5) {};
		\node [style=none, rotate=90] (67) at (0, 2.75) {...};
		\node [style=none] (68) at (4.5, -0.25) {};
		\node [style=none] (69) at (6.5, -0.25) {};
		\node [style=none] (70) at (6.5, -2.25) {};
		\node [style=none] (71) at (-6.5, -0.25) {};
		\node [style=none] (72) at (-4.5, -0.25) {};
		\node [style=none] (73) at (-6.5, -2.25) {};
		\node [style=none] (76) at (1.5, -1) {};
		\node [style=none] (77) at (6, -1) {};
		\node [style=none] (78) at (3.75, -1.5) {n-2};
	\end{pgfonlayer}
	\begin{pgfonlayer}{edgelayer}
		\draw [style=simple] (5.center) to (6.center);
		\draw [style=simple] (6.center) to (7.center);
		\draw [style=simple] (5.center) to (11.center);
		\draw [style=simple] (13.center) to (14.center);
		\draw [style=simple] (14.center) to (15.center);
		\draw [style=simple] (15.center) to (16.center);
		\draw [style=simple] (16.center) to (17.center);
		\draw [style=simple] (17.center) to (18.center);
		\draw [style=simple] (21.center) to (24.center);
		\draw [style=simple] (7.center) to (24.center);
		\draw [style=boldedge] (26.center) to (22.center);
		\draw [style=boldedge] (9.center) to (10.center);
		\draw [style=simple](32.center) to (33.center);
		\draw [style=simple](33.center) to (34.center);
		\draw [style=simple] (34.center) to (35.center);
		\draw [style=simple] (35.center) to (36.center);
		\draw [style=simple] (37.center) to (32.center);
		\draw [style=simple] (41.center) to (42.center);
		\draw [style=simple] (42.center) to (43.center);
		\draw [style=simple] (43.center) to (44.center);
		\draw [style=simple] (44.center) to (45.center);
		\draw [style=simple] (45.center) to (46.center);
		\draw [style=simple] (52.center) to (53.center);
		\draw [style=simple] (53.center) to (54.center);
		\draw [style=simple] (52.center) to (55.center);
		\draw [style=simple] (57.center) to (58.center);
		\draw [style=simple] (54.center) to (58.center);
		\draw [style=simple] (60.center) to (61.center);
		\draw [style=simple] (61.center) to (62.center);
		\draw [style=simple] (62.center) to (63.center);
		\draw [style=simple] (63.center) to (64.center);
		\draw [style=simple] (65.center) to (60.center);
		\draw [style=boldedge] (47.center) to (70.center);
		\draw [style=boldedge] (70.center) to (69.center);
		\draw [style=boldedge] (68.center) to (69.center);
		\draw [style=boldedge] (71.center) to (72.center);
		\draw [style=boldedge] (71.center) to (73.center);
		\draw [style=boldedge] (73.center) to (39.center);
		\draw [style=brace edge] (77.center) to (76.center);
	\end{pgfonlayer}
\end{tikzpicture}

%% file: figures/rmpu2.tikz
\begin{tikzpicture}
	\begin{pgfonlayer}{nodelayer}
		\node [style=box] (0) at (-2.25, -0.25) {$\delta$};
		\node [style=none] (1) at (-3.75, -0.25) {\dots};
		\node [style=none] (2) at (-3, 0) {};
		\node [style=none] (3) at (-3, 1.75) {};
		\node [style=none] (4) at (3, 1.75) {};
		\node [style=none] (5) at (-0.25, -0.5) {};
		\node [style=none] (6) at (-3.25, -0.5) {};
		\node [style=none] (7) at (-2.5, 0) {};
		\node [style=box] (8) at (0, -0.25) {$\mathbb{I}$};
		\node [style=none] (9) at (-0.25, 0) {};
		\node [style=none] (10) at (-1, 0) {};
		\node [style=none] (11) at (-1, 0.75) {};
		\node [style=none] (12) at (1, 0.75) {};
		\node [style=none] (13) at (1, 0) {};
		\node [style=none] (14) at (0.25, 0) {};
		\node [style=box] (15) at (2.25, -0.25) {$\mathbb{I}$};
		\node [style=none] (16) at (4, -0.25) {\dots};
		\node [style=none] (17) at (2.5, 0) {};
		\node [style=none] (18) at (3.25, -0.5) {};
		\node [style=none] (19) at (3, 0) {};
		\node [style=none] (20) at (0.25, -0.5) {};
		\node [style=Z dot] (21) at (0, 0.75) {};
		\node [style=Z dot] (24) at (0, 1.75) {};
		\node [style=none] (25) at (-1.5, 0) {};
		\node [style=none] (26) at (-1.5, 1.25) {};
		\node [style=none] (27) at (1.5, 1.25) {};
		\node [style=none] (28) at (1.5, 0) {};
		\node [style=none] (29) at (2, 0) {};
		\node [style=none] (30) at (-2, 0) {};
		\node [style=Z dot] (31) at (0, 1.25) {};
		\node [style=none] (32) at (-0.25, -2.5) {};
		\node [style=box] (33) at (0, -2.25) {$\mathbb{I}$};
		\node [style=none] (34) at (-0.25, -2) {};
		\node [style=none] (35) at (-1, -2) {};
		\node [style=none] (36) at (-1, -1.25) {};
		\node [style=none] (37) at (1, -1.25) {};
		\node [style=none] (38) at (1, -2) {};
		\node [style=none] (39) at (0.25, -2) {};
		\node [style=none] (40) at (0.25, -2.5) {};
		\node [style=Z dot] (41) at (0, -1.25) {};
		\node [style=box] (44) at (-5.25, -0.25) {$\mathbb{I}$};
		\node [style=none] (45) at (-6.5, 0) {};
		\node [style=none] (46) at (-6.5, 3.75) {};
		\node [style=none] (47) at (6.5, 3.75) {};
		\node [style=none] (48) at (-5.5, 0) {};
		\node [style=box] (49) at (5.25, -0.25) {$\mathbb{I}$};
		\node [style=none] (50) at (5.5, 0) {};
		\node [style=none] (51) at (6.5, 0) {};
		\node [style=Z dot] (52) at (0, 3.75) {};
		\node [style=none] (53) at (-4.5, 0) {};
		\node [style=none] (54) at (-4.5, 3.25) {};
		\node [style=none] (55) at (4.5, 3.25) {};
		\node [style=none] (56) at (4.5, 0) {};
		\node [style=none] (57) at (5, 0) {};
		\node [style=none] (58) at (-5, 0) {};
		\node [style=Z dot] (59) at (0, 3.25) {};
		\node [style=none, rotate=90] (60) at (0, 2.5) {...};
		\node [style=none] (61) at (4.5, -0.5) {};
		\node [style=none] (62) at (6.5, -0.5) {};
		\node [style=none] (63) at (6.5, -2.5) {};
		\node [style=none] (64) at (-6.5, -0.5) {};
		\node [style=none] (65) at (-4.5, -0.5) {};
		\node [style=none] (66) at (-6.5, -2.5) {};
		\node [style=none] (69) at (1.5, -1.25) {};
		\node [style=none] (70) at (6, -1.25) {};
	\end{pgfonlayer}
	\begin{pgfonlayer}{edgelayer}
		\draw [style=simple](2.center) to (3.center);
		\draw [style=simple](3.center) to (4.center);
		\draw [style=simple](2.center) to (7.center);
		\draw [style=simple] (9.center) to (10.center);
		\draw [style=simple] (10.center) to (11.center);
		\draw [style=simple] (11.center) to (12.center);
		\draw [style=simple] (12.center) to (13.center);
		\draw [style=simple] (13.center) to (14.center);
		\draw (17.center) to (19.center);
		\draw [style=simple] (4.center) to (19.center);
		\draw [style=boldedge] (20.center) to (18.center);
		\draw [style=boldedge] (5.center) to (6.center);
		\draw [style=simple](25.center) to (26.center);
		\draw [style=simple](26.center) to (27.center);
		\draw [style=simple] (27.center) to (28.center);
		\draw [style=simple] (28.center) to (29.center);
		\draw [style=simple] (30.center) to (25.center);
		\draw [style=simple] (34.center) to (35.center);
		\draw [style=simple] (35.center) to (36.center);
		\draw [style=simple] (36.center) to (37.center);
		\draw [style=simple] (37.center) to (38.center);
		\draw [style=simple] (38.center) to (39.center);
		\draw [style=simple](45.center) to (46.center);
		\draw [style=simple](46.center) to (47.center);
		\draw [style=simple](45.center) to (48.center);
		\draw [style=simple](50.center) to (51.center);
		\draw [style=simple] (47.center) to (51.center);
		\draw [style=simple](53.center) to (54.center);
		\draw [style=simple](54.center) to (55.center);
		\draw [style=simple] (55.center) to (56.center);
		\draw [style=simple] (56.center) to (57.center);
		\draw [style=simple] (58.center) to (53.center);
		\draw [style=boldedge] (40.center) to (63.center);
		\draw [style=boldedge] (63.center) to (62.center);
		\draw [style=boldedge] (61.center) to (62.center);
		\draw [style=boldedge] (64.center) to (65.center);
		\draw [style=boldedge] (64.center) to (66.center);
		\draw [style=boldedge] (66.center) to (32.center);
	\end{pgfonlayer}
\end{tikzpicture}

%% file: figures/chi1.tikz
\begin{tikzpicture}
	\begin{pgfonlayer}{nodelayer}
		\node [style=box] (0) at (0, 0) {$\delta$};
		\node [style=none] (1) at (-1, 0.25) {};
		\node [style=none] (2) at (1, -0.25) {};
		\node [style=none] (3) at (-1, -0.25) {};
		\node [style=none] (4) at (-0.25, 0.25) {};
		\node [style=none] (5) at (1, 0.25) {};
		\node [style=none] (6) at (0.25, 0.25) {};
		\node [style=none] (7) at (-1, -0.75) {};
		\node [style=none] (8) at (1, -0.75) {};
		\node [style=none] (9) at (-1, 1) {};
		\node [style=none] (10) at (1, 1) {};
		\node [style=Z dot] (11) at (0, 1) {};
	\end{pgfonlayer}
	\begin{pgfonlayer}{edgelayer}
		\draw (1.center) to (4.center);
		\draw [style=boldedge] (2.center) to (3.center);
		\draw [style=simple] (6.center) to (5.center);
		\draw [style=boldedge] (3.center) to (7.center);
		\draw [style=boldedge] (7.center) to (8.center);
		\draw [style=boldedge] (8.center) to (2.center);
		\draw [style=simple] (9.center) to (1.center);
		\draw [style=simple] (9.center) to (10.center);
		\draw [style=simple] (10.center) to (5.center);
	\end{pgfonlayer}
\end{tikzpicture}

%% file: figures/rmpu3.tikz
\begin{tikzpicture}
	\begin{pgfonlayer}{nodelayer}
		\node [style=box] (0) at (-2.25, -0.25) {$\mathbb{I}$};
		\node [style=none] (1) at (-3.75, -0.25) {\dots};
		\node [style=none] (2) at (-3, 0) {};
		\node [style=none] (3) at (-3, 1.75) {};
		\node [style=none] (4) at (3, 1.75) {};
		\node [style=none] (5) at (-0.25, -0.5) {};
		\node [style=none] (6) at (-3.25, -0.5) {};
		\node [style=none] (7) at (-2.5, 0) {};
		\node [style=box] (8) at (0, -0.25) {};
		\node [style=none] (9) at (-0.25, 0) {};
		\node [style=none] (10) at (-1, 0) {};
		\node [style=none] (11) at (-1, 0.75) {};
		\node [style=none] (12) at (1, 0.75) {};
		\node [style=none] (13) at (1, 0) {};
		\node [style=none] (14) at (0.25, 0) {};
		\node [style=box] (15) at (2.25, -0.25) {$\mathbb{I}$};
		\node [style=none] (16) at (4, -0.25) {\dots};
		\node [style=none] (17) at (2.5, 0) {};
		\node [style=none] (18) at (3.25, -0.5) {};
		\node [style=none] (19) at (3, 0) {};
		\node [style=none] (20) at (0.25, -0.5) {};
		\node [style=Z dot] (21) at (0, 0.75) {};
		\node [style=Z dot] (24) at (0, 1.75) {};
		\node [style=none] (25) at (-1.5, 0) {};
		\node [style=none] (26) at (-1.5, 1.25) {};
		\node [style=none] (27) at (1.5, 1.25) {};
		\node [style=none] (28) at (1.5, 0) {};
		\node [style=none] (29) at (2, 0) {};
		\node [style=none] (30) at (-2, 0) {};
		\node [style=Z dot] (31) at (0, 1.25) {};
		\node [style=none] (32) at (-0.25, -2.5) {};
		\node [style=box] (33) at (0, -2.25) {};
		\node [style=none] (34) at (-0.25, -2) {};
		\node [style=none] (35) at (-1, -2) {};
		\node [style=none] (36) at (-1, -1.25) {};
		\node [style=none] (37) at (1, -1.25) {};
		\node [style=none] (38) at (1, -2) {};
		\node [style=none] (39) at (0.25, -2) {};
		\node [style=none] (40) at (0.25, -2.5) {};
		\node [style=Z dot] (41) at (0, -1.25) {};
		\node [style=box] (44) at (-5.25, -0.25) {$\mathbb{I}$};
		\node [style=none] (45) at (-6.5, 0) {};
		\node [style=none] (46) at (-6.5, 3.75) {};
		\node [style=none] (47) at (6.5, 3.75) {};
		\node [style=none] (48) at (-5.5, 0) {};
		\node [style=box] (49) at (5.25, -0.25) {$\mathbb{I}$};
		\node [style=none] (50) at (5.5, 0) {};
		\node [style=none] (51) at (6.5, 0) {};
		\node [style=Z dot] (52) at (0, 3.75) {};
		\node [style=none] (53) at (-4.5, 0) {};
		\node [style=none] (54) at (-4.5, 3.25) {};
		\node [style=none] (55) at (4.5, 3.25) {};
		\node [style=none] (56) at (4.5, 0) {};
		\node [style=none] (57) at (5, 0) {};
		\node [style=none] (58) at (-5, 0) {};
		\node [style=Z dot] (59) at (0, 3.25) {};
		\node [style=none, rotate=90] (60) at (0, 2.5) {...};
		\node [style=none] (61) at (4.5, -0.5) {};
		\node [style=none] (62) at (6.5, -0.5) {};
		\node [style=none] (63) at (6.5, -2.5) {};
		\node [style=none] (64) at (-6.5, -0.5) {};
		\node [style=none] (65) at (-4.5, -0.5) {};
		\node [style=none] (66) at (-6.5, -2.5) {};
		\node [style=none] (69) at (1.5, -1.25) {};
		\node [style=none] (70) at (6, -1.25) {};
		\node [style=X dot] (71) at (1, -0.5) {};
		\node [style=X dot] (72) at (-1, -2.5) {};
		\node [style=none] (73) at (3.5, -1.75) {$j$};
	\end{pgfonlayer}
	\begin{pgfonlayer}{edgelayer}
		\draw [style=simple](2.center) to (3.center);
		\draw [style=simple](3.center) to (4.center);
		\draw [style=simple](2.center) to (7.center);
		\draw [style=simple] (9.center) to (10.center);
		\draw [style=simple] (10.center) to (11.center);
		\draw [style=simple] (11.center) to (12.center);
		\draw [style=simple] (12.center) to (13.center);
		\draw [style=simple] (13.center) to (14.center);
		\draw [style=simple](17.center) to (19.center);
		\draw [style=simple] (4.center) to (19.center);
		\draw [style=boldedge] (20.center) to (18.center);
		\draw [style=boldedge] (5.center) to (6.center);
		\draw [style=simple](25.center) to (26.center);
		\draw [style=simple](26.center) to (27.center);
		\draw [style=simple] (27.center) to (28.center);
		\draw [style=simple] (28.center) to (29.center);
		\draw [style=simple] (30.center) to (25.center);
		\draw [style=simple] (34.center) to (35.center);
		\draw [style=simple] (35.center) to (36.center);
		\draw [style=simple] (36.center) to (37.center);
		\draw [style=simple] (37.center) to (38.center);
		\draw [style=simple] (38.center) to (39.center);
		\draw [style=simple](45.center) to (46.center);
		\draw [style=simple](46.center) to (47.center);
		\draw [style=simple](45.center) to (48.center);
		\draw [style=simple](50.center) to (51.center);
		\draw [style=simple] (47.center) to (51.center);
		\draw [style=simple](53.center) to (54.center);
		\draw [style=simple](54.center) to (55.center);
		\draw [style=simple] (55.center) to (56.center);
		\draw [style=simple] (56.center) to (57.center);
		\draw [style=simple] (58.center) to (53.center);
		\draw [style=boldedge] (40.center) to (63.center);
		\draw [style=boldedge] (63.center) to (62.center);
		\draw [style=boldedge] (61.center) to (62.center);
		\draw [style=boldedge] (64.center) to (65.center);
		\draw [style=boldedge] (64.center) to (66.center);
		\draw [style=boldedge] (66.center) to (32.center);
		\draw [style=brace edge] (70.center) to (69.center);
	\end{pgfonlayer}
\end{tikzpicture}

%% file: figures/rmpu4.tikz
\begin{tikzpicture}
	\begin{pgfonlayer}{nodelayer}
		\node [style=none] (0) at (-0.25, 0.25) {};
		\node [style=box] (1) at (0, 0.5) {};
		\node [style=none] (2) at (-0.25, 0.75) {};
		\node [style=none] (3) at (-1.75, 0.75) {};
		\node [style=none] (4) at (-1.75, 1.5) {};
		\node [style=none] (5) at (1.75, 1.5) {};
		\node [style=none] (6) at (1.75, 0.75) {};
		\node [style=none] (7) at (0.25, 0.75) {};
		\node [style=none] (8) at (0.25, 0.25) {};
		\node [style=Z dot] (9) at (-1.25, 1.5) {};
		\node [style=X dot] (10) at (1.25, 0.25) {};
		\node [style=none] (14) at (-0.25, -1.75) {};
		\node [style=box] (15) at (0, -1.5) {};
		\node [style=none] (16) at (-0.25, -1.25) {};
		\node [style=none] (17) at (-1, -1.25) {};
		\node [style=none] (18) at (-1, -0.5) {};
		\node [style=none] (19) at (1, -0.5) {};
		\node [style=none] (20) at (1, -1.25) {};
		\node [style=none] (21) at (0.25, -1.25) {};
		\node [style=none] (22) at (0.25, -1.75) {};
		\node [style=Z dot] (23) at (0, -0.5) {};
		\node [style=X dot] (24) at (-1.25, -1.75) {};
		\node [style=none] (25) at (-1.25, 0.25) {};
		\node [style=none] (26) at (-1.25, -1.75) {};
		\node [style=Z dot] (27) at (1.25, 1.5) {};
		\node [style=Z dot] (28) at (-0.75, 1.5) {};
		\node [style=Z dot] (29) at (0.75, 1.5) {};
		\node [style=none] (30) at (-0.5, 1.5) {};
		\node [style=none] (31) at (0.5, 1.5) {};
		\node [style=none] (32) at (0, 1.5) {\dots};
		\node [style=none] (33) at (1.75, 0.25) {};
		\node [style=none] (34) at (1.75, -1.75) {};
		\node [style=none] (35) at (-1.75, 0.25) {};
		\node [style=none] (36) at (-1.75, -1.75) {};
		\node [style=none] (37) at (-1.5, 2) {};
		\node [style=none] (38) at (1.5, 2) {};
		\node [style=none] (39) at (0, 2.5) {$2j +1$};
	\end{pgfonlayer}
	\begin{pgfonlayer}{edgelayer}
		\draw [style=simple] (2.center) to (3.center);
		\draw [style=simple] (3.center) to (4.center);
		\draw [style=simple] (5.center) to (6.center);
		\draw [style=simple] (6.center) to (7.center);
		\draw [style=simple] (16.center) to (17.center);
		\draw [style=simple] (17.center) to (18.center);
		\draw [style=simple] (18.center) to (19.center);
		\draw [style=simple] (19.center) to (20.center);
		\draw [style=simple] (20.center) to (21.center);
		\draw [style=boldedge] (8.center) to (10);
		\draw [style=boldedge] (22.center) to (24);
		\draw [style=boldedge] (14.center) to (26.center);
		\draw [style=boldedge] (25.center) to (0.center);
		\draw [style=simple] (4.center) to (30.center);
		\draw [style=simple] (31.center) to (5.center);
		\draw [style=boldedge] (35.center) to (25.center);
		\draw [style=boldedge] (35.center) to (36.center);
		\draw [style=boldedge] (36.center) to (26.center);
		\draw [style=boldedge] (34.center) to (33.center);
		\draw [style=boldedge] (33.center) to (10);
		\draw [style=boldedge] (34.center) to (24);
		\draw [style=brace edge] (37.center) to (38.center);
	\end{pgfonlayer}
\end{tikzpicture}

%% file: figures/rmpu5.tikz
\begin{tikzpicture}
	\begin{pgfonlayer}{nodelayer}
		\node [style=none] (0) at (-1, -0.5) {};
		\node [style=none] (1) at (-0.25, -0.5) {};
		\node [style=box] (2) at (0, -0.25) {};
		\node [style=none] (3) at (-0.25, 0) {};
		\node [style=none] (4) at (-1, 0) {};
		\node [style=none] (5) at (-1, 0.75) {};
		\node [style=none] (6) at (1, 0.75) {};
		\node [style=none] (7) at (1, 0) {};
		\node [style=none] (8) at (0.25, 0) {};
		\node [style=none] (9) at (1.5, -0.5) {};
		\node [style=none] (10) at (0.25, -0.5) {};
		\node [style=Z dot] (11) at (0, 0.75) {};
		\node [style=X dot] (12) at (1, -0.5) {};
		\node [style=none] (13) at (-1.5, -0.75) {$\pi$};
		\node [style=none] (14) at (2, -0.75) {$\mu$};
	\end{pgfonlayer}
	\begin{pgfonlayer}{edgelayer}
		\draw [style=simple] (3.center) to (4.center);
		\draw [style=simple] (4.center) to (5.center);
		\draw [style=simple] (5.center) to (6.center);
		\draw [style=simple] (6.center) to (7.center);
		\draw [style=simple] (7.center) to (8.center);
		\draw [style=boldedge] (10.center) to (9.center);
		\draw [style=boldedge] (0.center) to (1.center);
	\end{pgfonlayer}
\end{tikzpicture}

%% file: figures/rmpu6.tikz
\begin{tikzpicture}
	\begin{pgfonlayer}{nodelayer}
		\node [style=box] (2) at (-2.25, 0) {};
		\node [style=none] (4) at (-3.75, 0) {\dots};
		\node [style=none] (5) at (-3, 0.25) {};
		\node [style=none] (6) at (-3, 2) {};
		\node [style=none] (7) at (3, 2) {};
		\node [style=none] (9) at (-0.25, -0.25) {};
		\node [style=none] (10) at (-3.25, -0.25) {};
		\node [style=none] (11) at (-2.5, 0.25) {};
		\node [style=box] (12) at (0, 0) {};
		\node [style=none] (13) at (-0.25, 0.25) {};
		\node [style=none] (14) at (-1, 0.25) {};
		\node [style=none] (15) at (-1, 1) {};
		\node [style=none] (16) at (1, 1) {};
		\node [style=none] (17) at (1, 0.25) {};
		\node [style=none] (18) at (0.25, 0.25) {};
		\node [style=box] (19) at (2.25, 0) {};
		\node [style=none] (20) at (4, 0) {\dots};
		\node [style=none] (21) at (2.5, 0.25) {};
		\node [style=none] (22) at (3.25, -0.25) {};
		\node [style=none] (24) at (3, 0.25) {};
		\node [style=none] (26) at (0.25, -0.25) {};
		\node [style=Z dot] (27) at (0, 1) {};
		\node [style=X dot] (28) at (1, -0.25) {};
		\node [style=X dot] (29) at (-1.5, -0.25) {};
		\node [style=Z dot] (30) at (0, 2) {};
		\node [style=none] (32) at (-1.5, 0.25) {};
		\node [style=none] (33) at (-1.5, 1.5) {};
		\node [style=none] (34) at (1.5, 1.5) {};
		\node [style=none] (35) at (1.5, 0.25) {};
		\node [style=none] (36) at (2, 0.25) {};
		\node [style=none] (37) at (-2, 0.25) {};
		\node [style=Z dot] (38) at (0, 1.5) {};
		\node [style=none] (39) at (-0.25, -2.25) {};
		\node [style=box] (40) at (0, -2) {};
		\node [style=none] (41) at (-0.25, -1.75) {};
		\node [style=none] (42) at (-1, -1.75) {};
		\node [style=none] (43) at (-1, -1) {};
		\node [style=none] (44) at (1, -1) {};
		\node [style=none] (45) at (1, -1.75) {};
		\node [style=none] (46) at (0.25, -1.75) {};
		\node [style=none] (47) at (0.25, -2.25) {};
		\node [style=Z dot] (48) at (0, -1) {};
		\node [style=X dot] (49) at (1, -2.25) {};
		\node [style=X dot] (50) at (-1.25, -2.25) {};
		\node [style=box] (51) at (-5.25, 0) {};
		\node [style=none] (52) at (-6.5, 0.25) {};
		\node [style=none] (53) at (-6.5, 4) {};
		\node [style=none] (54) at (6.5, 4) {};
		\node [style=none] (55) at (-5.5, 0.25) {};
		\node [style=box] (56) at (5.25, 0) {};
		\node [style=none] (57) at (5.5, 0.25) {};
		\node [style=none] (58) at (6.5, 0.25) {};
		\node [style=Z dot] (59) at (0, 4) {};
		\node [style=none] (60) at (-4.5, 0.25) {};
		\node [style=none] (61) at (-4.5, 3.5) {};
		\node [style=none] (62) at (4.5, 3.5) {};
		\node [style=none] (63) at (4.5, 0.25) {};
		\node [style=none] (64) at (5, 0.25) {};
		\node [style=none] (65) at (-5, 0.25) {};
		\node [style=Z dot] (66) at (0, 3.5) {};
		\node [style=none, rotate=90] (67) at (0, 2.75) {...};
		\node [style=none] (68) at (4.5, -0.25) {};
		\node [style=none] (69) at (6.5, -0.25) {};
		\node [style=none] (70) at (6.5, -2.25) {};
		\node [style=none] (71) at (-6.5, -0.25) {};
		\node [style=none] (72) at (-4.5, -0.25) {};
		\node [style=none] (73) at (-6.5, -2.25) {};
		\node [style=X dot] (74) at (3, -0.25) {};
		\node [style=X dot] (75) at (-3, -0.25) {};
		\node [style=none] (76) at (1.5, -1) {};
		\node [style=none] (77) at (6, -1) {};
		\node [style=none] (78) at (3.75, -1.5) {n-2};
		\node [style=none] (79) at (-1.25, 1.25) {};
		\node [style=none] (80) at (-1.25, -0.75) {};
		\node [style=none] (81) at (1.25, 1.25) {};
		\node [style=none] (82) at (1.25, -0.75) {};
	\end{pgfonlayer}
	\begin{pgfonlayer}{edgelayer}
		\draw [style=simple](5.center) to (6.center);
		\draw [style=simple](6.center) to (7.center);
		\draw [style=simple](5.center) to (11.center);
		\draw [style=simple] (13.center) to (14.center);
		\draw [style=simple] (14.center) to (15.center);
		\draw [style=simple] (15.center) to (16.center);
		\draw [style=simple] (16.center) to (17.center);
		\draw [style=simple] (17.center) to (18.center);
		\draw [style=simple](21.center) to (24.center);
		\draw [style=simple] (7.center) to (24.center);
		\draw [style=boldedge] (26.center) to (22.center);
		\draw [style=boldedge] (9.center) to (10.center);
		\draw [style=simple](32.center) to (33.center);
		\draw [style=simple](33.center) to (34.center);
		\draw [style=simple] (34.center) to (35.center);
		\draw [style=simple] (35.center) to (36.center);
		\draw [style=simple] (37.center) to (32.center);
		\draw [style=simple] (41.center) to (42.center);
		\draw [style=simple] (42.center) to (43.center);
		\draw [style=simple] (43.center) to (44.center);
		\draw [style=simple] (44.center) to (45.center);
		\draw [style=simple] (45.center) to (46.center);
		\draw [style=simple](52.center) to (53.center);
		\draw [style=simple](53.center) to (54.center);
		\draw [style=simple](52.center) to (55.center);
		\draw [style=simple](57.center) to (58.center);
		\draw [style=simple] (54.center) to (58.center);
		\draw [style=simple](60.center) to (61.center);
		\draw [style=simple](61.center) to (62.center);
		\draw [style=simple] (62.center) to (63.center);
		\draw [style=simple] (63.center) to (64.center);
		\draw [style=simple] (65.center) to (60.center);
		\draw [style=boldedge] (47.center) to (70.center);
		\draw [style=boldedge] (70.center) to (69.center);
		\draw [style=boldedge] (68.center) to (69.center);
		\draw [style=boldedge] (71.center) to (72.center);
		\draw [style=boldedge] (71.center) to (73.center);
		\draw [style=boldedge] (73.center) to (39.center);
		\draw [style=brace edge] (77.center) to (76.center);
		\draw [style=not edge] (80.center) to (79.center);
		\draw [style=not edge] (80.center) to (82.center);
		\draw [style=not edge] (82.center) to (81.center);
		\draw [style=not edge] (81.center) to (79.center);
	\end{pgfonlayer}
\end{tikzpicture}

%% file: figures/rmpu7.tikz
\begin{tikzpicture}
	\begin{pgfonlayer}{nodelayer}
		\node [style=box] (0) at (-2.25, 0) {};
		\node [style=none] (1) at (-3.75, 0) {\dots};
		\node [style=none] (2) at (-3, 0.25) {};
		\node [style=none] (3) at (-3, 2) {};
		\node [style=none] (4) at (3, 2) {};
		\node [style=none] (6) at (-3.25, -0.25) {};
		\node [style=none] (7) at (-2.5, 0.25) {};
		\node [style=box] (15) at (2.25, 0) {};
		\node [style=none] (16) at (4, 0) {\dots};
		\node [style=none] (17) at (2.5, 0.25) {};
		\node [style=none] (18) at (3.25, -0.25) {};
		\node [style=none] (19) at (3, 0.25) {};
		\node [style=X dot] (23) at (-1.5, -0.25) {};
		\node [style=Z dot] (24) at (0, 2) {};
		\node [style=none] (25) at (-1.5, 0.25) {};
		\node [style=none] (26) at (-1.5, 1.5) {};
		\node [style=none] (27) at (1.5, 1.5) {};
		\node [style=none] (28) at (1.5, 0.25) {};
		\node [style=none] (29) at (2, 0.25) {};
		\node [style=none] (30) at (-2, 0.25) {};
		\node [style=Z dot] (31) at (0, 1.5) {};
		\node [style=none] (32) at (-0.25, -2.25) {};
		\node [style=box] (33) at (0, -2) {};
		\node [style=none] (34) at (-0.25, -1.75) {};
		\node [style=none] (35) at (-1, -1.75) {};
		\node [style=none] (36) at (-1, -1) {};
		\node [style=none] (37) at (1, -1) {};
		\node [style=none] (38) at (1, -1.75) {};
		\node [style=none] (39) at (0.25, -1.75) {};
		\node [style=none] (40) at (0.25, -2.25) {};
		\node [style=Z dot] (41) at (0, -1) {};
		\node [style=X dot] (42) at (1, -2.25) {};
		\node [style=X dot] (43) at (-1.25, -2.25) {};
		\node [style=box] (44) at (-5.25, 0) {};
		\node [style=none] (45) at (-6.5, 0.25) {};
		\node [style=none] (46) at (-6.5, 4) {};
		\node [style=none] (47) at (6.5, 4) {};
		\node [style=none] (48) at (-5.5, 0.25) {};
		\node [style=box] (49) at (5.25, 0) {};
		\node [style=none] (50) at (5.5, 0.25) {};
		\node [style=none] (51) at (6.5, 0.25) {};
		\node [style=Z dot] (52) at (0, 4) {};
		\node [style=none] (53) at (-4.5, 0.25) {};
		\node [style=none] (54) at (-4.5, 3.5) {};
		\node [style=none] (55) at (4.5, 3.5) {};
		\node [style=none] (56) at (4.5, 0.25) {};
		\node [style=none] (57) at (5, 0.25) {};
		\node [style=none] (58) at (-5, 0.25) {};
		\node [style=Z dot] (59) at (0, 3.5) {};
		\node [style=none, rotate=90] (60) at (0, 2.75) {...};
		\node [style=none] (61) at (4.5, -0.25) {};
		\node [style=none] (62) at (6.5, -0.25) {};
		\node [style=none] (63) at (6.5, -2.25) {};
		\node [style=none] (64) at (-6.5, -0.25) {};
		\node [style=none] (65) at (-4.5, -0.25) {};
		\node [style=none] (66) at (-6.5, -2.25) {};
		\node [style=X dot] (67) at (3, -0.25) {};
		\node [style=X dot] (68) at (-3, -0.25) {};
		\node [style=none] (69) at (1.5, -1) {};
		\node [style=none] (70) at (6, -1) {};
		\node [style=none] (71) at (3.75, -1.5) {n-2};
	\end{pgfonlayer}
	\begin{pgfonlayer}{edgelayer}
		\draw [style=simple](2.center) to (3.center);
		\draw [style=simple](3.center) to (4.center);
		\draw [style=simple](2.center) to (7.center);
		\draw [style=simple](17.center) to (19.center);
		\draw [style=simple] (4.center) to (19.center);
		\draw [style=simple](25.center) to (26.center);
		\draw [style=simple](26.center) to (27.center);
		\draw [style=simple] (27.center) to (28.center);
		\draw [style=simple] (28.center) to (29.center);
		\draw [style=simple] (30.center) to (25.center);
		\draw [style=simple] (34.center) to (35.center);
		\draw [style=simple] (35.center) to (36.center);
		\draw [style=simple] (36.center) to (37.center);
		\draw [style=simple] (37.center) to (38.center);
		\draw [style=simple] (38.center) to (39.center);
		\draw [style=simple](45.center) to (46.center);
		\draw [style=simple](46.center) to (47.center);
		\draw [style=simple](45.center) to (48.center);
		\draw [style=simple](50.center) to (51.center);
		\draw [style=simple] (47.center) to (51.center);
		\draw [style=simple](53.center) to (54.center);
		\draw [style=simple](54.center) to (55.center);
		\draw [style=simple] (55.center) to (56.center);
		\draw [style=simple] (56.center) to (57.center);
		\draw [style=simple] (58.center) to (53.center);
		\draw [style=boldedge] (40.center) to (63.center);
		\draw [style=boldedge] (63.center) to (62.center);
		\draw [style=boldedge] (61.center) to (62.center);
		\draw [style=boldedge] (64.center) to (65.center);
		\draw [style=boldedge] (64.center) to (66.center);
		\draw [style=boldedge] (66.center) to (32.center);
		\draw [style=brace edge] (70.center) to (69.center);
		\draw [style=boldedge] (67) to (68);
		\draw [style=boldedge] (18.center) to (67);
	\end{pgfonlayer}
\end{tikzpicture}

%% file: figures/rmpu8.tikz
\begin{tikzpicture}
	\begin{pgfonlayer}{nodelayer}
		\node [style=box] (2) at (-2.25, 0) {};
		\node [style=none] (4) at (-3.75, 0) {\dots};
		\node [style=none] (5) at (-3, 0.25) {};
		\node [style=none] (6) at (-3, 2) {};
		\node [style=none] (7) at (3, 2) {};
		\node [style=none] (9) at (-0.25, -0.25) {};
		\node [style=none] (10) at (-3.25, -0.25) {};
		\node [style=none] (11) at (-2.5, 0.25) {};
		\node [style=box] (12) at (0, 0) {};
		\node [style=none] (13) at (-0.25, 0.25) {};
		\node [style=none] (14) at (-1, 0.25) {};
		\node [style=none] (15) at (-1, 1) {};
		\node [style=none] (16) at (1, 1) {};
		\node [style=none] (17) at (1, 0.25) {};
		\node [style=none] (18) at (0.25, 0.25) {};
		\node [style=box] (19) at (2.25, 0) {};
		\node [style=none] (20) at (4, 0) {\dots};
		\node [style=none] (21) at (2.5, 0.25) {};
		\node [style=none] (22) at (3.25, -0.25) {};
		\node [style=none] (24) at (3, 0.25) {};
		\node [style=none] (26) at (0.25, -0.25) {};
		\node [style=Z dot] (27) at (0, 1) {};
		\node [style=X dot] (28) at (1, -0.25) {};
		\node [style=X dot] (29) at (-1.5, -0.25) {};
		\node [style=Z dot] (30) at (0, 2) {};
		\node [style=none] (32) at (-1.5, 0.25) {};
		\node [style=none] (33) at (-1.5, 1.5) {};
		\node [style=none] (34) at (1.5, 1.5) {};
		\node [style=none] (35) at (1.5, 0.25) {};
		\node [style=none] (36) at (2, 0.25) {};
		\node [style=none] (37) at (-2, 0.25) {};
		\node [style=Z dot] (38) at (0, 1.5) {};
		\node [style=none] (39) at (-0.25, -2.25) {};
		\node [style=box] (40) at (0, -2) {};
		\node [style=none] (41) at (-0.25, -1.75) {};
		\node [style=none] (42) at (-1, -1.75) {};
		\node [style=none] (43) at (-1, -1) {};
		\node [style=none] (44) at (1, -1) {};
		\node [style=none] (45) at (1, -1.75) {};
		\node [style=none] (46) at (0.25, -1.75) {};
		\node [style=none] (47) at (0.25, -2.25) {};
		\node [style=Z dot] (48) at (0, -1) {};
		\node [style=X dot] (49) at (1, -2.25) {};
		\node [style=X dot] (50) at (-1.25, -2.25) {};
		\node [style=box] (51) at (-5.25, 0) {};
		\node [style=none] (52) at (-6.5, 0.25) {};
		\node [style=none] (53) at (-6.5, 4) {};
		\node [style=none] (54) at (6.5, 4) {};
		\node [style=none] (55) at (-5.5, 0.25) {};
		\node [style=box] (56) at (5.25, 0) {};
		\node [style=none] (57) at (5.5, 0.25) {};
		\node [style=none] (58) at (6.5, 0.25) {};
		\node [style=Z dot] (59) at (0, 4) {};
		\node [style=none] (60) at (-4.5, 0.25) {};
		\node [style=none] (61) at (-4.5, 3.5) {};
		\node [style=none] (62) at (4.5, 3.5) {};
		\node [style=none] (63) at (4.5, 0.25) {};
		\node [style=none] (64) at (5, 0.25) {};
		\node [style=none] (65) at (-5, 0.25) {};
		\node [style=Z dot] (66) at (0, 3.5) {};
		\node [style=none, rotate=90] (67) at (0, 2.75) {...};
		\node [style=none] (68) at (4.5, -0.25) {};
		\node [style=none] (69) at (6.5, -0.25) {};
		\node [style=none] (70) at (6.5, -2.25) {};
		\node [style=none] (71) at (-6.5, -0.25) {};
		\node [style=none] (72) at (-4.5, -0.25) {};
		\node [style=none] (73) at (-6.5, -2.25) {};
		\node [style=X dot] (74) at (3, -0.25) {};
		\node [style=X dot] (75) at (-3, -0.25) {};
		\node [style=none] (76) at (1.5, -1) {};
		\node [style=none] (77) at (6, -1) {};
		\node [style=none] (78) at (3.75, -1.5) {n-3};
	\end{pgfonlayer}
	\begin{pgfonlayer}{edgelayer}
		\draw [style=simple](5.center) to (6.center);
		\draw [style=simple](6.center) to (7.center);
		\draw [style=simple](5.center) to (11.center);
		\draw [style=simple] (13.center) to (14.center);
		\draw [style=simple] (14.center) to (15.center);
		\draw [style=simple] (15.center) to (16.center);
		\draw [style=simple] (16.center) to (17.center);
		\draw [style=simple] (17.center) to (18.center);
		\draw [style=simple](21.center) to (24.center);
		\draw [style=simple] (7.center) to (24.center);
		\draw [style=boldedge] (26.center) to (22.center);
		\draw [style=boldedge] (9.center) to (10.center);
		\draw [style=simple](32.center) to (33.center);
		\draw [style=simple](33.center) to (34.center);
		\draw [style=simple] (34.center) to (35.center);
		\draw [style=simple] (35.center) to (36.center);
		\draw [style=simple] (37.center) to (32.center);
		\draw [style=simple] (41.center) to (42.center);
		\draw [style=simple] (42.center) to (43.center);
		\draw [style=simple] (43.center) to (44.center);
		\draw [style=simple] (44.center) to (45.center);
		\draw [style=simple] (45.center) to (46.center);
		\draw [style=simple](52.center) to (53.center);
		\draw [style=simple](53.center) to (54.center);
		\draw [style=simple](52.center) to (55.center);
		\draw [style=simple](57.center) to (58.center);
		\draw [style=simple] (54.center) to (58.center);
		\draw [style=simple](60.center) to (61.center);
		\draw [style=simple](61.center) to (62.center);
		\draw [style=simple] (62.center) to (63.center);
		\draw [style=simple] (63.center) to (64.center);
		\draw [style=simple] (65.center) to (60.center);
		\draw [style=boldedge] (47.center) to (70.center);
		\draw [style=boldedge] (70.center) to (69.center);
		\draw [style=boldedge] (68.center) to (69.center);
		\draw [style=boldedge] (71.center) to (72.center);
		\draw [style=boldedge] (71.center) to (73.center);
		\draw [style=boldedge] (73.center) to (39.center);
		\draw [style=brace edge] (77.center) to (76.center);
	\end{pgfonlayer}
\end{tikzpicture}

%% file: figures/rmpu9.tikz
\begin{tikzpicture}
	\begin{pgfonlayer}{nodelayer}
		\node [style=box] (0) at (-2.25, -0.25) {$\mathbb{I}$};
		\node [style=none] (1) at (-3.75, -0.25) {\dots};
		\node [style=none] (2) at (-3, 0) {};
		\node [style=none] (3) at (-3, 1.75) {};
		\node [style=none] (4) at (3, 1.75) {};
		\node [style=none] (5) at (-0.25, -0.5) {};
		\node [style=none] (6) at (-3.25, -0.5) {};
		\node [style=none] (7) at (-2.5, 0) {};
		\node [style=box] (8) at (0, -0.25) {$\delta$};
		\node [style=none] (9) at (-0.25, 0) {};
		\node [style=none] (10) at (-1, 0) {};
		\node [style=none] (11) at (-1, 0.75) {};
		\node [style=none] (12) at (1, 0.75) {};
		\node [style=none] (13) at (1, 0) {};
		\node [style=none] (14) at (0.25, 0) {};
		\node [style=box] (15) at (2.25, -0.25) {$\mathbb{I}$};
		\node [style=none] (16) at (4, -0.25) {\dots};
		\node [style=none] (17) at (2.5, 0) {};
		\node [style=none] (18) at (3.25, -0.5) {};
		\node [style=none] (19) at (3, 0) {};
		\node [style=none] (20) at (0.25, -0.5) {};
		\node [style=Z dot] (21) at (0, 0.75) {};
		\node [style=Z dot] (24) at (0, 1.75) {};
		\node [style=none] (25) at (-1.5, 0) {};
		\node [style=none] (26) at (-1.5, 1.25) {};
		\node [style=none] (27) at (1.5, 1.25) {};
		\node [style=none] (28) at (1.5, 0) {};
		\node [style=none] (29) at (2, 0) {};
		\node [style=none] (30) at (-2, 0) {};
		\node [style=Z dot] (31) at (0, 1.25) {};
		\node [style=none] (32) at (-0.25, -2.5) {};
		\node [style=box] (33) at (0, -2.25) {$\delta$};
		\node [style=none] (34) at (-0.25, -2) {};
		\node [style=none] (35) at (-1, -2) {};
		\node [style=none] (36) at (-1, -1.25) {};
		\node [style=none] (37) at (1, -1.25) {};
		\node [style=none] (38) at (1, -2) {};
		\node [style=none] (39) at (0.25, -2) {};
		\node [style=none] (40) at (0.25, -2.5) {};
		\node [style=Z dot] (41) at (0, -1.25) {};
		\node [style=box] (44) at (-5.25, -0.25) {$\mathbb{I}$};
		\node [style=none] (45) at (-6.5, 0) {};
		\node [style=none] (46) at (-6.5, 3.75) {};
		\node [style=none] (47) at (6.5, 3.75) {};
		\node [style=none] (48) at (-5.5, 0) {};
		\node [style=box] (49) at (5.25, -0.25) {$\mathbb{I}$};
		\node [style=none] (50) at (5.5, 0) {};
		\node [style=none] (51) at (6.5, 0) {};
		\node [style=Z dot] (52) at (0, 3.75) {};
		\node [style=none] (53) at (-4.5, 0) {};
		\node [style=none] (54) at (-4.5, 3.25) {};
		\node [style=none] (55) at (4.5, 3.25) {};
		\node [style=none] (56) at (4.5, 0) {};
		\node [style=none] (57) at (5, 0) {};
		\node [style=none] (58) at (-5, 0) {};
		\node [style=Z dot] (59) at (0, 3.25) {};
		\node [style=none, rotate=90] (60) at (0, 2.5) {...};
		\node [style=none] (61) at (4.5, -0.5) {};
		\node [style=none] (62) at (6.5, -0.5) {};
		\node [style=none] (63) at (6.5, -2.5) {};
		\node [style=none] (64) at (-6.5, -0.5) {};
		\node [style=none] (65) at (-4.5, -0.5) {};
		\node [style=none] (66) at (-6.5, -2.5) {};
		\node [style=none] (69) at (1.5, -1.25) {};
		\node [style=none] (70) at (6, -1.25) {};
		\node [style=none] (71) at (1.5, -1.25) {};
		\node [style=none] (72) at (6, -1.25) {};
		\node [style=none] (73) at (3.75, -1.75) {n-2};
	\end{pgfonlayer}
	\begin{pgfonlayer}{edgelayer}
		\draw [style=simple](2.center) to (3.center);
		\draw [style=simple](3.center) to (4.center);
		\draw [style=simple](2.center) to (7.center);
		\draw [style=simple] (9.center) to (10.center);
		\draw [style=simple] (10.center) to (11.center);
		\draw [style=simple] (11.center) to (12.center);
		\draw [style=simple] (12.center) to (13.center);
		\draw [style=simple] (13.center) to (14.center);
		\draw [style=simple](17.center) to (19.center);
		\draw [style=simple] (4.center) to (19.center);
		\draw [style=boldedge] (20.center) to (18.center);
		\draw [style=boldedge] (5.center) to (6.center);
		\draw [style=simple](25.center) to (26.center);
		\draw [style=simple](26.center) to (27.center);
		\draw [style=simple] (27.center) to (28.center);
		\draw [style=simple] (28.center) to (29.center);
		\draw [style=simple] (30.center) to (25.center);
		\draw [style=simple] (34.center) to (35.center);
		\draw [style=simple] (35.center) to (36.center);
		\draw [style=simple] (36.center) to (37.center);
		\draw [style=simple] (37.center) to (38.center);
		\draw [style=simple] (38.center) to (39.center);
		\draw [style=simple](45.center) to (46.center);
		\draw [style=simple](46.center) to (47.center);
		\draw [style=simple](45.center) to (48.center);
		\draw [style=simple](50.center) to (51.center);
		\draw [style=simple] (47.center) to (51.center);
		\draw [style=simple](53.center) to (54.center);
		\draw [style=simple](54.center) to (55.center);
		\draw [style=simple] (55.center) to (56.center);
		\draw [style=simple] (56.center) to (57.center);
		\draw [style=simple] (58.center) to (53.center);
		\draw [style=boldedge] (40.center) to (63.center);
		\draw [style=boldedge] (63.center) to (62.center);
		\draw [style=boldedge] (61.center) to (62.center);
		\draw [style=boldedge] (64.center) to (65.center);
		\draw [style=boldedge] (64.center) to (66.center);
		\draw [style=boldedge] (66.center) to (32.center);
		\draw [style=brace edge] (72.center) to (71.center);
	\end{pgfonlayer}
\end{tikzpicture}

%% file: figures/rmpu10.tikz
\begin{tikzpicture}
	\begin{pgfonlayer}{nodelayer}
		\node [style=box] (0) at (-2.25, -0.25) {};
		\node [style=none] (1) at (-3.75, -0.25) {\dots};
		\node [style=none] (2) at (-3, 0) {};
		\node [style=none] (3) at (-3, 1.75) {};
		\node [style=none] (4) at (3, 1.75) {};
		\node [style=none] (5) at (-0.25, -0.5) {};
		\node [style=none] (6) at (-3.25, -0.5) {};
		\node [style=none] (7) at (-2.5, 0) {};
		\node [style=box] (8) at (0, -0.25) {$\delta$};
		\node [style=none] (9) at (-0.25, 0) {};
		\node [style=none] (10) at (-1, 0) {};
		\node [style=none] (11) at (-1, 0.75) {};
		\node [style=none] (12) at (1, 0.75) {};
		\node [style=none] (13) at (1, 0) {};
		\node [style=none] (14) at (0.25, 0) {};
		\node [style=box] (15) at (2.25, -0.25) {};
		\node [style=none] (16) at (4, -0.25) {\dots};
		\node [style=none] (17) at (2.5, 0) {};
		\node [style=none] (18) at (3.25, -0.5) {};
		\node [style=none] (19) at (3, 0) {};
		\node [style=none] (20) at (0.25, -0.5) {};
		\node [style=Z dot] (21) at (0, 0.75) {};
		\node [style=Z dot] (24) at (0, 1.75) {};
		\node [style=none] (25) at (-1.5, 0) {};
		\node [style=none] (26) at (-1.5, 1.25) {};
		\node [style=none] (27) at (1.5, 1.25) {};
		\node [style=none] (28) at (1.5, 0) {};
		\node [style=none] (29) at (2, 0) {};
		\node [style=none] (30) at (-2, 0) {};
		\node [style=Z dot] (31) at (0, 1.25) {};
		\node [style=none] (32) at (-0.25, -2.5) {};
		\node [style=box] (33) at (0, -2.25) {$\mathbb{I}$};
		\node [style=none] (34) at (-0.25, -2) {};
		\node [style=none] (35) at (-1, -2) {};
		\node [style=none] (36) at (-1, -1.25) {};
		\node [style=none] (37) at (1, -1.25) {};
		\node [style=none] (38) at (1, -2) {};
		\node [style=none] (39) at (0.25, -2) {};
		\node [style=none] (40) at (0.25, -2.5) {};
		\node [style=Z dot] (41) at (0, -1.25) {};
		\node [style=box] (44) at (-5.25, -0.25) {};
		\node [style=none] (45) at (-6.5, 0) {};
		\node [style=none] (46) at (-6.5, 3.75) {};
		\node [style=none] (47) at (6.5, 3.75) {};
		\node [style=none] (48) at (-5.5, 0) {};
		\node [style=box] (49) at (5.25, -0.25) {};
		\node [style=none] (50) at (5.5, 0) {};
		\node [style=none] (51) at (6.5, 0) {};
		\node [style=Z dot] (52) at (0, 3.75) {};
		\node [style=none] (53) at (-4.5, 0) {};
		\node [style=none] (54) at (-4.5, 3.25) {};
		\node [style=none] (55) at (4.5, 3.25) {};
		\node [style=none] (56) at (4.5, 0) {};
		\node [style=none] (57) at (5, 0) {};
		\node [style=none] (58) at (-5, 0) {};
		\node [style=Z dot] (59) at (0, 3.25) {};
		\node [style=none, rotate=90] (60) at (0, 2.5) {...};
		\node [style=none] (61) at (4.5, -0.5) {};
		\node [style=none] (62) at (6.5, -0.5) {};
		\node [style=none] (63) at (6.5, -2.5) {};
		\node [style=none] (64) at (-6.5, -0.5) {};
		\node [style=none] (65) at (-4.5, -0.5) {};
		\node [style=none] (66) at (-6.5, -2.5) {};
		\node [style=none] (69) at (1.5, -1.25) {};
		\node [style=none] (70) at (6, -1.25) {};
		\node [style=none] (71) at (1.5, -1.25) {};
		\node [style=none] (72) at (6, -1.25) {};
		\node [style=none] (73) at (3.75, -1.75) {n-2};
		\node [style=X dot] (75) at (-1.25, -0.5) {};
		\node [style=X dot] (77) at (-1, -2.5) {};
		\node [style=X dot] (78) at (3, -0.5) {};
		\node [style=X dot] (79) at (-3, -0.5) {};
	\end{pgfonlayer}
	\begin{pgfonlayer}{edgelayer}
		\draw [style=simple](2.center) to (3.center);
		\draw [style=simple](3.center) to (4.center);
		\draw [style=simple](2.center) to (7.center);
		\draw [style=simple] (9.center) to (10.center);
		\draw [style=simple] (10.center) to (11.center);
		\draw [style=simple] (11.center) to (12.center);
		\draw [style=simple] (12.center) to (13.center);
		\draw [style=simple] (13.center) to (14.center);
		\draw [style=simple](17.center) to (19.center);
		\draw [style=simple] (4.center) to (19.center);
		\draw [style=boldedge] (20.center) to (18.center);
		\draw [style=boldedge] (5.center) to (6.center);
		\draw [style=simple](25.center) to (26.center);
		\draw [style=simple](26.center) to (27.center);
		\draw [style=simple] (27.center) to (28.center);
		\draw [style=simple] (28.center) to (29.center);
		\draw [style=simple] (30.center) to (25.center);
		\draw [style=simple] (34.center) to (35.center);
		\draw [style=simple] (35.center) to (36.center);
		\draw [style=simple] (36.center) to (37.center);
		\draw [style=simple] (37.center) to (38.center);
		\draw [style=simple] (38.center) to (39.center);
		\draw [style=simple](45.center) to (46.center);
		\draw [style=simple](46.center) to (47.center);
		\draw [style=simple](45.center) to (48.center);
		\draw [style=simple](50.center) to (51.center);
		\draw [style=simple] (47.center) to (51.center);
		\draw [style=simple](53.center) to (54.center);
		\draw [style=simple](54.center) to (55.center);
		\draw [style=simple] (55.center) to (56.center);
		\draw [style=simple] (56.center) to (57.center);
		\draw [style=simple] (58.center) to (53.center);
		\draw [style=boldedge] (40.center) to (63.center);
		\draw [style=boldedge] (63.center) to (62.center);
		\draw [style=boldedge] (61.center) to (62.center);
		\draw [style=boldedge] (64.center) to (65.center);
		\draw [style=boldedge] (64.center) to (66.center);
		\draw [style=boldedge] (66.center) to (32.center);
		\draw [style=brace edge] (72.center) to (71.center);
	\end{pgfonlayer}
\end{tikzpicture}

%% file: figures/rmpu11.tikz
\begin{tikzpicture}
	\begin{pgfonlayer}{nodelayer}
		\node [style=box] (0) at (-2.25, -0.25) {};
		\node [style=none] (1) at (-3.75, -0.25) {\dots};
		\node [style=none] (2) at (-3, 0) {};
		\node [style=none] (3) at (-3, 1.75) {};
		\node [style=none] (4) at (3, 1.75) {};
		\node [style=none] (5) at (-0.25, -0.5) {};
		\node [style=none] (6) at (-3.25, -0.5) {};
		\node [style=none] (7) at (-2.5, 0) {};
		\node [style=box] (8) at (0, -0.25) {$\delta$};
		\node [style=none] (9) at (-0.25, 0) {};
		\node [style=none] (10) at (-1, 0) {};
		\node [style=none] (11) at (-1, 0.75) {};
		\node [style=none] (12) at (1, 0.75) {};
		\node [style=none] (13) at (1, 0) {};
		\node [style=none] (14) at (0.25, 0) {};
		\node [style=box] (15) at (2.25, -0.25) {};
		\node [style=none] (16) at (4, -0.25) {\dots};
		\node [style=none] (17) at (2.5, 0) {};
		\node [style=none] (18) at (3.25, -0.5) {};
		\node [style=none] (19) at (3, 0) {};
		\node [style=none] (20) at (0.25, -0.5) {};
		\node [style=Z dot] (21) at (0, 0.75) {};
		\node [style=Z dot] (24) at (0, 1.75) {};
		\node [style=none] (25) at (-1.5, 0) {};
		\node [style=none] (26) at (-1.5, 1.25) {};
		\node [style=none] (27) at (1.5, 1.25) {};
		\node [style=none] (28) at (1.5, 0) {};
		\node [style=none] (29) at (2, 0) {};
		\node [style=none] (30) at (-2, 0) {};
		\node [style=Z dot] (31) at (0, 1.25) {};
		\node [style=none] (32) at (6.5, -2.5) {};
		\node [style=box] (44) at (-5.25, -0.25) {};
		\node [style=none] (45) at (-6.5, 0) {};
		\node [style=none] (46) at (-6.5, 3.75) {};
		\node [style=none] (47) at (6.5, 3.75) {};
		\node [style=none] (48) at (-5.5, 0) {};
		\node [style=box] (49) at (5.25, -0.25) {};
		\node [style=none] (50) at (5.5, 0) {};
		\node [style=none] (51) at (6.5, 0) {};
		\node [style=Z dot] (52) at (6, 0) {};
		\node [style=none] (53) at (-4.5, 0) {};
		\node [style=none] (54) at (-4.5, 3.25) {};
		\node [style=none] (55) at (4.5, 3.25) {};
		\node [style=none] (56) at (4.5, 0) {};
		\node [style=none] (57) at (5, 0) {};
		\node [style=none] (58) at (-5, 0) {};
		\node [style=Z dot] (59) at (0, 3.25) {};
		\node [style=none, rotate=90] (60) at (0, 2.5) {...};
		\node [style=none] (61) at (4.5, -0.5) {};
		\node [style=none] (62) at (6.5, -0.5) {};
		\node [style=none] (63) at (6.5, -2.5) {};
		\node [style=none] (64) at (-6.5, -0.5) {};
		\node [style=none] (65) at (-4.5, -0.5) {};
		\node [style=none] (66) at (-6.5, -2.5) {};
		\node [style=none] (69) at (1.5, -1.25) {};
		\node [style=none] (70) at (6, -1.25) {};
		\node [style=none] (71) at (1.5, -1.25) {};
		\node [style=none] (72) at (6, -1.25) {};
		\node [style=none] (73) at (3.75, -1.75) {n-2};
		\node [style=X dot] (75) at (-1.25, -0.5) {};
		\node [style=X dot] (77) at (6, -0.5) {};
		\node [style=X dot] (78) at (3, -0.5) {};
		\node [style=X dot] (79) at (-3, -0.5) {};
	\end{pgfonlayer}
	\begin{pgfonlayer}{edgelayer}
		\draw [style=simple](2.center) to (3.center);
		\draw [style=simple](3.center) to (4.center);
		\draw [style=simple](2.center) to (7.center);
		\draw [style=simple] (9.center) to (10.center);
		\draw [style=simple] (10.center) to (11.center);
		\draw [style=simple] (11.center) to (12.center);
		\draw [style=simple] (12.center) to (13.center);
		\draw [style=simple] (13.center) to (14.center);
		\draw [style=simple](17.center) to (19.center);
		\draw [style=simple] (4.center) to (19.center);
		\draw [style=boldedge] (20.center) to (18.center);
		\draw [style=boldedge] (5.center) to (6.center);
		\draw [style=simple](25.center) to (26.center);
		\draw [style=simple](26.center) to (27.center);
		\draw [style=simple] (27.center) to (28.center);
		\draw [style=simple] (28.center) to (29.center);
		\draw [style=simple] (30.center) to (25.center);
		\draw [style=simple](45.center) to (46.center);
		\draw [style=simple](46.center) to (47.center);
		\draw [style=simple](45.center) to (48.center);
		\draw [style=simple](50.center) to (51.center);
		\draw [style=simple] (47.center) to (51.center);
		\draw [style=simple](53.center) to (54.center);
		\draw [style=simple](54.center) to (55.center);
		\draw [style=simple] (55.center) to (56.center);
		\draw [style=simple] (56.center) to (57.center);
		\draw [style=simple] (58.center) to (53.center);
		\draw [style=boldedge] (63.center) to (62.center);
		\draw [style=boldedge] (61.center) to (62.center);
		\draw [style=boldedge] (64.center) to (65.center);
		\draw [style=boldedge] (64.center) to (66.center);
		\draw [style=boldedge] (66.center) to (32.center);
		\draw [style=brace edge] (72.center) to (71.center);
	\end{pgfonlayer}
\end{tikzpicture}

%% file: figures/rmpu12.tikz
\begin{tikzpicture}
	\begin{pgfonlayer}{nodelayer}
		\node [style=box] (0) at (-2.25, -0.25) {};
		\node [style=none] (1) at (-3.75, -0.25) {\dots};
		\node [style=none] (2) at (-3, 0) {};
		\node [style=none] (3) at (-3, 1.75) {};
		\node [style=none] (4) at (3, 1.75) {};
		\node [style=none] (5) at (-0.25, -0.5) {};
		\node [style=none] (6) at (-3.25, -0.5) {};
		\node [style=none] (7) at (-2.5, 0) {};
		\node [style=box] (8) at (0, -0.25) {$\delta$};
		\node [style=none] (9) at (-0.25, 0) {};
		\node [style=none] (10) at (-1, 0) {};
		\node [style=none] (11) at (-1, 0.75) {};
		\node [style=none] (12) at (1, 0.75) {};
		\node [style=none] (13) at (1, 0) {};
		\node [style=none] (14) at (0.25, 0) {};
		\node [style=box] (15) at (2.25, -0.25) {};
		\node [style=none] (16) at (4, -0.25) {\dots};
		\node [style=none] (17) at (2.5, 0) {};
		\node [style=none] (18) at (3.25, -0.5) {};
		\node [style=none] (19) at (3, 0) {};
		\node [style=none] (20) at (0.25, -0.5) {};
		\node [style=Z dot] (21) at (0, 0.75) {};
		\node [style=Z dot] (24) at (0, 1.75) {};
		\node [style=none] (25) at (-1.5, 0) {};
		\node [style=none] (26) at (-1.5, 1.25) {};
		\node [style=none] (27) at (1.5, 1.25) {};
		\node [style=none] (28) at (1.5, 0) {};
		\node [style=none] (29) at (2, 0) {};
		\node [style=none] (30) at (-2, 0) {};
		\node [style=Z dot] (31) at (0, 1.25) {};
		\node [style=none] (32) at (-0.25, -2.5) {};
		\node [style=box] (33) at (0, -2.25) {};
		\node [style=none] (34) at (-0.25, -2) {};
		\node [style=none] (35) at (-1, -2) {};
		\node [style=none] (36) at (-1, -1.25) {};
		\node [style=none] (37) at (1, -1.25) {};
		\node [style=none] (38) at (1, -2) {};
		\node [style=none] (39) at (0.25, -2) {};
		\node [style=none] (40) at (0.25, -2.5) {};
		\node [style=Z dot] (41) at (0, -1.25) {};
		\node [style=box] (44) at (-5.25, -0.25) {};
		\node [style=none] (45) at (-6.5, 0) {};
		\node [style=none] (46) at (-6.5, 3.75) {};
		\node [style=none] (47) at (6.5, 3.75) {};
		\node [style=none] (48) at (-5.5, 0) {};
		\node [style=box] (49) at (5.25, -0.25) {};
		\node [style=none] (50) at (5.5, 0) {};
		\node [style=none] (51) at (6.5, 0) {};
		\node [style=Z dot] (52) at (0, 3.75) {};
		\node [style=none] (53) at (-4.5, 0) {};
		\node [style=none] (54) at (-4.5, 3.25) {};
		\node [style=none] (55) at (4.5, 3.25) {};
		\node [style=none] (56) at (4.5, 0) {};
		\node [style=none] (57) at (5, 0) {};
		\node [style=none] (58) at (-5, 0) {};
		\node [style=Z dot] (59) at (0, 3.25) {};
		\node [style=none, rotate=90] (60) at (0, 2.5) {...};
		\node [style=none] (61) at (4.5, -0.5) {};
		\node [style=none] (62) at (6.5, -0.5) {};
		\node [style=none] (63) at (6.5, -2.5) {};
		\node [style=none] (64) at (-6.5, -0.5) {};
		\node [style=none] (65) at (-4.5, -0.5) {};
		\node [style=none] (66) at (-6.5, -2.5) {};
		\node [style=none] (69) at (1.5, -1.25) {};
		\node [style=none] (70) at (6, -1.25) {};
		\node [style=none] (71) at (1.5, -1.25) {};
		\node [style=none] (72) at (6, -1.25) {};
		\node [style=none] (73) at (3.75, -1.75) {n-3};
		\node [style=X dot] (75) at (-1.25, -0.5) {};
		\node [style=X dot] (77) at (-1, -2.5) {};
		\node [style=X dot] (78) at (3, -0.5) {};
		\node [style=X dot] (79) at (-3, -0.5) {};
		\node [style=X dot] (80) at (1, -2.5) {};
	\end{pgfonlayer}
	\begin{pgfonlayer}{edgelayer}
		\draw [style=simple](2.center) to (3.center);
		\draw [style=simple](3.center) to (4.center);
		\draw [style=simple](2.center) to (7.center);
		\draw [style=simple] (9.center) to (10.center);
		\draw [style=simple] (10.center) to (11.center);
		\draw [style=simple] (11.center) to (12.center);
		\draw [style=simple] (12.center) to (13.center);
		\draw [style=simple] (13.center) to (14.center);
		\draw [style=simple](17.center) to (19.center);
		\draw [style=simple] (4.center) to (19.center);
		\draw [style=boldedge] (20.center) to (18.center);
		\draw [style=boldedge] (5.center) to (6.center);
		\draw [style=simple](25.center) to (26.center);
		\draw [style=simple](26.center) to (27.center);
		\draw [style=simple] (27.center) to (28.center);
		\draw [style=simple] (28.center) to (29.center);
		\draw [style=simple] (30.center) to (25.center);
		\draw [style=simple] (34.center) to (35.center);
		\draw [style=simple] (35.center) to (36.center);
		\draw [style=simple] (36.center) to (37.center);
		\draw [style=simple] (37.center) to (38.center);
		\draw [style=simple] (38.center) to (39.center);
		\draw [style=simple](45.center) to (46.center);
		\draw [style=simple](46.center) to (47.center);
		\draw [style=simple](45.center) to (48.center);
		\draw [style=simple](50.center) to (51.center);
		\draw [style=simple] (47.center) to (51.center);
		\draw [style=simple](53.center) to (54.center);
		\draw [style=simple](54.center) to (55.center);
		\draw [style=simple] (55.center) to (56.center);
		\draw [style=simple] (56.center) to (57.center);
		\draw [style=simple] (58.center) to (53.center);
		\draw [style=boldedge] (40.center) to (63.center);
		\draw [style=boldedge] (63.center) to (62.center);
		\draw [style=boldedge] (61.center) to (62.center);
		\draw [style=boldedge] (64.center) to (65.center);
		\draw [style=boldedge] (64.center) to (66.center);
		\draw [style=boldedge] (66.center) to (32.center);
		\draw [style=brace edge] (72.center) to (71.center);
	\end{pgfonlayer}
\end{tikzpicture}